\documentclass[a4paper,10pt]{article}
\usepackage{natbib}
\usepackage{amssymb}
\usepackage{amsmath}
\usepackage[pdftex]{graphicx}
\usepackage[a4paper]{geometry}

\newcommand{\eqnb}[1]{Eq.~(\ref{#1})}						
\newcommand{\sysnb}[1]{Eqs.~(\ref{#1})}						
\newcommand{\fignb}[1]{Fig.~\ref{#1}}						
\newcommand{\mfrac}[2]{\scriptstyle \frac{#1}{#2}}

\newcommand{\colvec}[3]{					
	\begin{pmatrix}
		#1\\
		#2\\
		#3
	\end{pmatrix}
}
\newcommand{\colvecb}[3]{					
	\begin{bmatrix}
		#1\\
		#2\\
		#3
	\end{bmatrix}_{(r,\theta,x)}
}

\newcommand{\as}{\overline{s}}
\newcommand{\asig}{\overline{\sigma}}
\newcommand{\aet}{\overline{\eta}}

\newcommand{\aU}{\overline{U}}
\newcommand{\aR}{\overline{R}}

\newcommand{\eX}{\mathbf{e}_x}
\newcommand{\eY}{\mathbf{e}_y}
\newcommand{\eZ}{\mathbf{e}_z}

\newcommand{\ar}{\overline{\mathbf{r}}}
\newcommand{\ax}{\overline{x}}
\newcommand{\ay}{\overline{y}}
\newcommand{\az}{\overline{z}}
\newcommand{\at}{\overline{\mathbf{t}}}
\newcommand{\an}{\overline{\mathbf{n}}}
\newcommand{\ap}{\overline{\mathbf{p}}}
\newcommand{\am}{\overline{\mathbf{m}}}

\newcommand{\hx}{\hat{x}}
\newcommand{\hy}{\hat{y}}
\newcommand{\hz}{\hat{z}}

\newcommand{\hxt}{\hat{x}_T}
\newcommand{\hyt}{\hat{y}_T}
\newcommand{\hzt}{\hat{z}_T}
\newcommand{\hwt}{\hat{w}_T}

\newcommand{\vth}{\mathbf{e}_\theta}
\newcommand{\vr}{\mathbf{e}_r}
\newcommand{\vx}{\mathbf{e}_x}
\newcommand{\htu}{\hat{u}}
\newcommand{\htv}{\hat{v}}
\newcommand{\hxl}{\hat{x}_L}
\newcommand{\hyl}{\hat{y}_L}
\newcommand{\hzl}{\hat{z}_L}

\newcommand{\hxb}{\hat{x}_B}
\newcommand{\hyb}{\hat{y}_B}

   \title{Matched asymptotic expansions\\ for twisted elastic knots:
    a self-contact problem with non-trivial contact topology}

    \author{N. Clauvelin\thanks{UPMC Univ Paris 06, UMR 7190, Institut Jean Le Rond d'Alembert, F-75005 Paris, France}, B. Audoly and S. Neukirch}

\begin{document}

	\maketitle

    \begin{abstract}
	We derive solutions of the Kirchhoff equations for a knot tied
	on an infinitely long elastic rod subjected to combined
	tension and twist.  We consider the case of simple (trefoil)
	and double (cinquefoil) knots; other knot topologies can be
	investigated similarly.  The rod model is based on Hookean
	elasticity but is geometrically non-linear.  The problem is
	formulated as a non-linear self-contact problem with unknown
	contact regions.  It is solved by means of matched asymptotic
	expansions in the limit of a loose knot.  Without any a priori
	assumption, we derive the topology of the contact set, which
	consists of an interval of contact flanked
	by two isolated points of contacts.  We study the influence of
	the applied twist on the equilibrium.

    \end{abstract}

\section{Introduction}

Knots are found in everyday life, shoe lacing being probably the
most common example. They are also essential in a number of activities
such as climbing and sailing.  In science, knots have long been
studied in the field of mathematics, the main motivation being to
propose a topological classification of the various knot types, see
the review
by~\citet{TaborKlapper-The-dynamics-of-knots-and-curves-part-I-1994}.
Recently, there has been an upsurge of interest in knots in the
biological context: knots form spontaneously in many long polymers
chains such as DNA~\citep{Katritch:Geometry-and-physics-of-knots:1996}
or proteins, and have been tied on biological filaments
\citep{Arai:Tying-a-molecular-knot-with-optical-tweezers:1999}.
Knotted filaments have a lower resistance to tension than unknotted ones and break preferably at the
knot~\citep{Saitta:Influence-of-a-knot-on-the-strength-of-a-polymer-strand:1999,Pieranski:Localization-of-breakage-points-in-knotted-strings:2001}.
Despite a wide range of potential applications, the mechanics of knots
is little advanced. The present paper is an attempt to approach knots
from a mechanical perspective by using a 
well-established model of thin elastic rods.

The problem of finding so-called ideal knot shapes has received
much attention in the past
decade~\citep{Katritch:Geometry-and-physics-of-knots:1996,%
Stasiak-Katritch-EtAl-Ideal-Knots-1998}.  In this geometrical
description of tight knots, a impenetrable tube with constant radius is drawn
around an inextensible curve in Euclidean space and one seeks, for
each knot type, the configurations of the curve such that the radius 
of the tube is maximum.
The case of open knots, where the
curve does not close upon itself, has been studied
by~\citet{Pieranski:Tight-open-knots:2001} in connection with the
breakage of knotted filaments under
tension~\citep{Pieranski:Localization-of-breakage-points-in-knotted-strings:2001}.

To go beyond a purely geometrical description of knots, it is natural
to formulate the problem in the framework of the theory of elasticity.
The case of tight knots, or even of moderately tight knots, leads to a
problem of 3D elasticity with geometrical nonlinearities (finite
rotations), finite strains, and self-contact along an unknown surface:
there is no hope to
derive analytical solutions.  Numerical solution of this problem
raises considerable difficulties too, which have not yet been tackled
to the best of our knowledge.  In the present paper, we study the
limit of \emph{loose} knots, when the total contour length captured in
the knot is much larger than the radius of the filament.  In this
limit, it is possible to use a Cosserat type model and describe the
rod as an inextensible curve embedded with a material frame, obeying
Kirchhoff equations; as we show, the equilibria of open knots can
be solved analytically in this limit.

Self-contact in continuum mechanics, and in the theory of elastic rods
in particular, leads to problems that are both interesting and
difficult.  This comes from the fact that the set of points in contact
is not known in advance --- in fact, not even the topology of this set
is known.  This paper builds up on prior work by
\citet{Von-Der-M:Elastic-knots-in-Euclidean-3-space:1999,%
Schuricht-Mosel-Euler-Lagrange-Equations-for-Nonlinearly-2003}, who
characterizes the smoothness of the contact force in equilibria of
elastic rods, and by
\citet{ColemaSwigon-Theory-of-Supercoiled-Elastic-Rings-2000}, who write down
the Kirchhoff equations for rods in self-contact explicitly,
including the unknown contact force.  These equations have been
solved by numerical continuation
in specific geometries by~\citet{ColemaSwigon-Theory-of-Supercoiled-Elastic-Rings-2000,%
Heijden-Neukirch-EtAl-Instability-and-self-contact-phenomena-2003,%
Neukirch-Extracting-DNA-Twist-Rigidity-2004}.
In these papers, the authors simultaneously solve for the non-linear
Kirchhoff equations and for the unknown contact forces.  In the present
paper, we shown that, \emph{under the same set of assumptions that
warrant applicability of the Kirchhoff equations}, one can in fact
neglect the geometrical nonlinearities in the region of self-contact.  As a
result, nonlinearities and contact can be addressed in well separated
spatial domains.  This brings in an important simplification and, as
the result, we are able for the first time to derive analytical
solutions of a self-contact problem for rods undergoing finite
displacement, exhibiting a non-trivial contact set topology.

Our solution is constructed with matched asymptotic expansions 
with respect to
a small parameter $\epsilon$ which is zero for a perfectly thin rod.
As is done routinely in boundary layer analysis, we use qualitative
reasonings (dimensional analysis) to justify how the various
quantities scale with the small parameter $\epsilon$.  We emphasize
that our final solution is exact and does not involve any other
assumption than the smallness of the parameter $\epsilon$ : it
is \emph{asymptotically} exact.  Our presentation is based on formal
expansions; proofs of convergence are beyond the scope of the present
paper and can hopefully be established in the future.

In the present paper, we consider a knot loaded under mixed tension
$T$ and twist $U$.  In a previous short paper
\citep{Audoly:Elastic-Knots:2007}, we have announced some of the
results reported here, for the case of a purely tensile loading, $U=0$.
In addition to presenting a justification of these results, we
address here the influence of twist on the knot shape.

The outline of the present paper is as follows.  In
Section~\ref{sec:Model}, we introduce the Kirchhoff equations for rods
in equilibrium, including the contact forces relevant for the knotted
geometry; we discuss the equivalent formulation as a minimization
problem with topological constraints.  In
Section~\ref{sec:ZeroThickness}, we consider the case of an elastic
curve with vanishing thickness and show that the region of contact
collapses to a point connecting a circular loop and two straight
tails.  In Section~\ref{sec:SmallThicknessDimensionalAnalysis}, we
carry out the dimensional analysis of the solution with small but
nonzero thickness.  We show that the equilibrium solution is composed
of three types of regions, namely a loop and two tails connected by a
braid.  The scaling of the unknowns with the small dimensionless
parameter $\epsilon$ are identified.  Following the general
methodology of matched asymptotic expansion, we use these scalings to
devise a perturbation scheme of the original equations in
(non-integer) powers of $\epsilon$.  The resulting equations are
written down and solved in the various regions: the tails are solved
in Section~\ref{sec:TailSolution}, the loop in
Section~\ref{sec:LoopSolution}. The solution in the braid region is the
most challenging as this is where contact occurs, and in
Section~\ref{sec:BraidSolution} we obtain a universal solution
describing the shape of the rod in this region.  In Section~\ref{sec:Matching}
we build a global solution by matching the solutions derived
previously in each region.  We obtain a unique equilibrium solution
for any given value of the loading parameters (force and twist).  In
Section~\ref{sec:Experiments}, this theory is validated by
experiments.  Appendix~\ref{app:Topologies} discusses the topology of
the contact set in more details.

\section{Model}
\label{sec:Model}

\begin{figure}[tbp]
    \centering
    \includegraphics[width=.99\columnwidth]{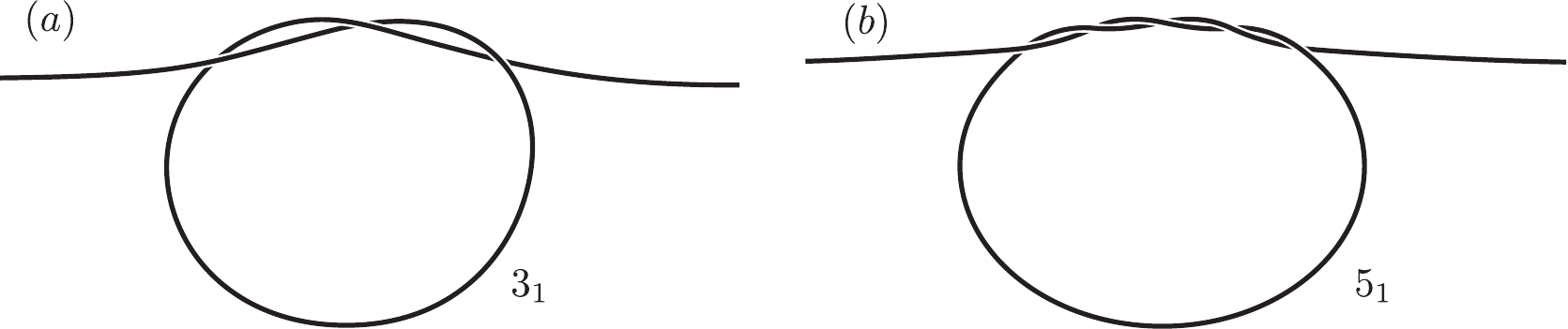}
    \caption{Two knot types are considered here: (a) simple open knot,
    also known as trefoil knot, noted $3_{1}$, and (b) double open
    knot, also known as cinquefoil knot, noted $5_{1}$.  The theory can
    be extended to other knot types.}
    \label{fig:SimpleAndDoubleKnots}
\end{figure}

We seek equilibrium solutions of a thin elastic rod bent into an
open~\footnote{In topology, a knot is defined as a closed, non
self-intersecting curve.  Here we consider curves having two infinite
tails, hence the name `open knots'.} knot with a prescribed type, and
subjected to tensile end force and torsional end moment, as shown on
\fignb{fig:KnotGeometry}.  In the present paper, we focus on two
specific knot types, which are open trefoil knots, also called
\emph{simple} knot and noted $3_{1}$, and open cinquefoil knots, also
called \emph{double} knot and noted $5_{1}$, see
\fignb{fig:SimpleAndDoubleKnots}. 
Other knot types can be handled similarly.  The rod is infinitely long
and the loading is applied at infinity.  

Our model is based on the Kirchhoff equations for the
mechanical equilibrium of elastic rods.  We consider the case of an
unshearable~\footnote{By unshearable, we mean that the rod satisfies
the Navier-Bernoulli kinematical hypothesis. }, inextensible rod with
circular cross-section --- this is the standard model for elastic
rods, which can be derived under fairly general hypotheses from 3D
elasticity theory~\footnote{Extensions of the present results to
different rod models do not raise any fundamental difficulty.}.
Contact of the rod with itself is assumed to be frictionless.  The
mathematical formulation of the problem is based on classical models
and is relatively straightforward; the challenge of the present
analysis is to deal with geometrical nonlinearities and self-contact
--- one of our contributions is to determine the topology of the
contact set which is not known in advance.

In the present section, we recall the Kirchhoff equations for rods 
and show how they can be applied to the geometry considered. 
We emphasize the minimization problem underlying the equations of 
equilibrium, and put the equations in a dimensionless form.

\subsection{Kinematics}
We consider an infinite isotropic elastic rod, bent into an open
knot as shown in \fignb{fig:KnotGeometry}, with a circular
cross section of radius $h$, a bending modulus $B$ and a twisting
modulus $C$.
Centerline of the rod is parameterized by the arc-length $s$ and is
defined by its Cartesian equation,
\begin{equation}
    \mathbf{r}(s) = \left( x(s),y(s),z(s) \right)
    \textrm{,}
    \nonumber
\end{equation}
where the orientation of the axes is specified below.  The tangent to
the centerline is noted
\begin{equation}
    \mathbf{t}(s) = \frac{\mathrm{d}\mathbf{r}}{\mathrm{d}s}
    \textrm{.}
    \label{eq:TangentDefinition}
\end{equation}
Since the rod is assumed inextensible, the tangent is a unit vector,
\begin{equation}
    |\mathbf{t}(s)| = 1
    \label{eqn:UnitTangent}
\end{equation}
for all $s$.  We note $\mathbf{m}(s)$ the
internal moment in the rod and $\mathbf{n}(s)$ the internal force ---
these variables describe stress distribution in the cross-section in
Kirchhoff theory of rods.

\subsection{Constitutive relations}

We assume a linear elastic response (Hookean elasticity), which is
consistent with the small strain approximation underlying Kirchhoff
theory.  The constitutive law for a rod with symmetric (e.  g.
circular) cross-section can be conveniently written in vector 
form~\citep{LandauLifshitz-Theory-of-Elasticity-Course-of-Theoretical-Physics-1981}
:
\begin{equation}
	\mathbf{m}(s)=B\,\mathbf{t}(s)\times\mathbf{t}'(s)
	+C\,\tau(s)\,\mathbf{t}(s),
	\label{eqn:ConstitutiveRelations}
\end{equation}
where $\tau(s)$ is the material twist of the rod.  The above
expression is a condensed form of the constitutive relations for a rod
that are usually written in coordinates in the material frame.  The
first term in the right-hand side is the bending moment and lies in
the cross-section; for a symmetric rod, this bending moment is the
binormal, $\mathbf{t}\times \mathbf{t}'$, times the bending stiffness
$B$.  The second term in the right-hand side is the twisting moment and
is along the tangent: the twisting moment is the material twist, $\tau$,
times the twist stiffness $C$.  Since the rod is considered inextensible
and unshearable, the internal force
is the Lagrange multiplier associated with these kinematical
constraints, and it not given by a constitutive law.

\subsection{Loading}

At the end of the rod corresponding to $s\to+\infty$, a tensile force
$\mathbf{T}$ and a torsional moment $\mathbf{U}$ are applied, see
\fignb{fig:KnotGeometry}.  These two vectors are assumed to be
collinear, and are used to define the axis $z$.  Global mechanical
equilibrium requires that an opposite force $-\mathbf{T}$ and moment
$-\mathbf{U}$ are applied at the other end, $s\to-\infty$.  At
equilibrium, the two long tails of the rod will be aligned with the
direction $z$ of the force. Owing to our choice of axis, we write
\begin{equation}
    \mathbf{T} = T\,\eZ\quad
    \textrm{and}\quad
    \mathbf{U} = U\,\eZ
    \textrm{.}
    \nonumber
\end{equation}
Stability of the long tails require $T>0$ but the twist $U$ can be 
positive or negative.
%

\subsection{Symmetry}
\label{ssec:Symmetry}

Given the symmetry of the loading, we focus~\footnote{Since the
equations are non-linear, one could argue that some solutions having
no symmetry at all could exist, as happens in buckling problems.  We
would miss such solutions since we restrict the analysis to the
symmetric case from the beginning.  This remark applies to the case of
a finite thickness $h$, but to the problem addressed here, which
concerns the limit of a small thickness $h$.  In this limit, we show
that the equations can be linearized; using the principle of linear
superposition, one can focus on symmetric solutions.} on equilibrium
solutions that are symmetric.  More accurately, we assume that the
knotted rod is invariant by rotation with angle $\pi$ about an axis
perpendicular to the axis $z$ defined by the loading.  This is
consistent as the endpoints at infinity are swapped by this
transformation, and so the loading is globally invariant under this
transformation~\footnote{Note that the knot is not invariant
by a reflection
with respect to a plane perpendicular to the $z$ axis: this reflection
leaves the loading globally invariant but changes the knot
type, turning a left-handed knot into a right-handed one.}.

\begin{figure}
	\begin{center}
	\includegraphics[width=0.7\columnwidth]{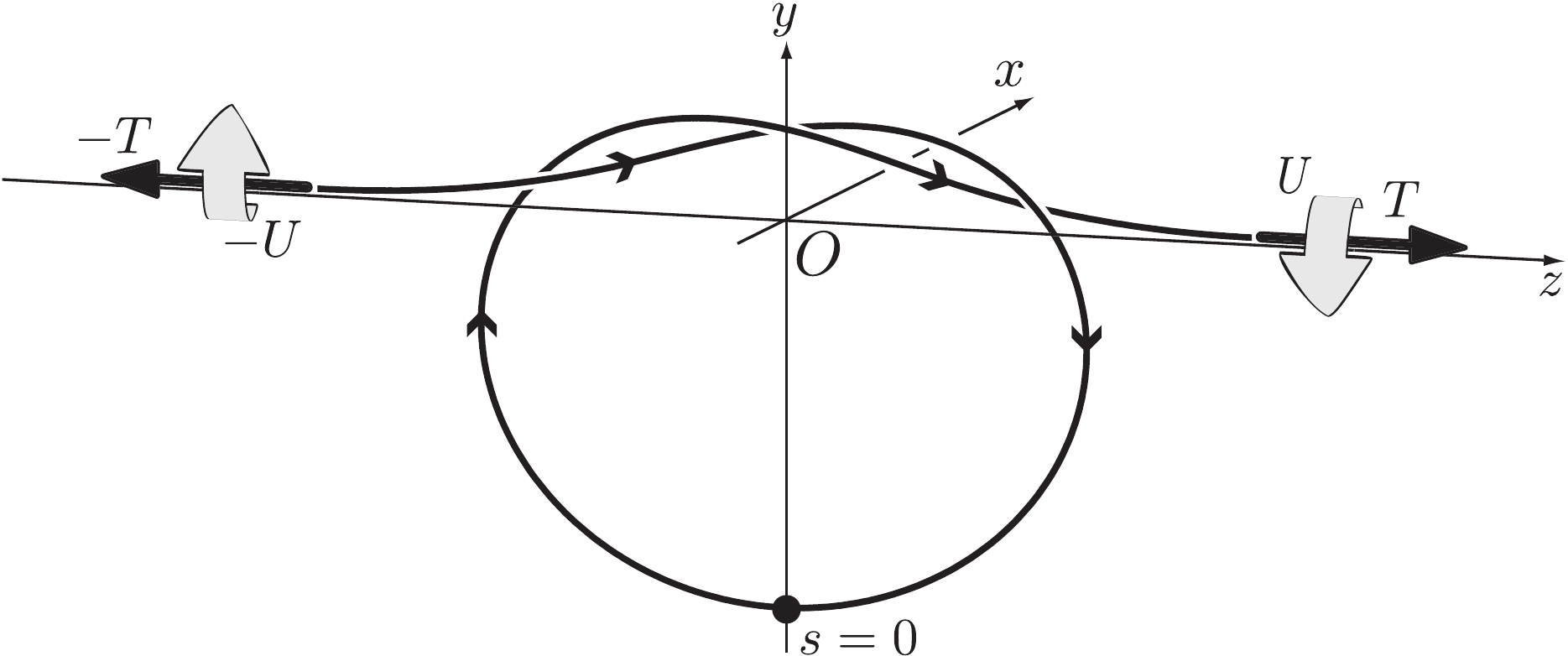}
	\caption{An infinitely long rod is bent into a knot with a
	given type, here a trefoil knot ($3_{1}$), and loaded with
	combined twist $U$ and axial force $T$.  In this paper, we
	derive equilibrium solutions for this non-linear self-contact
	problem.}
	\label{fig:KnotGeometry}
	\end{center}
\end{figure}

Let us call $y$ the axis defining this symmetry by rotation with an
angle $\pi$.  The intersection of the perpendicular axes $z$ and $y$
defined so far will be the origin $O$ of our Cartesian coordinates.
The direction perpendicular to $y$ and $z$ defines the third axis $x$,
in such a way that $(x,y,z)$ is direct and orthonormal.  Intersection
of the axis of symmetry $y$ with the centerline defines what can be
called the midpoint of the rod --- intuitively, this is the bottom of
the loop in \fignb{fig:KnotGeometry}.  This midpoint is taken as the
origin of the arc-length coordinate, $s=0$.  With this convention, the
symmetry by rotation about $y$ with angle $\pi$ maps a point on the
centerline with coordinate $s$ onto the point with opposite
coordinate $(-s)$. Using this property, it is sufficient to find the 
equilibrium shape of the rod over one half, say the positive half 
$0\leq s< +\infty$: the other half can be found by applying the 
symmetry.

\subsection{Variational formulation, constraints}

The equilibrium shape of the knotted rod can be found be solving a
minimization problem: this equilibrium shape is a minimizer of the
total energy of the rod (potential energy associated with loading at
endpoints plus elastic energy) under the combined constraints of
inextensibility, non-penetration and prescribed knot topology. This 
variational view of the problem will be useful later for solving the 
braid region in Section~\ref{sec:BraidSolution}.

The total energy of the rod is defined as:
\begin{equation}
	E=\int_{-\infty}^{+\infty}\left(\frac{B}{2}\,\kappa^2 +
	\frac{C}{2}\,\tau^2 \right)ds +T\,D_\infty-U\,R_\infty ,
	\label{eqn:TotalEnergy}
\end{equation}
where $\kappa$ and $\tau$ stand for the curvature and the twist of the
rod. The integral term in the right-hand side is the elastic energy 
associated with the constitutive 
law~(\ref{eqn:ConstitutiveRelations}). The bending term depends on 
the scalar curvature
\begin{equation}
    \kappa = |\mathbf{t}'(s)|
    \textrm{.}
    \label{eqn:ScalarCurvatureDefinition}
\end{equation}
The last two terms in equation~(\ref{eqn:TotalEnergy}) represent the
work of the applied tensile force $T\eZ$, related to the end-to-end
shortening $D_{\infty}$, and of the applied torsional moment $U\eZ$,
related to the relative rotation $R_{\infty}$ of the ends.

The minimization of this energy is subjected to a series of
constraints.  First, the inextensibility constraint is expressed by
equation~(\ref{eqn:UnitTangent}).  Second, the topology of the knot is
prescribed (this topological constraint cannot be written down easily
in the general case; it will be shown to impose the value of a winding
index in the braid when we focus on loose knots later on).  Third and
lastly, one has to consider the non-penetration constraint which can
be expressed as:
\begin{equation}
	\left| \mathbf{r}(s_1)-\mathbf{r}(s_2) \right| \geq 2\,h,
	\label{eqn:PenetrationConstraint}
\end{equation}
for any $s_1$ and $s_2$ such that $\left|s_1-s_2\right|>4h$.  Note
that the radius of the rod $h$ enters in the equation at this point in
the right-hand side of equation~(\ref{eqn:PenetrationConstraint}).
The trick of restricting the penetration test to couples of points
$(s_{1},s_{2})$ separated by a curvilinear distance greater than
$4\,h$ is due to
\cite{Von-Der-M:Elastic-knots-in-Euclidean-3-space:1999}, and avoids
mistaking close neighbors on the centerline for points violating the
non-penetration condition --- it is given for mathematical consistency
but is not needed in the following: for the problem we consider, we
know \emph{a priori} that the arc-length separation of two points in
contacts is large, namely of order $2\pi\,R$ where the radius $R$ of
the loop is a known quantity of order 1.

\subsection{Equilibrium: Kirchhoff equations}

The equilibrium equations for a rod can be derived from the
energy~(\ref{eqn:TotalEnergy}) by the Euler-Lagrange
method~\citep{Bourgat-Tallec-EtAl-Modelisation-et-calcul-des-grands-1988,%
Steigmann-Faulkner-Variational-theory-for-spatial-1993}.
This leads to the following equations:
\begin{subequations}    
    \label{sys:KirchhoffEquations}
\begin{gather}
    \mathbf{r}'(s)  = \mathbf{t}(s)
    \label{eqn:KirchhoffEquations-Tangent} \\
    \mathbf{t}'(s) = \frac{\mathbf{m}(s)}{B}\times \mathbf{t}(s) 
    \label{eqn:KirchhoffEquations-Constit} \\
    \mathbf{m}'(s) + \mathbf{t}(s)\times \mathbf{n}(s) = \mathbf{0}
    \label{eqn:KirchhoffEquations-MomentsEquil} \\
    \mathbf{n}'(s) + \mathbf{p}(s) = \mathbf{0}
    \label{eqn:KirchhoffEquations-ForcesEquil}
\end{gather}
\end{subequations}
where primes denote derivation with respect to arc-length $s$.
The first equation is the definition of the tangent, already
encountered in equation~(\ref{eq:TangentDefinition}).  The second
equation combines the constitutive
relations~(\ref{eqn:ConstitutiveRelations}) with the 
definition~(\ref{eqn:ScalarCurvatureDefinition}) of 
curvature.
The last two equations express the equilibrium of moments and forces 
on an infinitesimal rod element, and are known as the Kirchhoff
equations~\citep{LandauLifshitz-Theory-of-Elasticity-Course-of-Theoretical-Physics-1981}.
The vector $\mathbf{p}(s)$ is the density of distributed force per
unit length applied on the rod, sometimes referred to as the contact
pressure.  In the present problem, gravity is neglected and the only
force $\mathbf{p}(s)$ to be considered is the one arising from the
contact pressure in the regions of contact --- if there is no contact,
$\mathbf{p}(s) = \mathbf{0}$.

\subsection{Contact set, contact force}

Let us define the contact set as the set of
couples of arc-lengths, $(s_{1},s_{2})$, defining
cross-sections that are in contact:
\begin{equation}
    \mathfrak{C}=\big\{(s_1,s_2) \quad\textrm{such that} \quad \left|
    s_1-s_2\right| > 4h\textrm{ and }\left|
    \mathbf{r}(s_1)-\mathbf{r}(s_2)\right| = 2\,h\big\}
    \textrm{,}
    \label{eq:ContactSetDefinition}
\end{equation}
where the first inequality, $\left| s_1-s_2\right| > 4h$, is to avoid
mistaking close neighbors for regions of penetrations, as explained
earlier, and the second inequality $\left|
\mathbf{r}(s_1)-\mathbf{r}(s_2)\right| = 2\,h$ is the contact criterion.

We are touching here the main challenge of self-contact problems: the
profile of the contact pressure $\mathbf{p}(s)$ is required to compute
the centerline by integration of the Kirchhoff equations, but it
depends itself non-linearly on the geometry of contact, that is on the
shape of the centerline.  In other words, $\mathbf{p}(s)$ and the
contact set $\mathfrak{C}$ must be determined in a self-consistent way
but none is known \emph{a priori}.
In particular, the topology of the contact set is not known in
advance.  It will be obtained later as an outcome of our calculations.

For any couple $(s_1,s_2)$ in the set $\mathfrak{C}$, the
corresponding cross-section are in contact.  By the action-reaction
principle, we have $\mathbf{p}(s_1)=-\mathbf{p}(s_2)$.  In addition
we assume frictionless contact: the force $\mathbf{p}$ has
to be normal to the rod surface, and so is aligned with $(\mathbf{r}(s_{2}) 
- \mathbf{r}(s_{1}))$.  This implies that the contact force is aligned
with the vector joining the points $\mathbf{r}(s_{1})$ and
$\mathbf{r}(s_{2})$:
\begin{equation}
	\mathbf{p}(s_1)=p(s_1)\frac{\left(\mathbf{r}(s_1)-\mathbf{r}(s_2)\right)}{2h}=p(s_1)
	\colvec{(x(s_1)-x(s_2))/(2h)}{(y(s_1)-y(s_2))/(2h)}{(z(s_1)-z(s_2))/(2h)}
	\textrm{,}
	\label{eqn:ContactPressureEquations}
\end{equation}
for $(s_{1},s_{2})\in\mathfrak{C}$.  In this equation we have
introduced the scalar contact pressure $p(s)$; since the rod is a 1D
object, the contact force has the dimension of a force per unit
\emph{length} but we shall nevertheless call it a contact pressure.
For the solution to be physical, the pressure must be
positive:
\begin{equation}
    p(s) \geq 0
    \textrm{.}
    \label{eq:ContactPressureIsPositive}
\end{equation}
In terms of the scalar contact pressure, the action-reaction principle 
can be rewritten as
\begin{equation}
    p(s_1)=p(s_2)
    \textrm{.}
    \label{eqn:ContactForceActionReaction}
\end{equation}

\subsection{Boundary conditions}

Thanks to the symmetry introduced in Section~\ref{ssec:Symmetry}, the
equations of equilibrium~(\ref{sys:KirchhoffEquations}) need be solved
over half the rod only, that is for $0\leq s<+\infty$.  These
equations form a boundary value problem as there are conditions to
be satisfied at both ends of the interval. The following conditions 
must be satisfied at the endpoint $s=+\infty$:
\begin{subequations}
    \label{sys:BoundaryConditions}
\begin{gather}
    \mathbf{m}(+\infty) = U\,\eZ
    \textrm{,} \label{eqn:BoundaryConditionsM}\\
    \mathbf{n}(+\infty) = T\,\eZ
    \textrm{,} \label{eqn:BoundaryConditionsN}\\
    \mathbf{t}(+\infty) =\eZ
    \textrm{,} \label{eqn:BoundaryConditionsT}\\
    \mathbf{r}(+\infty)\times\mathbf{t}(+\infty) =\mathbf{0}
    \textrm{.} \label{eqn:BoundaryConditionsR}
\end{gather}
\end{subequations}
The asymptotic conditions holding at the opposite end of
the rod, $s\to -\infty$, can be found by symmetry.  The first two
asymptotic conditions above enforce the loading applied at infinity.
Condition~(\ref{eqn:BoundaryConditionsT}) imposes that the tangent is
asymptotically aligned with the applied force, which is an obvious
necessary condition for minimizing the energy.  The last
condition~(\ref{eqn:BoundaryConditionsR}) defines the $z$ axis as the
asymptote of the centerline far away from the knot --- without this
convention there would be infinitely many solutions, corresponding to
the invariance of the system under rigid-body translations
perpendicular to the $z$ axis.

The boundary conditions at the midpoint $s=0$ of the rod derive from  
the invariance of the solution by a rotation of angle $\pi$ about the 
$y$ axis:
\begin{subequations}
    \label{sys:SymmetryLoopBottom}
    \begin{align}
	\mathbf{t}(0)\cdot\eY & = 0 \label{eq:SymmetryLoopBottom-T}\\
	\mathbf{m}(0)\cdot\eY & = 0\\
	\mathbf{n}(0)\cdot\eY & = 0\\
	\mathbf{r}(0)\times \eY & = \mathbf{0}
    \end{align}
\end{subequations}
The justification for each of these three equations is similar, and
will be given here for the first one only.  Let us write the Cartesian
coordinates of the tangent $\mathbf{t}(0)$ at midpoint as
$\mathbf{t}(0)= (t_{0}^x, t_{0}^y, t_{0}^z)$.  According to our
symmetry assumption, rotating the system by an angle $\pi$ about the
$y$ axis is equivalent to reversing the orientation of the centerline.
The first operation changes the tangent to $(-t_{0}^x, +t_{0}^y,
-t_{0}^z)$, while the second one changes it to its opposite,
$(-t_{0}^x, -t_{0}^y, -t_{0}^z)$.  Equality of these two vectors
imposes $t^y_{0} = 0$, which yields
equation~(\ref{eq:SymmetryLoopBottom-T}).

\subsection{Invariants}

Due to their variational nature, the equilibrium
equations~(\ref{sys:KirchhoffEquations}) are associated with several
invariants as discussed
by~\cite{Maddocks:Conservation-laws-in-the-dynamics-of-rods:1994}.
These invariants are known to be conserved in the absence of
distributed force, $\mathbf{p}=0$.  In the present case, the 
distributed
force can be nonzero but remains everywhere perpendicular to the 
tangent $\underline{t}(s)$.
Under this assumption, it is straightforward to check that the
following expressions are still invariants:
\begin{equation}
    I_{1} = \mathbf{m}(s)\cdot \mathbf{t}(s)\quad
    \textrm{and}
    \quad
    I_{2} = \frac{|\mathbf{m}(s)|^2}{2\,B}+\mathbf{n}(s)\cdot\mathbf{t}(s)
    \textrm{.}
    \label{eqn:TwoInvariants}
\end{equation}
The first invariant is directly proportional to the material twist
$\tau(s) = I_{1}/C$, and is known to be uniform for a rod with
symmetric cross-section in equilibrium in the absence of distributed
torques.  The famous Kirchhoff analogy identifies the equations of
equilibrium of a symmetric rods to the equations of motion of a table
top subjected to gravity.  According to this analogy, the first
invariant expresses conservation of the angular moment about the axis
of the top; the second invariant expresses conservation of the energy
of the top.

The constant values of these invariants are imposed by the boundary 
conditions~(\ref{sys:BoundaryConditions}):
\begin{subequations}
    \label{eqns:Invariants}
    \begin{equation}
	    I_1 = U = C \tau(s)
	    \quad\textrm{and therefore }
	    \tau(s)=\frac{U}{C}
	    \textrm{,}
	    \label{eqn:TwistInvariant}
    \end{equation}
    and
    \begin{equation}
	    I_2 =\frac{U^{2}}{2\,B}+T
	    \textrm{.}
	    \label{eqn:ForceInvariant}
    \end{equation}
\end{subequations}
Note that conservation of these invariants requires frictionless 
contact.

\subsection{Dimensionless form}
\label{ssec:DimensionlessEquations}

The problem has been formulated so far in terms of the loading
parameters $T$ and $U$, of the thickness $h$, and of the elastic
stiffnesses $B$ and $C$.  In this section, we use dimensional analysis
and rewrite the equations in a form that depends on two dimensionless
parameters, $\aU$ and $\epsilon$, only. 

To this end, we introduce the
characteristic length $L^\star$, force $F^\star$ and moment $M^\star$ as follows:
\begin{equation}
    L^\star = \sqrt{\frac{B}{T}},
    \qquad
    F^\star = T,
    \qquad
    M^\star = \sqrt{B\,T}
    \textrm{.}
    \label{eqn:CharacteristicQuantities}
\end{equation}
These quantities are used to define dimensionless variables, noted 
with a bar. For instance, we define the dimensionless arc-length 
$\as$
and position $\ar$ as
\begin{equation}
    \as = \frac{s}{L^\star} = s\,\sqrt{\frac{T}{B}},
    \qquad
    \ar(\as) = 
    \frac{\mathbf{r}(\as\,L^\star)}{L^\star} = 
    \mathbf{r}(s)\,\sqrt{\frac{T}{B}}
    \textrm{.}
    \label{eqn:RescaledLengths}
\end{equation}
Note that rescaled functions, such as $\ar$, are always considered to
be a function of a rescaled argument, here $\as$ and not $s$: a prime 
on a barred 
function implies that derivation is with respect to the rescaled 
arc-length. For instance, the rescaled tangent is defined by
\begin{equation}
    \at(\as) = \ar' (\as)= \frac{\mathrm{d}\ar}{\mathrm{d}\as} = 
    \frac{\mathrm{d}(\mathbf{r}/L^\star)}{\mathrm{d}(s/L^\star)} = 
    \mathbf{r}'(s) = \mathbf{t}(s) =\mathbf{t}(\as\,L^\star)
    \textrm{,}
    \nonumber
\end{equation}
and happens to be the same unit vector as the physical tangent 
$\mathbf{t}$, evaluated at the corresponding point $s = \as\,L^\star$. 
Similarly, the rescaled curvature is defined as
$\overline{\kappa} = |\at'|$:
\begin{equation}
	\overline{\kappa}(\as)=\frac{\kappa(\as\,L^\star)}{1/L^\star}
	\textrm{.}
	\label{eqn:RescaledCurvature}
\end{equation}

The rescaled internal moment and torsional couple 
are:
\begin{equation}
    \am(\as) = \frac{\mathbf{m}(\as\,L^\star)}{M^\star} = 
    \frac{\mathbf{m}(\as\,L^\star)}{\sqrt{B\,T}}
    \quad
    \textrm{and}
    \quad
    \aU = \frac{U}{M^\star} = \frac{U}{\sqrt{B\,T}}
    \textrm{.}
    \label{eqn:RescaledMoments}
\end{equation}

The internal force $\mathbf{n}$ is naturally rescaled using the
typical force $F^\star = T$, while the contact force per unit length,
$\mathbf{p}$, is rescaled using the dimension $F^{\star}/L^\star$:
\begin{equation}
    \an(\as)=\frac{\mathbf{n}(\as\,L^\star)}{T}
    \quad\mathrm{and}
    \quad\ap(\as)=\frac{\mathbf{p}(\as\,L^\star)}{T}\,\sqrt{\frac{B}{T}}
    \textrm{.}
    \label{eqn:RescaledForces}
\end{equation}

Having defined the rescaled form of the various quantities, we 
proceed to
rewrite the equations of the problem in dimensionless form.  We start
with the constitutive relation~(\ref{eqn:ConstitutiveRelations}):
\begin{equation}
    \am(\as) = \at(\as)\times\at'(\as) + \aU\,\at(\as)
    \textrm{.}
    \label{eqn:RescaledConstitutiveRelations}
\end{equation}
The kinematical relations and equilibrium
equations~(\ref{sys:KirchhoffEquations}) write:
\begin{subequations}
    \label{sys:RescaledKirchhoffEquations}
    \begin{gather}
	\ar'(\as) = \mathbf{t}(\as), 
	\label{eqn:RescaledKirchhoffEquationsR}\\
	\mathbf{\at}'(\as) = \am(\as)\times\mathbf{t}(\as),
	\label{eqn:RescaledKirchhoffEquationsT}\\
	\am'(\as)+ \mathbf{t}(\as)\times\an(\as) = \mathbf{0},
	\label{eqn:RescaledKirchhoffEquationsM}\\
	\an'(\as)+ \ap(\as) = \mathbf{0}
	\textrm{.}
	\label{eqn:RescaledKirchhoffEquationsN} 
    \end{gather}
\end{subequations}
For the asymptotic conditions~(\ref{sys:BoundaryConditions}) we obtain:
\begin{subequations}
    \label{sys:RescaledBoundaryConditions}
    \begin{gather}
	\am(+\infty)=\aU\,\eZ,
	\label{eqn:RescaledBoundaryConditionsM}\\
	\an(+\infty)=\eZ,
	\label{eqn:RescaledBoundaryConditionsN}\\
	\at(+\infty)=\eZ,
	\label{eqn:RescaledBoundaryConditionsT}\\
	\ar(+\infty)\times\at(+\infty)=\mathbf{0}
	,
	\label{eqn:RescaledBoundaryConditionsR}
    \end{gather}
\end{subequations}
while the midpoint conditions take the same form as the original
expressions~(\ref{sys:SymmetryLoopBottom}), with original variables
replaced by barred ones:
\begin{subequations}
    \label{sys:RescaledSymmetryLoopBottom}
    \begin{align}
	\at(0)\cdot\eY & = 0\textrm{,}\\
	\am(0)\cdot\eY & = 0\textrm{,}\\
 	\an(0)\cdot\eY & = 0.\textrm{}\\
	\ar(0)\times \eY & = \mathbf{0}
    \end{align}
\end{subequations}
In rescaled form, the
invariants~(\ref{eqns:Invariants}) read:
\begin{subequations}
    \label{sys:RescaledInvariants}
    \begin{align}
	\overline{I}_1 & = \am(\as)\cdot\at(\as) = \aU ,
	\label{eqn:RescaledTwistInvariant}\\
	\overline{I}_2 & = \frac{\am^2(\as)}{2} +
	\an(\as)\cdot\at(\as) = \frac{\aU^2}{2}+1 \textrm{.}
	\label{eqn:RescaledForceInvariant} 
    \end{align}
\end{subequations}

We now turn to the non-penetration constraint.  In terms of $\ar =
\mathbf{r}/L^\star$, \eqnb{eqn:PenetrationConstraint} can be rewritten as
\begin{equation}
    | \ar(\as_1)-\ar(\as_2) | \geq 2\,\frac{h}{L^\star}
    \nonumber
\end{equation}
In the right-hand side a fundamental dimensionless number of the problem
has appeared, namely the ratio of the rod thickness to the
characteristic length built from the traction force $T$ and the
bending stiffness $B$ of the rod.  It will be convenient to deal with
this dimensionless number using an auxiliary number $\epsilon$
defined as
\begin{equation}
    \epsilon = 2^{1/4} \,\sqrt{\frac{h}{L^{\star}}} = 
    \left(\frac{2\,h^2\,T}{B}\right)^{1/4}
    \textrm{.}
    \label{eqn:EpsilonDefinition}
\end{equation}
This details of the present definition of $\epsilon$ will be motivated
later in Eq.~(\ref{eqn:EpsilonDefinitionInTermsOfHR}). In terms of 
$\epsilon$, the non-penetration condition can be written as
\begin{equation}
   \big(\ar(\as_1)-\ar(\as_2) \big)^2 \geq 2\,\epsilon^4
    \label{eqn:RescaledPenetrationConstraint}
\end{equation}

In equations~(\ref{eqn:RescaledConstitutiveRelations})
to~(\ref{eqn:RescaledPenetrationConstraint}), we have rewritten all
the equations of the problem in terms of two dimensionless parameters
only, $\aU$ and $\epsilon$, defined in
Eqs.~(\ref{eqn:RescaledMoments}) and~(\ref{eqn:EpsilonDefinition})
respectively.  The first parameter $\aU$ is the rescaled torsional
moment; the second parameter $\epsilon$ is the aspect-ratio of the
rod.  In this paper, we focus on the limit of thin rod, or a loose
knot, $\epsilon\to 0$, and build an asymptotic solution of the set of
equations above, for arbitrary values of $\aU$.

\section{Zero radius solution}
\label{sec:ZeroThickness}

Before we build a solution of the equations for small $\epsilon$, it
is useful to consider the limit of an infinitely thin rod, $h=0$, that
is the problem of tying a knot on an elastic \emph{curve}.  This case
corresponds to $\epsilon=0$, and is the subject of the present
section. 

As we shall show later, the limit $\epsilon\to 0$ is singular: the
contact region has nonzero length for $\epsilon>0$ but shrinks to a
point in the limit $\epsilon=0$.  It follows that the solution of the
limit problem ($\epsilon = 0$) is less regular than the solutions for
$\epsilon>0$ and must be defined in a weak sense.  In addition, the
topological constraint on the knot type is not easy to write down for
the limit problem.  We shall not study convergence for $\epsilon\to 0$
in a mathematically rigorous way.  Following a pragmatic approach, a
solution for $\epsilon = 0$ is constructed in the present Section
based on some assumptions (planarity, existence of a point-like
contact).  These assumptions will be validated later when we show that
it is possible to extend this $\epsilon=0$ solution into a family of
smooth solutions indexed by $\epsilon>0$.

\subsection{Explicit solution}

\begin{figure}
	\begin{center}
	\includegraphics[width=0.6\columnwidth]{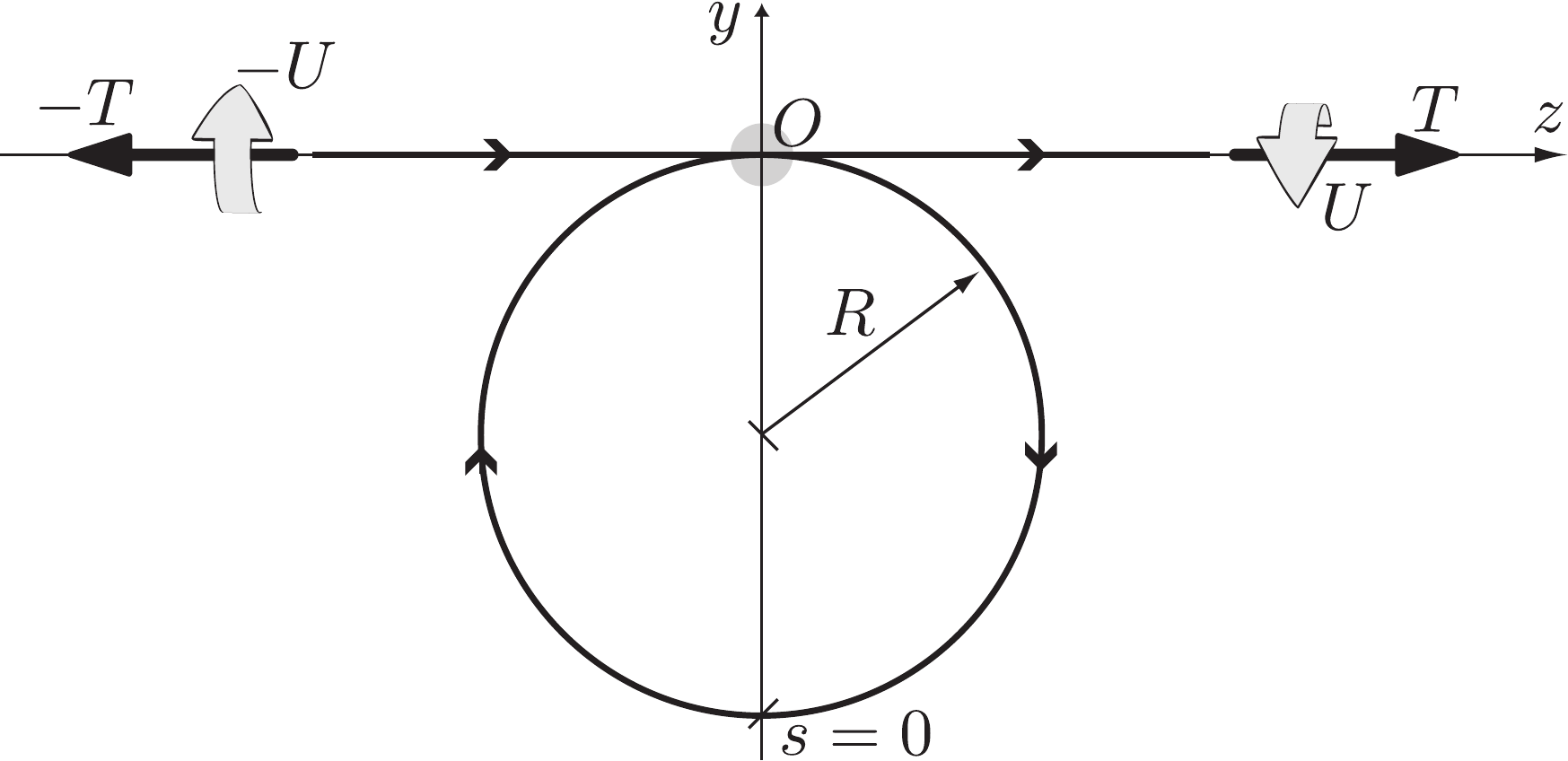}
	\caption{Case of zero thickness, $h=0$.  The equilibrium
	solution is planar and made up of two semi-infinite straight
	tails connected to a perfectly circular loop with radius $R$. 
	The top of the loop is connected to the tails across a singular 
	point $O$, shown in gray, where both the internal force and 
	moment are discontinuous.}
	\label{fig:ZeroRadiusSolution}
	\end{center}
\end{figure}

In the limit of zero thickness $h=0$, we consider a solution made up
of a circular loop connected to two straight, semi-infinite tails.
Owing to the assumed symmetry of the solution, we focus on one half
the rod and consider a half-circle starting from the midpoint,
connected to a single straight, semi-infinite tail.  There is no
contact, except at the singular point $O$ where the loop and tail
merge, see \fignb{fig:ZeroRadiusSolution}.  By our previous definition
of axes, this point $O$ is the origin of the Cartesian frame, the loop
is contained in the $(y,z)$ plane and the tails lie along the $z$
axis.

Any quantity pertaining to the limit of a zero thickness, $h=0$ or
$\epsilon=0$, introduced in the present Section will be denoted with a
superscript `$0$'.  A subscript `$L$' refers to the loop region, and
`$T$' to the tail region.  Let $R$ be the radius of the loop, which
will be given in Eq.~(\ref{eqn:AraiRadius}), and $\aR = R/L^\star$ the
rescaled radius.  The loop region is given by the classical circular
solution of the Kirchhoff equations.  The centerline is a circle given
in parametric equation:
\begin{subequations}
    \label{sys:LoopZeroRadius}
    \begin{align}
	\ar^0_L(\as) &= \left( 0,
	\,-\left( \aR +\aR\,\cos(\as/\aR)\right), \,-\aR\,\sin(\as/\aR)
	\right), \label{eqn:LoopZeroRadiusR} \\
	\at^0_L(\as) &= \left( 0, \,\sin(\as/\aR), \,-\cos(\as/\aR)
	\right). \label{eqn:LoopZeroRadiusT}
    \end{align}
Note that the constants of integration are such that the top of the
loop, $\as = \pi\,\aR$ is at the origin:
$\ar^0_L(\pi\,\aR)=\mathbf{0}$.  The tangent at midpoint is
$\at^0_L(0)=-\eZ$, and that at the top is $\at^0_L(\pi\,\aR)=+\eZ$.
The internal force and moment in the loop can be found by plugging
these expressions into
equations~(\ref{eqn:RescaledConstitutiveRelations}--%
\ref{sys:RescaledSymmetryLoopBottom}) with $\ap = \mathbf{0}$:
	\begin{align}
	    \am^0_L(\as) &= \left( 1/\aR, \,\aU\,\sin(\as/\aR),
	    \,-\aU\,\cos(\as/\aR) \right),
	    \label{eqn:LoopZeroRadiusM}\\
	    \an^0_L(\as) &= ( \aU/\aR,\,0,\,0 )
	    \textrm{.}
	    \label{eqn:LoopZeroRadiusN}
	\end{align}    
\end{subequations}
These solutions describe the loop region, $-\pi\,\aR \leq \as \leq 
\pi\,\aR$.

The solution in the tail is even simpler and describes a straight rod
under combined axial tension and twist:
\begin{subequations}
    \label{sys:TailsZeroRadius}
    \begin{align}
	\ar^0_T(\as)&=(\as-\pi\aR)\eZ,
	\label{eqn:TailsZeroRadiusR} \\
	\at^0_T(\as)&=\eZ, \label{eqn:TailsZeroRadiusT} \\
	\am^0_T(\as)&=\aU\,\eZ, \label{eqn:TailsZeroRadiusM} \\
	\an^0_T(\as)&=\eZ, \label{eqn:TailsZeroRadiusN}
    \end{align}
\end{subequations}
these expressions being applicable for $\as \geq \pi\,\aR$.

\subsection{Singular braid point}

At the point connecting the loop and tail regions, $\as=\pi\,\aR$, the
solution is discontinuous.  Across this point, the internal force
jumps from $\an_{L}^0(\pi\,\aR) = \aU\,\eX/\aR$ to
$\an_{T}^0(\pi\,\aR)=\eZ$, while the bending moment, defined as the
cross-sectional projection of $\am$, drops from $\eX/\aR$ to
$\mathbf{0}$.  These discontinuities point to the presence of contact
forces in this region, and will be explained by our analysis of the
braid region for finite $\epsilon$, see
Section~\ref{sec:BraidSolution}.
%

For this solution to be complete, there remains to compute the radius
$R$ of the loop.  This can be done by writing the conservation of the
second invariant given by \eqnb{eqn:RescaledForceInvariant} across the
singular point:
\begin{equation}
	\frac{1}{2}\,\left(\frac{1}{\aR^2}+\aU^2\right)
	=
	\frac{\aU^2}{2}+1
	\textrm{,} \nonumber
\end{equation}
where the left-hand side comes from the loop region and the 
right-hand side from the tail. This implies
\begin{equation}
    \aR = \frac{1}{\sqrt{2}}
    \textrm{,}\quad\textrm{that is }
    R=\sqrt{\frac{B}{2\,T}}
    \textrm{.}
    \label{eqn:AraiRadius}
\end{equation}
This result was previously obtained by
\cite{Arai:Tying-a-molecular-knot-with-optical-tweezers:1999} based on
energy minimization of the energy~(\ref{eqn:TotalEnergy}) with respect
to $R$.

\section{Perturbation scheme}
\label{sec:SmallThicknessDimensionalAnalysis}

In Section~\ref{ssec:DimensionlessEquations}, the equilibrium of a
knotted rod has been written as a system of coupled, non-linear,
ordinary differential equations depending on two dimensionless
parameters, $\aU$ and $\epsilon$.  In this paper, we consider the
limit of a small $\epsilon$, $\epsilon\ll 1$.  This limit is in fact
the only one consistent with Kirchhoff (or Cosserat) description of
the rod as a 1D elastic object.  Indeed, Kirchhoff theory comes from a
reduction of 3D elasticity, and is justified when the thickness $h$ is
much smaller than the typical radius of curvature of the centerline.
This typical radius of curvature is $L^\star$, meaning that Kirchhoff
approximation makes sense in the limit $h\ll L^\star$, that is
$\epsilon\ll 1$. 
The opposite limit of a perfectly tight knot defines a geometrical 
problem which has extensively been 
studied, see~\cite{Pieranski:Tight-open-knots:2001,%
Katritch:Geometry-and-physics-of-knots:1996,%
Cantarell:Visualizing-the-tightening-of-knots:2005}.

The limit $\epsilon \ll 1$ under consideration corresponds to a rod
whose radius $h$ becomes infinitely small while its elastic moduli are
kept constant, or equivalently to the case of a fixed radius $h$ and
elastic moduli when the pulling force becomes very small.  We refer to
this limit generically as the limit of a \emph{loose knot}.  Our
somewhat arbitrary definition of the perturbation parameter $\epsilon$
in Eq.~(\ref{eqn:EpsilonDefinition}) has in fact been motivated by the
simple relation
\begin{equation}
	\epsilon=\sqrt{\frac{h}{R}},
	\label{eqn:EpsilonDefinitionInTermsOfHR}
\end{equation}
where $h$ is the rod thickness and $R$ the loop radius defined in
\eqnb{eqn:AraiRadius}. The limit of a loose knot corresponds to 
$\epsilon\to 0$.

\begin{figure}
	\begin{center}
	\includegraphics{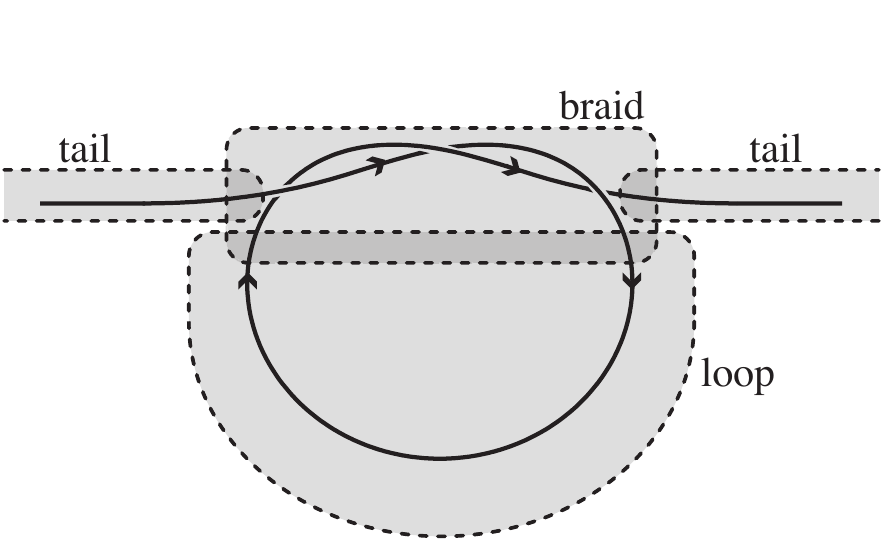}
	\caption{In the limit of a loose knot considered here,
	$\epsilon\ll 1$, the equilibrium solution can be decomposed
	into three domains: an almost circular loop, two almost 
	straight tails, and a braid region where self-contact takes 
	place. In the vocabulary of asymptotic analysis, the braid 
	region is an inner layer, with typical length $\ell$ much 
	smaller than the typical size $R$ of the loop and tail 
	regions (outer layers). Note the existence of 
	so-called intermediate region, in darker gray, at the overlap 
	between braid and tails, and between loop and braid.}
	\label{fig:KnotDomains}
	\end{center}
\end{figure}

In Section~\ref{sec:ZeroThickness}, we introduced a solution
corresponding to the limit $h=0$, that is to $\epsilon=0$.  This
solution features a singular point where some contact occurs.  One of
the main contributions of the present paper is to come up with a
detailed description of this contact region, called the braid later
on, for small but finite $h$.  A key remark, formulated
by~\cite{Gallotti:Stiff-knots:2007}, is that the contact region
remains very localized for small $\epsilon$.  Together with the
explicit solution for $h=0$ given in Section~\ref{sec:ZeroThickness},
this suggests the decomposition of the knot solution in three domains
shown in \fignb{fig:KnotDomains}: a quasi-circular loop, two
quasi-rectilinear tails and a braid region in between.
We shall now study the orders of magnitudes of the displacement
relevant to these different regions. This is an important preliminary 
step for the quantitative analysis presented in the following sections.
A simple scaling argument, given in our preliminary
paper~\citep{Audoly:Elastic-Knots:2007}, shows that the size $\ell$ of the contact
region is of order of the geometric mean of the loop radius $R$ and
the rod thickness $h$, which we write
\begin{equation}
    \ell \sim \sqrt{h\,R}
    \textrm{.}
    \nonumber
\end{equation}
This defines an intermediate length scale, between the `large' length 
$R$ and the small length $h$. To justify this scaling, we note that 
the transverse displacement in the braid is fixed by contact and is 
of order $h$; over a typical length $\ell$, this yields a typical 
curvature $h/\ell^2$. At the exit of the braid, this curvature has to 
be matched with that in the loop, of order $1/R$. Balancing $h/\ell^2$ 
with $1/R$ yields $\ell \sim \sqrt{h\,R}$ as proposed above.

By this argument the solution features three widely
different length scales, $R\gg \ell\gg h$.  The large scale $R$ is
relevant in the loop and tail regions.  In the braid region, the
relevant length scale is $\ell$ along the braid axis, and $h$ in the
perpendicular direction.  Defining the rescaled, typical braid
length $\overline{\ell}$ and the rescaled radius $\overline{h}$ by
\begin{equation}
    \overline{\ell} = \frac{\ell}{L^\star},
    \qquad
    \overline{h} = \frac{h}{L^\star}
    \textrm{,}
    \nonumber
\end{equation}
we note the orders of magnitude associated with the three fundamental
lengths in rescaled form:
\begin{equation}
    \overline{h}\sim \epsilon ^2\quad \ll \quad
    \overline{\ell} \sim \epsilon^1 \quad \ll \quad
    \aR\sim \epsilon^0
    \nonumber
\end{equation}
In the vocabulary of inner or boundary layer analysis, the loop and
tail regions are both called \emph{outer regions}, while the braid is
called the \emph{inner region}~\footnote{Outer regions are those that are
present in the zero thickness solution of
Section~\ref{sec:ZeroThickness}: they undergo a regular perturbation
for small but nonzero $\epsilon$.  In contrast, the inner region
(braid) is undefined in the $\epsilon=0$ solution and so has to be built
from scratch when $\epsilon>0$.}.

The above argument clearly shows that the limit $\epsilon\to 0$ is
singular and that a uniform expansion of the solution with the
parameter $\epsilon$ is not possible.  This is typical of boundary
layer problem --- or \emph{inner} layer problems in the present case.
The classical approach to such problems is to use matched asymptotic
expansions, that is to build a solution domain by domain using
different approximations in the outer and inner domains, and to match
these solutions in the regions of overlap between two adjacent domains.

%
As mentioned earlier, the outer regions undergo a regular
perturbation.  This suggests the following, simple expansions in the 
tail region (subscript $T$) and in the loop region (subscript $L$):
\begin{equation}
	\ar_T(\as) = \ar^0_T(\as) +
	\epsilon\,\colvec{\hxt(\as)}{\hyt(\as)}{\hzt(\as)} + \dots
	\quad \mbox{and} \quad
	\ar_L(\as) = \ar^0_L(\as) +
	\epsilon\,\colvec{\hx_L(\as)}{\hy_L(\as)}{\hz_L(\as)} +
	\dots
	\label{eqn:TailAndLoopPerturbations}
\end{equation}
The functions $\ar^0_T$ and $\ar^0_L$ relevant to the zero thickness
case have been given in Eqs.~(\ref{eqn:LoopZeroRadiusR})
and~(\ref{eqn:TailsZeroRadiusR}).  The six unknown functions
$\hxt(\as)$, $\hyt(\as)$, \ldots, $\hz_L(\as)$ describe the
first-order perturbation in the tail and loop regions, and will be
found later by solving the linearized Kirchhoff equations.

For the inner region (braid, with subscript $B$), there is no solution
available in the limit of zero thickness and the above scaling 
argument suggests an expansion
of the form:
\begin{equation}
	\ar_B(\asig) = 
	\colvec{0}{\epsilon\,\tau_{B}}{0}
	+
	\colvec{\epsilon^2\,\hx_B(\asig)}
	{\epsilon^2\,\hy_B(\asig)}
	{\epsilon\,\asig} + \dots
	\label{eqn:BraidPerturbation}
\end{equation}
The first term in the right-hand side represents an infinitesimal
rigid-body translation along the $y$ axis: the center of symmetry of
the braid does not need to remain at the origin when $\epsilon$ is
nonzero, but can only move along the $y$ axis due
to the symmetry.  The second term is not a rigid-body motion. It 
makes use of the stretched coordinates
$\hx = \ax / \epsilon^2$, $\hy = (\ay - \epsilon \, \tau_B) / \epsilon^2$ and $\asig =
\az/\epsilon$ suggested by the above scaling analysis:
the axial dilation factor $1/\epsilon$ and the transverse one
$1/\epsilon^2$ comes from the lengths scales
$\overline{\ell} \sim \epsilon$ and
and
$\overline{h}\sim \epsilon^2$ found earlier.  The use of stretched
variables is classical in problems of elasticity with a small parameter, such as
those that arise in the analysis of slender elastic bodies.  The
number $\tau_{B}$ and the functions $\hx_B$ and $\hy_B$ are unknowns
which will be determined later.

Note that we use the stretched coordinate $\asig$ as the parameter for
the centerline in the braid region, and this $\asig= \az/\epsilon $ is
not the arc-length (in the absence of ambiguity it is common to use
the same notation $\ar_{B}$ for the functions mapping arc-length $\as$
to centerline position $\mathbf{r}$, or stretched axial variable
$\asig$ to centerline position $\mathbf{r}$).  The above scalings
imply that the braid is almost parallel to the $z$ axis, and so the
tangent can nowhere be perpendicular to $\eZ$: in the braid region,
there is a one-to-one mapping between the arc-length and the parameter
$\asig$, which is proportional to $z$, and it makes sense to use 
$\asig$ as a parameter along the braid.

The equations~(\ref{eqn:TailAndLoopPerturbations})
and~(\ref{eqn:BraidPerturbation}) provide a starting point for our
analysis.  These expansions will be plugged into the general equations
for the knot derived earlier in Section~\ref{sec:Model}.  The
resulting equations for the perturbed tail will be solved in
Section~\ref{sec:TailSolution}; those for the perturbed loop in
Section~\ref{sec:LoopSolution}; finally, the leading-order braid
solution, which is more difficult to derive, will be given
Section~\ref{sec:BraidSolution}.  As implied by the name `matched
asymptotics', the last step is to match the solutions obtained in the
different regions; this is done in Section~\ref{sec:Matching} by
requiring consistency of the expansions coming from the two adjacent
domains in the regions of overlap.  This provides a smooth solution of
the Kirchhoff equations over the entire domain, which will be shown to
be unique under some hypotheses.

\section{Tail solution}
\label{sec:TailSolution}

In this section we solve the linearized Kirchhoff equations in the
tails, which are given by an infinitesimal perturbation near the
straight solution, see Eq.~(\ref{eqn:TailAndLoopPerturbations}).
These linearized equations are classical and arise in the analysis of
linear stability of a straight, twisted rod under helical buckling.
We characterize the first-order perturbation to the straight
configuration due to the presence of the knot by computing the
functions $(\hxt,\hyt,\hzt)$.  As explained earlier, we focus on the
tail located on the positive side of the $z$ axis.

\subsection{Linearized Kirchhoff equations near a straight 
configuration}

To start with, let us plug 
Eq.~(\ref{eqn:TailAndLoopPerturbations}) into the 
definition~(\ref{eqn:KirchhoffEquations-Tangent}) of the tangent and 
compute
\begin{equation}
    |\at_{T}(\as)|^2 = |\eZ|^{2} + 
    2\,\epsilon\,\eZ\cdot(\hxt'(\as),\hyt'(\as),\hzt'(\as)) + \cdots
    = 1 + 2\,\epsilon\,\hzt'(\as)+\cdots
    \nonumber
\end{equation}
where the dots stand for higher order terms in $\epsilon$. By the 
inextensibility condition~(\ref{eqn:UnitTangent}), the left-hand side 
has to be equal to $1$ for any value of $\epsilon$ and so 
$\hzt'(\as)=0$ in the right-hand side: 
\begin{equation}
	\hzt(\as) = \hat{z}^c_T,
	\label{equa:z_tail_ordre_un}
\end{equation}
where $\hat{z}^c_T$ is a real constant.  This constant can be
interpreted as an infinitesimal rigid-body translation of the tail
along its axis, accommodating the change of curvilinear length captured
in the loop and braid regions.

There is no contact in the tail regions and so the contact force
$\mathbf{p}$ is zero.  By
Eq.~(\ref{eqn:RescaledKirchhoffEquationsN}), the internal force $\an$
is then uniform over the whole tail.  Now, this constant value of the
internal force is set by the asymptotic
condition~(\ref{eqn:RescaledBoundaryConditionsN}), and so
\begin{equation}
    \an_{T}(\as) = \eZ
    \label{eqn:TailInternalForce}
\end{equation}
a quantity that does not depend on $\epsilon$.  This makes is possible
to integrate the equation for the equilibrium of moments:
\begin{equation}
	\am_T(\as) = \overline{\mathbf{m}}_T^K
	+  \eZ \times \ar_T(\as) 
	\nonumber
	\textrm{,}
\end{equation}
where $\overline{\mathbf{m}}_T^K$ is a constant of integration whose
value, $\overline{\mathbf{m}}_K=\aU\,\mathbf{e}_z$, is provided by the
boundary conditions~(\ref{sys:RescaledBoundaryConditions}):
\begin{equation}
	\am_T(\as) = \colvec{0}{0}{\aU} +
	\epsilon\,\colvec{-\hyt(\as)}{\hxt(\as)}{0} + \dots
	\label{eqn:TailInternalMoment}
\end{equation}

Equation~(\ref{eqn:RescaledKirchhoffEquationsT}) for the rate of
rotation of the tangent is automatically satisfied at order zero; at 
first order in $\epsilon$, it writes:
\begin{equation}
	0 + \epsilon\,\colvec{\hxt''(\as)}{\hyt''(\as)}{0} = 
	\aU\,\eZ \times
	\epsilon\,\colvec{\hxt'(\as)}{\hyt'(\as)}{0} 
	+
	\epsilon \colvec{-\hyt(\as)}{\hxt(\as)}{0}  \times
	\eZ
	\textrm{.} \nonumber
\end{equation}
This vector equation is automatically satisfied along the $z$ axis. 
Projection along $x$ and $y$ axis yields a system of two equations:
\begin{subequations}
    \label{sys:TailsEquations}
    \begin{align}
        \hxt''(\as)-\hxt(\as)+\aU\,\hyt'(\as) &= 0, 
	\\
	\hyt''(\as)-\hyt(\as)-\aU\,\hxt'(\as) &= 0
	\textrm{.}
    \end{align}
\end{subequations}
Here we have written the equations for a rod in the small deflection
approximation.  In the problem at hand, it turns out that the tension
term~\footnote{%
The value of the tension $T\neq 0$ does not appear in the equation as
it has been effectively set to $1$ by our choice of dimensionless
variables.}
dominates over the bending term in the balance of transverse forces;
this explains why we have a second-order equation rather than the
classical fourth-order equation of beam problems.

These equations for a twisted rod linearized near a straight
configuration are identical to the ones obtained in the linear
analysis of helical buckling,
see~\cite{HeijdenThompson-Helical-and-Localised-Buckling-in-Twisted-Rods:-2000}.
Eqs.~(\ref{sys:TailsEquations}) can be put in a compact form when
expressed in terms of the complex variable
$\hwt(\as)=\hxt(\as)+i\,\hyt(\as)$:
\begin{equation}
	\hwt(\as)''-\hwt(\as)-i\,\aU\,\hwt'(\as)=0 \, ,
	\label{eqn:TailsCplxEquation}
\end{equation}
where $i^2=-1$.
We seek solutions of this linear differential equation with constant
coefficients in the form of exponential functions
$\hwt(\as)=\Gamma\,e^{k\,(\as-\pi\,\aR)}$, where $k$ and $\Gamma$ are
complex constants.  Note that we are free to incorporate the constant
term $-k\,\pi\,\aR$ in the argument of the exponential; this
amounts to change the definition of the undetermined constant
$\Gamma$ and will turn out to be convenient later on.
The possible values of the complex number $k$ are given by
the roots of the characteristic polynomial of
\eqnb{eqn:TailsCplxEquation}:
\begin{equation}
	k^2 -i\,\aU\,k-1=0\textrm{.} \nonumber
\end{equation}
These roots are noted $k_1=-a+i\,b$ and $k_2=a+i\,b$ where:
\begin{equation}
	a(\aU)=\sqrt{1-\left(\frac{\aU}{2}\right)^2}
	\quad\mathrm{and}\quad b(\aU)=\frac{\aU}{2}
	\textrm{,}
	\label{eqn:TailsRoots}
\end{equation}
The general solution of the equation for $\hwt$ could be 
written as a linear combination of the functions 
\begin{equation}
	e^{-a\,\as}\,
	\left(
	\cos (b\,\as)
	+
	i\,\sin (b\,\as)
	\right) \quad \textrm{and}\quad
	e^{+a\,\as}\,
	\left(
	\cos (b\,\as)
	+i\,\sin (b\,\as)
	\right)
	\nonumber
\end{equation}
but, as said above, it is more convenient to use $\as-\pi\,\aR$ as argument. Without no loss of generality, we write the general solution as
\begin{multline}
	\hwt(\as)=
	\Gamma_{-} \, e^{-a\,(\as-\pi\,\aR)}\,
	\left(
	\cos [b\,(\as-\pi\,\aR)]
	+
	i\,\sin [b\,(\as-\pi\,\aR)]
	\right) \\
	{} + \Gamma_+ \, e^{+a\,(\as-\pi\,\aR)}\,
	\left(
	\cos[b\,(\as-\pi\,\aR)]
	+i\,\sin[b\,(\as-\pi\,\aR)]
	\right)
	\textrm{.}
\end{multline}
Here, $\Gamma_-$ and $\Gamma_+$ are two complex constants of
integration.  The exponentially large solutions are incompatible with
the boundary conditions~(\ref{sys:RescaledBoundaryConditions}) and so
are discarded: $\Gamma_+=0$.

Noting $\lambda$ and $\mu$ the real and imaginary parts of the unknown
complex amplitude $\Gamma_- =\lambda +i\,\mu$, we
can write the general solution for the displacement as
\begin{align}
    \hxt(\as) &= \Re(\hwt(\as)) = \left(
    \lambda\,\cos[b\,(\as-\pi\,\aR)]
    -\mu\,\sin[b\,(\as-\pi\,\aR)]
    \right)\,e^{-a\,(\as-\pi\,\aR)},
    \nonumber\\
    \hyt(\as) &= \Im(\hwt(\as)) = \left( 
    \mu\,\cos[b\,(\as-\pi\,\aR)]
    +\lambda\,\sin[b\,(\as-\pi\,\aR)]
    \right)\,e^{-a\,(\as-\pi\,\aR)}
    \textrm{.}
    \nonumber
\end{align}

By equations~(\ref{eqn:TailsZeroRadiusR}),
~(\ref{eqn:TailAndLoopPerturbations}),
and~(\ref{equa:z_tail_ordre_un})
 the arc-length $\as$ is 
related to the $\az$ coordinate by
\begin{equation}
    \az = \az_T(\as) = (\as - \pi\,\aR) + \epsilon\,\hat{z}^c_T+\dots
    \nonumber
\end{equation}
We can use this relation to introduce a change of variable and
parameterize the centerline by $\az$ instead of $\as$.  For instance,
the above expression for $\hxt(\as)$ can be rewritten $\hxt(\az)
=(\lambda\,\cos(b\,\az) -\mu\,\sin(b\,\az)
)\,e^{-a\,\az}+\mathcal{O}(\epsilon)$, where the
$\mathcal{O}(\epsilon)$ notation means that the equality is exact up
to terms of order $\epsilon$.  This leads to the following
parameterization of the tail, which is valid to first order in
$\epsilon$ included:
\begin{subequations}
    \label{eqn:TailsFullSolution}
\begin{equation}
    	\ar_T(\az)=
	\az\,\eZ + \epsilon\,
	(\hxt(\az)\,\eX + \hyt(\az)\,\eY)
	+\cdots \nonumber
\end{equation}
where
\begin{align}
    \hxt(\az) & = 
    (\lambda\,\cos(b\,\az)-\mu\,\sin(b\,\az))\,e^{-a\,\az}+\cdots 
    \label{eqn:TailsFullSolution-Hxt}\\
    \hyt(\az) & = 
    (\mu\,\cos(b\,\az)+\lambda\,\sin(b\,\az))\,e^{-a\,\az}+\cdots
    \label{eqn:TailsFullSolution-Hyt}
\end{align}
\end{subequations}
This solution depends on two real parameters, $\lambda$ and $\mu$, 
which we call the internal parameters of the tail. They are referred 
to collectively as $\mathbf{\Psi}_T$:
\begin{equation}
    \mathbf{\Psi}_T=\left(\lambda,\mu\right)
    \textrm{,}
    \label{eqn:PsiTail}
\end{equation}
and will be determined later by matching with the other regions.

\subsection{Asymptotic expansion near junction with braid}
\label{sssec:AsymptoticExpansionLoop}

The matching problem, studied later in Section~\ref{sec:Matching}, is
based on the expansion of the tail solution given above in
Eqs.~(\ref{eqn:TailsFullSolution}) near the junction with the braid,
that is near the origin $\az = 0$.  This expansion is computed here.

The rescaled $x$ and $y$ coordinates of a current point $\ar_{T}$ on
the centerline are noted $\ax_{T}$ and $\ay_{T}$.  By
Eqs.~(\ref{eqn:TailsFullSolution}), their expansion is of the form
\begin{subequations}
    \label{sys:TailsAsymptoticExpansions}
    \begin{align}
	\ax_T(\az) &= \epsilon\,X_T + \epsilon\,X'_T\,\az 
	+ \mathcal{O}(\epsilon^2,\epsilon\,\az^2)
	\label{eqn:TailsAsymptoticExpansions-X}
	\\
	\ay_T(\az) &= \epsilon\,Y_T + \epsilon\,Y'_T\,\az +
	\mathcal{O}(\epsilon^2,\epsilon\,\az^2)
	\textrm{.}
	\label{eqn:TailsAsymptoticExpansions-Y}
    \end{align}
\end{subequations}
As implied by the $\mathcal{O}(.)$ notation, the right-hand side is
the beginning of an expansion where we have neglected terms of order
$\epsilon^2$ coming from the next order in the global expansion with
respect to $\epsilon$, and of order $\epsilon\,\az^2$ coming from
quadratic terms in the expansion of
Eqs.~(\ref{eqn:TailsFullSolution-Hxt})
and~(\ref{eqn:TailsFullSolution-Hyt}) with respect to $\az$.

In equations~(\ref{sys:TailsAsymptoticExpansions}) above, the four
coefficients $(X_{T},X'_{T},Y_{T},Y'_{T})$ are found by identification
with the series expansion of $\hxt(\az)$ and $\hyt(\az)$ given in
Eqs.~(\ref{eqn:TailsFullSolution}) near $\az = 0$:
\begin{align}
    X_{T} & = \hxt(\az = 0) = \lambda\nonumber\\ 
    Y_{T} & = \hyt(\az = 0) = \mu\nonumber\\ 
    X'_{T} & = \hxt'(\az = 0) = -a\,\lambda-b\,\mu\nonumber\\ 
    Y'_{T} & = \hyt'(\az = 0) = b\,\lambda - a\,\mu
    \textrm{.}
    \nonumber
\end{align}
For the matching problem studied later, it is convenient to put these
expressions into matrix form:
\begin{equation}
	\begin{pmatrix}
		X_T\\
		X'_T\\
		Y_T\\
		Y'_T
	\end{pmatrix}
	= 
	\mathbf{M}_T(\aU)\cdot\mathbf{\Psi}_T \qquad\textrm{where }
	\mathbf{M}_T(\aU) =
	\begin{pmatrix}
	    1 & 0\\
	    -a(\aU) & -b(\aU)\\
	    0 & 1\\
	    b(\aU) & -a(\aU) 
	\end{pmatrix}.
	\label{eqn:TailsMatrixExpansion}
\end{equation}
Note that this equation defines the matrix $\mathbf{M}_T(\aU)$
explicitly as a function of the loading parameter $\aU =
U/\sqrt{B\,T}$.  This matrix $\mathbf{M}_T(\aU)$ captures the elastic
response of the tail to perturbations applied at its end $\az = 0$; it
is the only quantity relevant to the tail that will be used in the
matching problem of Section~\ref{sec:Matching}.

\subsection{Helical instability}
\label{ssec:TailHelicalInstability}

We mentioned that the linearized
equations~(\ref{eqn:TailsFullSolution}) arise in the classical
analysis of linear stability of a straight, twisted rod under helical buckling.
For $\aU = \pm 2$, $a(\aU) = 0$ in Eq.~(\ref{eqn:TailsRoots}) and the
two complex roots $k_{1}$ and $k_{2}$ collide: this is the threshold
of linear stability for this helical buckling mode.  This instability 
will show up in the analysis of the knot later on.

\section{Loop solution}
\label{sec:LoopSolution}

In this section, we solve the Kirchhoff equation linearized near the
planar, circular configuration relevant for the loop region.  This
problem comes from applying
perturbation~(\ref{eqn:TailAndLoopPerturbations}) to the zero
thickness solution~(\ref{eqn:LoopZeroRadiusR}) for the loop.  It is
somewhat similar to the classical analysis of stability of a circular
rod under twist, known as Michell's instability,
see~\cite{Michell:On-the-stability-of-a-bent-and-twisted-wire:1889}. 
We focus on one half loop, corresponding to
the interval $0\leq \as\leq \pi\,\aR$.  Deformation of the other half 
can be found
by symmetry.  The point with arc-length $\as=0$ is the bottom of the
loop and that with coordinate $\as = \pi\,\aR$ describes the junction
with the braid (up to first order corrections in arc-length, as
discussed below).

\subsection{Linearized Kirchhoff equations near a circular configuration}
\label{ssec:LoopLinearizedEqns}

For the loop solution, it is convenient to use the following
cylindrical basis in the $(y,z)$ plane:
\begin{equation}
	\begin{cases}
	\vr(\theta) = -\cos\theta\,\eY - \sin\theta\,\eZ\\
	\vth(\theta)  = \sin\theta\,\eY - \cos\theta\,\eZ\\
	\end{cases}
	\textrm{.}
	\label{eqn:LoopBasis}
\end{equation}
These vectors defines an orthonormal frame $(\vr,\,\vth,\,\vx)$ for any 
value of $\theta$. With the choice
\begin{equation}
    \theta(\as ) = \frac{\as}{\aR}
    \textrm{,}
    \nonumber
\end{equation}
this basis is adapted to the zero thickness solution in the sense that
$\vth(\theta(\as)) = \at_L^0(\as)$.  In the absence of ambiguity, the
dependence of $\theta$ on $\as$ is not always written explicitly In
the rest of this Section.  The following derivation rules apply:
\begin{equation}
	\frac{\mathrm{d} \vr}{\mathrm{d}\as} = 
	\frac{\vth(\theta)}{\aR},
	\qquad
	\frac{\mathrm{d}\vth}{\mathrm{d}\as} = 
	- \frac{\vr(\theta)}{\aR}
	\label{eqn:CylindricalFrameDerivatives}
	\textrm{.}
\end{equation}
Vectors decomposed in the basis $(\vr,\,\vth,\,\vx)$ are denoted with 
\emph{square} brackets.

We introduce the first order perturbation of the tangent of the loop
in the moving frame using two functions $\htu$ and $\htv$:
\begin{equation}
	\at_L(\as) = \at_L^0(\as) +
	\epsilon\,\colvecb{\htu(\as)}{0}{\htv(\as)} = \vth(\as) +
	\epsilon\,\left( \htu(\as)\,\vr(\as) + \htv(\as)\,\vx \right)
	\textrm{.}
	\label{eqn:LoopTangentInCylindricalFrame}
\end{equation}
By the inextensibility constraint~(\ref{eqn:UnitTangent}), the
perturbation to $\at_L(\as)$ along $\vth$ vanishes.
The functions $\htu$ and $\htv$ are related to the functions $\hxl$, $\hyl$ and $\hzl$ introduced in Eq.~(\ref{eqn:TailAndLoopPerturbations}):
\begin{equation}
    \hxl'(\as) = \htv(\as),\quad
    \hyl'(\as) = -\cos(\as/\aR)\,\htu(\as),\quad
    \hzl'(\as) = -\sin(\as/\aR)\,\htu(\as)
    \textrm{.}\quad
    \label{eqn:FromHtvToHxl}
\end{equation}
These equations will be used later to compute to $\hxl$, $\hyl$ and
$\hzl$.

To use the constitutive
relation~(\ref{eqn:RescaledConstitutiveRelations}), we first need to
compute the derivative of the perturbed tangent given by
Eq.~(\ref{eqn:LoopTangentInCylindricalFrame}).  Using the derivatives
of the cylindrical vectors in
Eq.~(\ref{eqn:CylindricalFrameDerivatives}), we find
\begin{equation}
    \at'_L(\as) =
    -\frac{\vr}{\aR} +
    \epsilon\,\colvecb{\htu'(\as)}{\htu(\as)/\aR}{\htv'(\as)}
    +\cdots
    \textrm{.}
    \label{eqn:LoopTangentDerivatives}
\end{equation}
Plugging this expression into the constitutive equation, we obtain:
\begin{equation}
	\am_L(\as) = \colvecb{0}{\aU}{1/\aR} +
	\epsilon\,\colvecb{\htv'(\as) +
	\aU\,\htu(\as)}{-\htv(\as)/\aR}{-\htu'(\as)+\aU\,\htv(\as)}
	\textrm{,}
	\label{eqn:LoopMoment}
\end{equation}
an expression which is valid up to first order in $\epsilon$.

Like the tails, the loop is free of contact.  As a result, the contact
force vanishes, $\mathbf{p}= \mathbf{0}$, and the internal force
$\an(\as)$ takes on a constant value over the whole loop.  At dominant
order, this value has to match that given in
Eq.~(\ref{eqn:LoopZeroRadiusN}) for the zero thickness solution.  In
addition, its linear correction in $\epsilon$ has to be consistent
with the symmetry conditions~(\ref{sys:RescaledSymmetryLoopBottom}). 
This shows that the internal force in the loop is of the form
\begin{equation}
    \an_L(\as) = \frac{\aU}{\aR}\,\eX+ \epsilon\,\left( \alpha\,\eX +
	\beta\,\eZ \right)
	\textrm{,}
    \label{eqn:LoopInternalForce}
\end{equation}
where $\alpha$ and $\beta$ are two constants to be determined.

Combining Eqs.~(\ref{eqn:LoopMoment})
and~(\ref{eqn:LoopInternalForce}), we find that the equilibrium of
moments~(\ref{eqn:RescaledKirchhoffEquationsM}) can be expressed as a
set of linear equations for the loop perturbation $(\htu,\htv)$:
\begin{subequations}
    \label{sys:LoopEquations}
    \begin{align}
	\htv''(\as)
	+ \aU\,\htu'(\as) + \frac{\htv(\as)}{\aR^2}  & = - \alpha
	\\
	-\htu''(\as) + \aU\,\htv'(\as)  &=  - \beta\,\sin
	\frac{\as}{\aR}
	\textrm{.}
    \end{align}    
\end{subequations}
To integrate this differential system of total order four, we need to 
find four initial conditions.

To this end, we proceed as in Eq.~(\ref{eqn:LoopInternalForce}) and
write down the centerline position, tangent and internal moment at the
bottom of the loop which are compatible both with the zero radius
solution, see Eqs.~(\ref{sys:LoopZeroRadius}), and with the symmetry
conditions~(\ref{sys:SymmetryLoopBottom}):
\begin{subequations}
    \label{sys:LoopParameters}
    \begin{alignat}{2}
	\ar_L(0) &= -2\,\aR\,\eY & & +
	\epsilon\,\rho\,\eY, \label{sys:LoopParameters-R}\\
	\at_L(0) &= -\eZ & &- \epsilon\,\phi\,\eX,\\
	\am_L(0) &= \left(\frac{1}{\aR}\,\eX - \aU\,\eZ\right) & &+
	\epsilon\,\left(\gamma\,\eX + \delta\,\eZ \right) 
    \end{alignat}
\end{subequations}
The constants $\rho$, $\phi$, $\gamma$ and $\delta$ introduced here
will be determined later: $\rho$ represents an
infinitesimal motion of the bottom of the loop along the axis $y$ of
symmetry; $\phi$ represents an infinitesimal rotation of
the bottom of the loop about the axis $y$.  Writing the two invariants,
given in \sysnb{sys:RescaledInvariants} at $\as=0$ we can eliminate 
to two other constants:
\begin{equation}
	\gamma =\beta\,\aR \quad
	\textrm{and}
	\quad
	\delta =- \frac{\phi}{\aR}
	\textrm{.}
    \label{eqn:LoopInvariantsResults}
\end{equation} 
There remain four internal parameters for the loop, namely $\alpha$ 
and $\beta$ introduced in Eq.~(\ref{eqn:LoopInternalForce}), and 
$\rho$ and $\phi$ in Eq.~(\ref{sys:LoopParameters}). Using a notation 
similar to that for the tails, we collect these unknown parameters into 
a vector $\mathbf{\Psi}_L$:
\begin{equation}
    \mathbf{\Psi}_L=(\alpha,\,\beta,\,\rho,\,\phi)
    \label{eqn:DefinePsiL}
\end{equation}

The initial condition for the set of differential 
equations~(\ref{sys:LoopEquations}) can now be written as a function 
of the loop parameters. They read
\begin{subequations}
    \label{sys:LoopInitialConditions}
    \begin{align}
	\htu(0) &=
	- \frac{1}{\epsilon}\left( \at_L(0)-\at^0_L(0) \right)\cdot\eY
	&&= 0,\\
	\htu'(0) &= -\frac{1}{\epsilon}\left( \am_L(0)-\am^0_L(0)
	\right)\cdot\eX + \aU\,\htv(0)
	&&= -\beta\,\aR - \aU\,\phi ,\\
	\htv(0) &= \frac{1}{\epsilon}\left( \at_L(0)-\at_L^0(0)
	\right)\cdot\eX 
	&&= -\phi ,\\
	\htv'(0) &= -\frac{1}{\epsilon}\left( \am_L(0)-\am^0_L(0)
	\right)\cdot\eY - \aU\,\htu(0)
	&&= 0 
    \end{align}
\end{subequations}

%
%
Equations~(\ref{sys:LoopEquations}) with initial
conditions~(\ref{sys:LoopInitialConditions}) are linear differential 
equations with constant coefficients. Their solution reads
\begin{subequations}
    \label{sys:LoopSolutionUV}
\begin{align}
	\htu(\as) & = \alpha\,\frac{\aR^3\,\aU}{\overline{K}_L^2(\aU)}
	\left(\frac{1}{\overline{K}_L(\aU)}\sin\left(\frac{\as}{\aR}\,
	\overline{K}_L(\aU)\right)-\frac{\as}{\aR}\right)
	\label{eqn:LoopUSolution}
	\\
	& \hspace{5cm}
	{}-\frac{\aR}{\overline{K}_L(\aU)}\,\sin\left(\frac{\as}{\aR}\,\overline{K}_L(\aU)\right)
	\left(\beta\,\aR+\phi\,\aU\right)
	\nonumber
	\\
	\htv(\as) & =
	\frac{\beta\,\aR}{\aU}\cos\left(\frac{\as}{\aR}\right)
	+\alpha\,\frac{\aR^2}{\overline{K}^2_L(\aU)}\left(\cos\left(\frac{\as}{\aR}\,
	\overline{K}_L(\aU)\right)-1\right)
	\label{eqn:LoopVSolution}
	\\ 
	& \hspace{5cm}
	{} -\cos\left(\frac{\as}{\aR}\,\overline{K}_L(\aU)\right)\left(\frac{\beta\,\aR}{\aU}+\phi\right)
	\textrm{,}
	\nonumber
\end{align}
\end{subequations}
where we have introduced the auxiliary function $\overline{K}_L(\aU)=(
1+\aR^2\,\aU^2)^{1/2}$.  Integrating Eq.~(\ref{eqn:FromHtvToHxl}), one
can find an explicit expression for the functions $\hxl(\as)$,
$\hyl(\as)$ and $\hzl(\as)$ (the calculation is not difficult but the
final expressions are long and the result is not given here).  The
constants of integration are provided by
Eq.~(\ref{sys:LoopParameters-R}) and are $\hxl(0)=0$, $\hyl(0)=\rho$
and $\hzl(0)=0$.


\subsection{Asymptotic expansion near junction with braid}

To match this solution with the braid, we shall need an asymptotic expansion of
this solution near the top of the loop.  In principle, this step is
not difficult as it involves computing series expansion of the
explicit solution just derived; in practice, the calculation is too
tedious to be tractable by hand and was carried out with the help of a
symbolic calculation language.

It is convenient to describe the asymptotic shape of the top of
the loop using a Cartesian equation.  To do so, we eliminate
the variable $\as$ in favor of $\az$ and expand the previous solution
in series when $\as$ is close to $\pi\,\aR$.  Let us consider a
current point on the centerline near the top of the loop with
arc-length coordinate $\as = \pi\,\aR + \aet$, where $\eta$ is a small
quantity.  We shall make a fundamental 
assumption, justified at the end, namely that $\aet$ is at most of 
order $\sqrt{\epsilon}$:
\begin{equation}
    |\aet| \stackrel{<}{\sim} \epsilon^{1/2}
    \textrm{.}
    \label{eqn:OrderOfMagnitudeEtaBar}
\end{equation}

To prepare the change of variable, we work out the relation
between $\az$ and $\as$
\begin{equation}
	\begin{split}
		\az = \az_L(\as) &= \az_{L}^0(\as)+\epsilon\,\hzl(\as)+ \mathcal{O}(\epsilon^2),\\
		& = \az_{L}^0(\pi\,\aR+\aet)+\epsilon\,\hzl(\pi\,\aR+\aet)+ \mathcal{O}(\epsilon^2),\\
		& = -\aR\,\sin\left(\pi+\frac{\aet}{\aR}\right) +
		\epsilon\,\hzl(\pi\,\aR+\aet) + \mathcal{O}(\epsilon^2) \\
		&= \aet + \mathcal{O}(\aet^3) + \epsilon\,
		\left(\hzl(\pi\,\aR) + \aet\,\hzl'(\pi\,\aR) +
		\mathcal{O}(\aet^2)
		\right) + \mathcal{O}(\epsilon^2) \\
		& = \aet + \epsilon\,\hzl(\pi\,\aR) + 
		\mathcal{O}(\epsilon^{3/2})
	\end{split}
	\nonumber
\end{equation}
In the last line we have used $\hzl'(\pi\,\aR)=0$, see
Eq.~(\ref{eqn:FromHtvToHxl}). We  have also collected
all the $\mathcal{O}$ terms into
a dominant contribution, of order $\epsilon^{3/2}$ or smaller, using 
Eq.~(\ref{eqn:OrderOfMagnitudeEtaBar}).  Elimination of the
arc-length variables $\as$ and $\aet$ is then possible using the
equality
\begin{equation}
    \aet = \as - \pi\,\aR = \az - \epsilon\,\hzl(\pi\,\aR)
    + 
    \mathcal{O}(\epsilon^{3/2})
    \textrm{.}
    \label{eqn:LoopEliminatearc-lengthNearTop}
\end{equation}
The term $\hzl(\pi\,\aR)$ is known explicitly from the last section.
Setting $\az=0$ in Eq.~(\ref{eqn:LoopEliminatearc-lengthNearTop})
above yields the arc-length $\as_{O}$ of the point on the rod closest
to the origin $O$:
\begin{equation}
    \as_{O} = \pi\,\aR- \epsilon\,\hzl(\pi\,\aR)
    \textrm{,}
    \label{eqn:SODefinition}
\end{equation}
where the quantity
$\hzl(\pi\,\aR)$ in the right-hand side will be given
at the end of Section~\ref{sec:Matching}.
%

We expand $\ax_{L}$ and $\ay_{L}$ similarly to $\az_L$:
\begin{equation}
    \label{eqn:LoopAsymptoticExpansionsWithEtaVariable}
    \begin{split}
    \ax_L(\pi\,\aR+\aet) & =
    \epsilon\,\left(\hxl(\pi\,\aR) +
    \aet\,\hxl'(\pi\,\aR)
    \right)
    + \mathcal{O}(\epsilon^2)
    \\
    \ay_L(\pi\,\aR+\aet) & = 
    -\frac{\aet^2}{2\,\aR} + 
    \epsilon\,
    \left(
    \hyl(\pi\,\aR) +
    \aet\,\hyl'(\pi\,\aR)
    \right) + \mathcal{O}(\epsilon^2)
    \textrm{.}
\end{split}
\end{equation}
%
The first term in the right-hand side of the second equation,
$-\aet^2/(2\,\aR)$, arises from curvature of the loop in the zero
thickness solution.  This term has to be retained as it is of the same
order of magnitude as the other terms when $\aet$ is of order
$\sqrt{\epsilon}$.

We can now use Eq.~(\ref{eqn:LoopEliminatearc-lengthNearTop}) to
eliminate the arc-length $\aet$.  This leads to a Cartesian equation
of the top of the loop in the form:
\begin{subequations}
    \label{sys:LoopAsymptoticExpansions}
    \begin{align}
	\ax_L(\az) &= \epsilon\,X_L +
	\epsilon\,\az\,X'_L + \mathcal{O}(\epsilon^2), 
	\label{eqn:LoopAsymptoticExpansions-X}	
	\\
	\ay_L(\az) &= -\frac{\az^2}{2\,\aR}+\epsilon\,Y_L +
	\epsilon\,\az\,Y'_L + \mathcal{O}(\epsilon^2),
	\label{eqn:LoopAsymptoticExpansions-Y}
    \end{align}
Because of Eq.~(\ref{eqn:OrderOfMagnitudeEtaBar}), this expansion is
valid when $\az$ is of order $\sqrt{\epsilon}$ or smaller:
\begin{equation}
    |\az| \stackrel{\sim}{<} \epsilon^{1/2} 
    \textrm{.}
    \label{eqn:LoopAsymptoticExpansionMaxZ}
\end{equation}
\end{subequations}
 The coefficients $(X_L,X'_L, Y_L,Y'_L)$ of the asymptotic expansion
 are found by identification with
 Eq.~(\ref{eqn:LoopAsymptoticExpansionsWithEtaVariable}):
\begin{align}
    X_L & = \hxl(\pi\,\aR)   
    &  
    Y_L & = \hyl(\pi\,\aR)
    \nonumber \\
    X'_L & = \hxl'(\pi\,\aR)
    &  
    Y'_L & = \hyl'(\pi\,\aR) + 
    \frac{\hzl(\pi\,\aR)}{\aR}
    \nonumber
\end{align}
The quantities in the right-hand side are all known explicitly from
the analysis of the loop given in
Section~\ref{ssec:LoopLinearizedEqns}.  Being solutions of a set of
linearized differential equations, they all depend linearly on the
loop parameters $\mathbf{\Psi}_L=(\alpha,\,\beta,\,\rho,\,\phi)$, as
revealed by Eqs.~(\ref{sys:LoopSolutionUV}).  The linear mapping 
giving the expansion coefficients as a function of the loop 
parameters reads
\begin{equation}
    \left(
    \begin{array}{c}
	X_{L} \\ X'_{L} \\ Y_{L} \\ Y'_{L}
    \end{array}
    \right)
     = \mathbf{M}_L(\aU)\cdot\mathbf{\Psi}_L
    \textrm{, where }
	\mathbf{M}_L(\aU)=
	\begin{pmatrix}
%
%
%
		\frac{-\pi\,\overline{K}_L+\overline{K}_L^s}{(\overline{K}_L/\aR)^3}
		&
		-\frac{\aR^2\,\overline{K}_L^s}{\aU\,\overline{K}_L}
		& 0 &
		-\frac{\aR\,\overline{K}_L^s}{\overline{K}_L}
		\\
		\frac{-1+\overline{K}_L^c}{(\overline{K}_L/\aR)^2}
		&
		-\frac{\left(1+\overline{K}_L^c\right)}{\aU/\aR}
		& 0 & -\overline{K}_L^c 
		\\
		-\frac{1+2\aR^2\aU^2+\overline{K}_L^c}{\aU\,(\overline{K}_L/\aR)^2}
		&
		\frac{\left(1+\overline{K}_L^c\right)}{\aU^2/\aR}
		& 1 & \frac{1+\overline{K}_L^c}{\aU}
		\\
		\frac{\aR\,\overline{K}_L^s}{\aU\,\overline{K}_L}
		&
		-\frac{\overline{K}_L\overline{K}_L^s}{\aU^2}
		& 0 &
		-\frac{\overline{K}_L\,\overline{K}_L^s}{\aR\,\aU}
		\end{pmatrix}
		\textrm{.}
	\label{eqn:LoopMatrix}
\end{equation}
To keep the notations compact, we have noted $\overline{K}_L =
\overline{K}_L(\aU) = ( 1+\aR^2\,\aU^2)^{1/2}$ and introduced the
shorthand notations $\overline{K}_L^c = \overline{K}_L^c (\aU)=
\cos(\pi\,\overline{K}_L(\aU)) $ and $\overline{K}_L^s =
\overline{K}_L^s (\aU)= \sin(\pi\,\overline{K}_L(\aU)) $.  This
explicit expression for the matrix $\mathbf{M}_L(\aU)$ comes from the
analytical solution for the loop given in the previous section.
Recall also that $\aR = 1/\sqrt{2}$ by Eq.~(\ref{eqn:AraiRadius}).
The matrix $\mathbf{M}_L(\aU)$ defined above~\footnote{Note that the
matrix $\mathbf{M}_T(\aU)$ has a smooth limit for $\aU\to 0$.
Even though there are some powers of $\aU$ in the denominators, the
following expressions are smooth near $\aU = 0$:
$\frac{1+\overline{K}_L^c}{\aU^2} \to 0$ and
$\frac{\overline{K}_L^s}{\aU^2} \to -\frac{\pi}{4}$.
} plays a
role similar to $\mathbf{M}_T(\aU)$ for the tails: it captures the elastic
response of the loop to the perturbation induced by the presence of
the braid.

\section{Braid solution}
\label{sec:BraidSolution}

\begin{figure}
	\begin{center}
	\includegraphics[width=0.6\columnwidth]{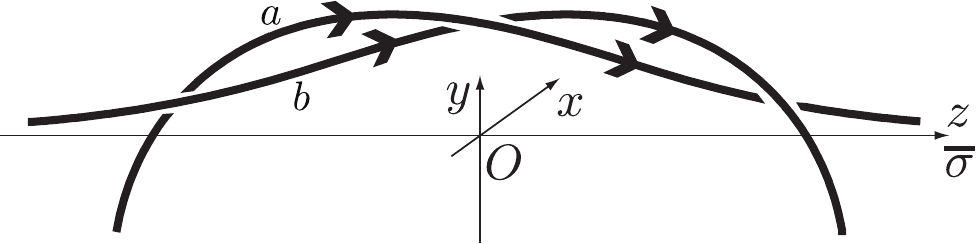}
	\caption{Braid geometry.  We call $a$ the strand having
	positive arc length $\as$ ($\as \approx \pi \aR$ in the center
	of the braid), and $b$ the strand with negative arc length
	($\as \approx -\pi\aR$ in the center).}
	\label{fig:BraidGeometry}
	\end{center}
\end{figure}

The solutions in the outer domain (loop and tails) have been derived
in the two previous sections.  We now proceed to solving the internal
region (braid), which is more difficult as it involves self-contact.
A key remark is that the scaling relations expressed in
Eq.~(\ref{eqn:BraidPerturbation}) imply that the tangent deflects from
the $z$ axis by a small angle, of order $\overline{h}/\overline{\ell}
\sim \epsilon \ll 1$.  As a result, the approximation of small
displacements hold and the Kirchhoff equations can be linearized; by
linearity, the braid problem can then be decomposed into an average
problem without contact, and a difference problem where contact takes
place with a \emph{fixed}, virtual cylinder.
As earlier, a subscript $B$ denotes quantities associated with the
braid.  The two strands composing the braid are labeled with
superscripts $a$ and $b$, as shown in \fignb{fig:BraidGeometry}.

\subsection{Centerlines}

We have introduced a rescaled axial displacement consistent with the 
scalings for the braid:
\begin{equation}
    \asig = \frac{\az}{\epsilon} = \frac{z}{\epsilon\,L^{\star}}
    \label{eqn:SigmaBar}
\end{equation}
The leading order term of the expansion in the braid was given in
equation~(\ref{eqn:BraidPerturbation}).  For the
first strand, labeled $a$, it reads
\begin{equation}
    \label{eq:BraidParametricEquationOfSigmaBar}
    \ar^a_{B}(\asig)
    = 
    \left(
    \begin{array}{c}
	\ax^a_{B}(\asig) \\
	\ay^a_{B}(\asig) \\
	\az^a_{B}(\asig)
    \end{array}
    \right)
    = 
    \left(
    \begin{array}{c}
	\epsilon^2\,\hx_{B}^a(\asig) \\
	\epsilon\,\tau_{B}+\epsilon^2\,\hy_{B}^a(\asig) \\
	\epsilon\,\asig
    \end{array}
    \right)
    \textrm{.}
\end{equation}
The centerline of the other strand is defined by a similar formula
with $a$ replaced by $b$ (note that the constant $\tau_{B}$, which
represents a global translation of the braid, is common to both
strands and so has no index $a$ or $b$).  Our unknowns for the braid
problem are the translation $\tau_{B}$ and the four functions
$\hx_{B}^a$, $\hy_{B}^a$, $\hx_{B}^b$ and $\hy_{B}^b$, defined in
terms of stretched coordinates.

Recall that the strands $a$ and $b$ are mapped onto each other by a
symmetry of angle $\pi$ about the $z$ axis.  By our choice of axes,
$\asig=0$ is the center of the braid: the symmetry is expressed
by the following relations,
\begin{subequations}
	\label{eqn:BraidSymmetry}
    \begin{align}
	\hx^b_B(\asig) & = -\hx^a_B(-\asig) 
	\\
	\hy^b_B(\asig) & = +\hy^a_B(-\asig)
	\textrm{,}
    \end{align}
\end{subequations}
which implies that there are only two independent functions to be
determined, say $\hx^a $ and $\hy^a$. 

It is useful to introduce
an auxiliary quantity, the velocity $c^a(\asig)$ at which the
centerline is swept out in this parameterization --- we 
do not use arc-length parameterization here:
\begin{equation}
    c^a(\asig) = |{\ar^a_{B}}'(\asig)| = \epsilon + \mathcal{O}(\epsilon^3)
    \textrm{.}
    \nonumber
\end{equation}
The unit tangent is then defined by
\begin{equation}
    \at_{B}^a(\asig) = 
    \frac{{\ar^a_{B}}'(\asig)}{c^a(\asig)}
    = 
    \eZ+\epsilon\,(\hxb^a{}'(\asig)\,\eX + 
    \hyb^b{}'(\asig)\,\eY)+\cdots
    \label{eqn:BraidTangent}
\end{equation}
and a similar equation for the other strand.  Note that primes applied
to functions such as $\hx_{B}$, $\hy_{B}$ or $\at_{B}^a$ denote
derivatives with respect to their argument, $\asig$ here.

%
%

\subsection{Contact}

For any pair of points in contact in the braid, let $\asig^a$ and 
$\asig^b$ be the rescaled coordinates of the point on braid $a$ and 
on braid $b$, respectively. The contact condition writes
\begin{equation}
    \left| \ar^a_{B}(\asig^a)-\ar^b_{B}(\asig^b) \right| = \frac{2h}{L^{\star}} =
	\sqrt{2}\,\epsilon^2,
    \label{eqn:BraidContactEqualitySquared}
\end{equation}
We square both sides of the equation and use the the centerline
parameterization given in
equation~(\ref{eq:BraidParametricEquationOfSigmaBar}):
\begin{equation}
    \epsilon^4\,\big[(\hxb^a(\asig^a)-\hxb^b(\asig^b))^2
    +
    (\hyb^a(\asig^a)-\hyb^b(\asig^b))^2\big]
    +
    \epsilon^2\,(\asig^a - \asig^b)^2
    =
    (\sqrt{2}\,\epsilon^2)^2
    \nonumber
\end{equation}
In this equation, there is only one term of order $\epsilon^2$ and no
lower order term; this term has to cancel, which implies:
\begin{equation}
    \asig^a = \asig^b
    \quad\textrm{(for points in contact)}
    \label{eqn:BraidPointsInContactHaveSameZ}
\end{equation}
At next order, we obtain
\begin{equation}
    (\hxb^a(\asig)-\hxb^b(\asig))^2
    +
    (\hyb^a(\asig)-\hyb^b(\asig))^2
    =
    2
    \quad
    \textrm{(for points in contact)}
    \label{eqn:TransverseDistanceForContact}
\end{equation}

By equation~(\ref{eqn:BraidPointsInContactHaveSameZ}), contact occurs
only between points lying in the same plane perpendicular to the $z$
axis at the leading order in $\epsilon$.  Let us define the locus of
the contact in physical space as
\begin{equation}
    \mathfrak{D}=\big\{\asig \quad\textrm{such that}\quad
    \left| \ar^a_{B}(\asig)-\ar^b_{B}(\asig) \right|  =
	\sqrt{2}\,\epsilon^2
    \big\}
    \nonumber
\end{equation}
This set $\mathfrak{D}$ is composed of the rescaled $z$ coordinate,
called $\asig$, of the points in contact, unlike the original set
$\mathfrak{C}$ which describes contact points based on pairs of
arc-lengths.  In the limit $\epsilon \ll 1$, the set $\mathfrak{D}$
provides a description of contact much simpler than the generic one,
based on $\mathfrak{C}$.  In Eq.~(\ref{eqn:ContactSetTopology}), we
shall compute the set $\mathfrak{D}$ explicitly, and show that it has
non-trivial topology (it has `holes' in it).  For reference, we
mention that the initial set $\mathfrak{C}$ can be reconstructed by
\begin{equation}
    \mathfrak{C}=\bigcup_{\{s_{1},s_{2}\}\in\mathfrak{C}'}\{(s_{1},s_{2}),(s_{2},s_{1})\} 
    \quad\textrm{where }
    \mathfrak{C}'=
    \bigcup_{\asig\in\mathfrak{D}}\{
    ((\as_{O}+ \epsilon\,\asig)L^\star,(-\as_{O}+ \epsilon\,\asig)L^\star)
    \}
    \nonumber
\end{equation}
and $\as_{O}$ was defined in Eq.~(\ref{eqn:SODefinition}) as the
arc-length of the point on strand $a$ closest to origin, that is such
that $\az = 0$.

We shall now express the contact pressure $\mathbf{p}$.  First
note that the action-reaction principle in
Eq.~(\ref{eqn:ContactForceActionReaction}) can be rewritten as $p^a(\asig) =
p^b(\asig)$; we can therefore omit the
superscript and note $p(\asig)$ the scalar contact pressure
associated with contact occurring at coordinate $\az =
\epsilon\,\asig$.  According to
equation~(\ref{eqn:ContactPressureEquations}), the (vector)
contact pressure is the scalar pressure $p(\asig)$ times the unit vector
joining the barycenters of two cross-sections that are in contact; in 
rescaled form this reads, see Eq.~(\ref{eqn:RescaledForces}),
\begin{equation}
    \overline{\mathbf{p}}^a(\asig) 
    =
    \overline{p}(\asig)\,
    \frac{\ar_{B}^a(\asig) - \ar_{B}^b(\asig)}{\sqrt{2}\,\epsilon^2}
    =
    \frac{\overline{p}(\asig)}{\sqrt{2}}\,\left(
    \begin{array}{c}
	\hxb^a(\asig) - \hxb^b(\asig) \\
	\hyb^a(\asig) - \hyb^b(\asig) \\
	0
    \end{array}
    \right) + \cdots
    \label{eqn:BraidContactForceSpelledOut}
\end{equation}
This quantity appears to be orthogonal to the $z$ axis at this order, as
expected.

\subsection{Equations of equilibrium at leading order}

Combining the constitutive 
relation~(\ref{eqn:RescaledConstitutiveRelations}) with the 
formula~(\ref{eqn:BraidTangent}) for the tangent, we obtain the 
internal moment as:
\begin{equation}
    \am_B^a(\asig) 
    = \at_{B}^a(\asig) \times
    \frac{\at_{B}^a{}'(\asig)}{c^a(\asig)} + \aU\,\at_{B}^a(\asig) \\
    =
    \colvec{-\hyb''(\asig)}{\hxb''(\asig)}{\aU} + 
    \cdots
%
	\label{eqn:BraidMoment}
\end{equation}
Note the normalization factor $1/c^a$ in the above expression for the
normal curvature vector, $\mathrm{d}\at/\mathrm{d}s =
(\mathrm{d}\at/\mathrm{d}\asig) / c^a$, which is required since
parameterization does not use arc-length.

For any vectors $\mathbf{a}$ and $\mathbf{u}$ such that $\mathbf{u}$
has unit length ($\mathbf{u}^2 = 1$), the following identity holds
$    \mathbf{a} = (\mathbf{a}\cdot \mathbf{u})\,\mathbf{u} - 
    \mathbf{u} \times (\mathbf{u} \times \mathbf{a})
    \textrm{.}
$
With $\mathbf{a} =
\an_B^a(\asig)$ and $\mathbf{u} = \at_B^a(\asig)$, it can be used to 
compute the internal force:
\begin{equation}
    \an_B^a(\asig) = (\at_B^a(\asig)\cdot\an_B^a(\asig))\,\at_B^a(\asig)
    -\at_B^a(\asig)\times(\at_B^a(\asig)\times \an_B^a(\asig))
    \textrm{.}
    \nonumber
\end{equation}
The factor $(\at_B^a(\asig)\cdot\an_B^a(\asig))$ can be expressed
using the second invariant $\overline{I}_{2} = \aU^2/2+1$ given in
Eq.~(\ref{eqn:RescaledForceInvariant}), while the balance of
moments~(\ref{eqn:RescaledKirchhoffEquationsM}) allows one to rewrite
the vector in the last term as $(\at_B^a(\asig)\times \an_B^a(\asig))
= -\frac{\am_B^a{}'(\asig)}{c^a(\asig)} $, this right-hand side being given itself by
Eq.~(\ref{eqn:BraidMoment}).  This yields the following expression for
the internal force in the braid
\begin{equation}
    \an_B^a (\asig)=
    \left(
    \frac{\aU^2}{2}+1  - \frac{{\am_B^a}^2(\asig)}{2}
    \right)\,\at_B^a(\asig) + 
    \at_B^a(\asig)\times \frac{\am_B^a{}'(\asig) }{c^a(\asig)}
    \label{eqn:InternalForceInBraid}
\end{equation}
The first term in the right-hand side is of order $\epsilon^0$ (it
is bounded for small $\epsilon$) and is
dominated by the second term, of order $1/\epsilon$ because of the 
denominator $c^a(\asig) = \epsilon + \cdots$.  This yields the leading 
order term for the internal force:
\begin{equation}
	\an_B^a(\asig) =
	- \frac{\hxb^a{}'''(\asig)\,\eX + \hyb^a{}'''(\asig)\,\eY}{\epsilon}
	+ \cdots
	\label{eqn:BraidForce}
\end{equation}
where the ellipsis stands for negligible terms that are bounded for
small $\epsilon$.  The internal force $\an_B^b$ in the other braid is
given by a similar formula.

The balance of forces~(\ref{eqn:RescaledKirchhoffEquationsN}) writes
$\an_B^a{}'(\asig)/c^a(\asig) + \overline{\mathbf{p}} (\asig) =
\mathbf{0}$ in the current parameterization.  Using
Eq.~(\ref{eqn:BraidContactForceSpelledOut}) for the contact force,
this yields
\begin{equation}
    -\frac{1}{\epsilon^2}\hxb^a{}'''' + 
    \frac{\overline{p}}{\sqrt{2}}\,(\hxb^a - \hxb^b) = 0
    \textrm{,}
    \nonumber
\end{equation}
and similar equations for the $y$ direction and for the other strand.
This equation shows that the rescaled contact
pressure $\overline{p}$ has to be of order $1/\epsilon^2$ in order to
balance bending.  Therefore, we define the final rescaling for the
contact pressure by:
\begin{equation}
    \hat{p}(\asig) = \epsilon^2\,\overline{p}(\asig) = 
    \frac{\epsilon^2\,B^{1/2}}{T^{3/2}}\,p(\asig)
    \textrm{.}
    \nonumber
\end{equation}
By the previous argument, this $\hat{p}(\asig)$ has a finite limit 
for $\epsilon\to 0$.  The
equations of equilibrium then take the form
%
%
\begin{subequations}
    \label{sys:BraidEquations}
    \begin{align}
	\label{sys:BraidEquations-X}
        \hxb^{a}{}''''(\asig)&
	= 
	\frac{1}{\sqrt{2}}\,\hat{p}(\asig)\,\left(\hxb^a(\asig)-\hxb^b(\asig)\right)
	\\
	\hyb^{a}{}''''(\asig)&
	=
	\frac{1}{\sqrt{2}}\,\hat{p}(\asig)\,\left(\hyb^a(\asig)-\hyb^b(\asig)\right)
	\textrm{.}
    \end{align}
\end{subequations}
This is a set of
fourth-order differential equations for the deflection, which are
coupled through contact.  The contact pressure $\hat{p}(\asig)$ is the
Lagrange multiplier associated with the non-penetration condition, and
is not known in advance.  The rest of Section~\ref{sec:BraidSolution}
is devoted to solving these equations with appropriate boundary
conditions.

Equations~(\ref{sys:BraidEquations}) stand for an elastic
rod in the small deflection approximation, subjected to the normal distributed
force given by the right-hand side.  Note that the twist loading
parameter $\aU$ does not appear in the equations for the braid at the
leading order, as bending effects dominate twist.  A nice consequence
is that the braid problem is universal: unlike the tail and loop
problems studied earlier, the formulation of the rescaled braid
problem involves no parameter (except for the knot type which is a
discrete parameter).

\subsection{Decomposition into average and difference problems}

Taking advantage of the linearity, we can combine
\sysnb{sys:BraidEquations} and the two similar equations for
$\hxb^{b}$ and $\hxb^{b}$ into a difference and an average problem.
%
The average variables $(f,g)$ and the difference variables $(u,v)$ 
are defined as follows:
\begin{subequations}
    \label{eqn:BraidAverageAndDiffVariables}
\begin{align}
    f(\asig)& =\frac{1}{\sqrt{2}}\left(\hx^a_B(\asig)+\hx^b_B(\asig)\right)
    &
    u(\asig)&=\frac{1}{\sqrt{2}}\left(\hx^b_B(\asig)-\hx^a_B(\asig)\right)
    \\
    g(\asig)& =\frac{1}{\sqrt{2}}\left(\hy^a_B(\asig)+\hy^b_B(\asig)\right)
    &
    v(\asig)&=\frac{1}{\sqrt{2}}\left(\hy^b_B(\asig)-\hy^a_B(\asig)\right)
    \label{eqn:BraidAverageAndDiffVariables-GV}
\end{align}
\end{subequations}
Summing Eq.~(\ref{sys:BraidEquations-X}) and the similar equation for
strand $b$, namely $\hxb^{b}{}'''' =
\hat{p}\,(\hxb^b-\hxb^a)/\sqrt{2}$, we find $f''''=0$.  By the same
argument, $g''''=0$.  The unknown contact force disappears from the
average problem:
\begin{subequations}
    \label{sys:BraidAverageEquations}
    \begin{align}
        f''''(\asig) & = 0 
        \\
        g''''(\asig) & = 0 
	\textrm{.}
    \end{align}
\end{subequations}
In the next section, we derive the asymptotic conditions associated
with these equations, which are then solved in
Section~\ref{ssec:BraidAveragePb}.

For the difference problem we obtain
\begin{subequations}
    \label{sys:BraidDifferenceEquations}
    \begin{align}
	u(\asig)'''' & = (\sqrt{2}\,\hat{p}(\asig))\,u(\asig)
	\\
	v(\asig)'''' & = (\sqrt{2}\,\hat{p}(\asig))\,v(\asig)
   \end{align}
\end{subequations}
%
%
Although the contact force $\hat{p}(\asig)$ is still present in the
difference problem, the contact
condition~(\ref{eqn:TransverseDistanceForContact}) takes a very simple
form when formulated as a function of the difference variables:
$u^2+v^2 = 1$.  As a result the non-penetration condition is
expressed by the inequality
\begin{equation}
    u^2(\asig) + v^2(\asig) \geq 1
    \textrm{,}
    \label{eqn:DifferenceConstraint}
\end{equation}
and the problem is much easier to solve.
The average and difference variables are subjected to the following
parity conditions, deriving from
equation~(\ref{eqn:BraidSymmetry}):
\begin{subequations}
    \label{eqn:BraidMoreSymmetry}
    \begin{align}
	f(-\asig)& =-f(\asig) & u(-\asig) &= u(\asig)\\
	g(-\asig)& = g(\asig) & v(-\asig) &=-v(\asig) 
    \end{align}
\end{subequations}

\subsection{Asymptotic conditions}

We have derived in the previous section the differential equations for
the braid.  These equations make use of a stretched variable $\asig$.
In the present section we derive the asymptotic conditions the
solutions must satisfy for large values of the stretched variable
$\asig$.  As usual in matched asymptotic analysis, the asymptotic
conditions for the inner problem are required for the inner solution
to match the outer solutions in the region of overlap (intermediate
region), where both the inner and outer solutions are valid --- see
Section~\ref{sec:Matching} for a detailed discussion of this matching
procedure.  Given our conventions, summarized in
\fignb{fig:BraidGeometry}, the strand $a$ of the braid connects to the
loop for $\asig \to -\infty$, and to the tail for $\asig \to +\infty$.

Let us start with the condition for matching the internal moment.  At
the top of the loop $\theta \simeq \pi$ and so $\vth \simeq \eZ$;
the internal moment,  given by Eq.~(\ref{eqn:LoopMoment}), then reads
$\am_L(\pi\,\aR) =\eX/\aR + \aU\,\eZ + \ldots$.  Comparison with the braid moment, given by
Eq.~(\ref{eqn:BraidMoment}), yields the asymptotic condition 
$\hxb^{a}{}'' \to 0$ and $\hyb^{a}{}'' \to -\frac{1}{\aR}$ as
$\asig\to -\infty$.  Using the value of $\aR = 1/\sqrt{2}$ given by
Eq.~(\ref{eqn:AraiRadius}), we write
\begin{alignat}{2}
    \hxb^{a}{}'' & \to 0
    &\quad&
    \textrm{for } \asig\to -\infty \textrm{,}
    \nonumber \\ 
    \hyb^{a}{}'' & \to -\sqrt{2}
    &\quad&
    \textrm{for } \asig\to -\infty \textrm{.}
    \nonumber
\end{alignat}
The asymptotic condition for $\asig\to +\infty$ is obtained by matching
the moment at the origin of the tail, $\am_T = \aU\,\eZ$, with 
Eq.~(\ref{eqn:BraidMoment}). It yields:
\begin{alignat}{2}
    \hxb^{a}{}'' & \to 0
    &\quad&
    \textrm{for } \asig\to +\infty \textrm{,}
    \nonumber \\ 
    \hyb^{a}{}'' & \to 0
    &\quad&
    \textrm{for } \asig\to +\infty \textrm{.}
    \nonumber
\end{alignat}

A similar argument gives the matching condition for the internal
force.  The force is bounded for small $\epsilon$ in the two outer
regions, see Eqs.~(\ref{eqn:TailInternalForce})
and~(\ref{eqn:LoopInternalForce}).
In contrast the force in the braid diverges for small $\epsilon$,
as shown by the term of order $(1/\epsilon)$ in Eq.~(\ref{eqn:BraidForce}).
This term must therefore vanish when strand $a$ reaches the loop ($\asig \to -\infty$) or the tail ($\asig \to +\infty$):
\begin{alignat}{2}
    \hxb^{a}{}''' & \to 0
    &\quad&
    \textrm{for } \asig\to \pm\infty\textrm{,}
    \nonumber \\ 
    \hyb^{a}{}''' & \to 0
    &\quad&
    \textrm{for } \asig\to \pm\infty\textrm{.}
    \nonumber
\end{alignat}
%
Using the parity conditions~(\ref{eqn:BraidSymmetry}), we obtain the same 
equations for the other strand. The asymptotic conditions are then
expressed in terms of the average variables:
\begin{subequations}
    \label{sys:BraidAverageBoundaryConditions}
    \begin{align}
	f''(\pm\infty) & \to  0  & g''(\pm\infty) & \to -1\\
	f'''(\pm\infty) & \to 0 &
	g'''(\pm\infty) &\to 0
    \end{align}
\end{subequations}
and of the difference variables:
\begin{subequations}
    \label{sys:BraidDifferenceBoundaryConditions}
    \begin{align}
	u''(\pm\infty) & \to  0  & v''(\pm\infty) & \to \mp 1
	\label{eqn:BraidDifferenceBoundaryConditions-Order2}\\
	u'''(\pm\infty) & \to 0 &
	v'''(\pm\infty) &\to 0
	\label{eqn:BraidDifferenceBoundaryConditions-Order3}
	\textrm{,}
    \end{align}
\end{subequations}
where we used a shorthand notation meaning that $v''$ goes to
$-1$ for $\asig\to+\infty$, and to
$+1$ for $\asig\to-\infty$.

\subsection{Solution of average problem}
\label{ssec:BraidAveragePb}

The average problem being insensitive to contact forces, its
solution is straightforward. The general
solution of Eqs.~(\ref{sys:BraidAverageEquations}) yields for $f(\asig)$
and $g(\asig)$ polynomials of order 3.  
To satisfy the parities, see Eq.~(\ref{eqn:BraidMoreSymmetry}), we write:
\begin{align}
    f(\asig) &=
    c_1\,\asig +
    c_3\,\asig^3 ,
    \nonumber \\
    g(\asig) &=  c_0 + c_2\,\asig^2 
    \textrm{,}
    \nonumber
\end{align}
where $c_{0}$, $c_{1}$, $c_{2}$ and $c_{3}$ are real constants.
%
Two of these four constants are set by the asymptotic
conditions~(\ref{sys:BraidAverageBoundaryConditions}) and we have:
\begin{subequations}
    \label{sys:BraidAverageSolution}
    \begin{align}
	f(\asig) &= c_1\,\asig,\\
	g(\asig) &= c_0 -\frac{\asig^2}{2}
	\textrm{.}
    \end{align}
\end{subequations}
The two remaining constants $c_{0}$ and  $c_{1}$ will be found by 
solving the matching problem, see Section~\ref{sec:Matching}.

\subsection{Solution of difference problem}
\label{ssec:SolutionOfDifferenceProblem}

We proceed to solve the equations for the difference
problem~(\ref{sys:BraidDifferenceEquations}), subjected to the
asymptotic conditions~(\ref{sys:BraidDifferenceBoundaryConditions}).
In the right-hand sides of Eqs.~(\ref{sys:BraidDifferenceEquations}),
the unknown contact pressure $\hat{p}(\asig)$ has to be determined, in
a way that is consistent with the contact set $\mathfrak{D}$:
$\hat{p}(\asig)$ can be nonzero for those $\asig$ that are elements of
$\mathfrak{D}$ only.  By Eq.~(\ref{eqn:DifferenceConstraint}) the
contact set $\mathfrak{D}$ depends on the difference variables only:
\begin{equation}
    \mathfrak{D} = \big\{\asig \quad\textrm{such that}\quad
    u^2(\asig) + v^2(\asig) = 1 \big\}
    \textrm{.}
    \label{eqn:Contact}
\end{equation}
This set will be found as an outcome of the solution of the difference
problem.

\subsubsection{Variational formulation}

Solution of the difference problem is greatly eased by pointing
out the simple variational structure underlying the equations.
We shall now show that solutions of the difference problems are 
minimizers of the following energy:
\begin{equation}
	E =
	\int_{-W}^{+W}\frac{{u''}^2(\asig)+{v''}^2(\asig)}{2}
	\,\mathrm{d}\asig
	+ v'(W) + v'(-W)
	\textrm{.}
	\label{eqn:BraidDifferenceEnergy}
\end{equation}
Minimization is done with respect to the functions $u(\asig)$ and
$v(\asig)$ which are defined over the interval $[-W,W]$, are twice
differentiable, and are subjected to the non-penetration
condition~(\ref{eqn:DifferenceConstraint}) --- here,
$W$ is large but fixed number. We shall also include an
additional constraint, related to the knot type:
the parametric curve $(u(\asig),v(\asig))$ has to make a prescribed
number of turns around the origin; this topological constraint is
discrete and so does not affect the Euler-Lagrange equations.  

Before we show that solutions of the braid
equations~(\ref{sys:BraidDifferenceEquations}) subject to asymptotic
conditions and constraints are minimizers of the
energy~(\ref{eqn:BraidDifferenceEnergy}) for large enough real numbers
$W$, we shall first give a physical interpretation of this energy.  To
this end, we define a virtual rod, called the difference rod, by the
following Cartesian equation: $\{x=u(\asig),y=v(\asig),z=\asig/\epsilon\}$.
Being based on the difference variables $u$ and $v$, this
difference rod winds around the $z$ axis exactly in the same way as
the strand $b$ winds around the strand $a$ in the original problem.
Note that this difference rods extends from $z=-W/\epsilon$ to
$z=+W/\epsilon$: because of the
factor $1/\epsilon$ in the definition the rod deviates only slightly
from the $z$ axis and its arc length is approximately $z\approx \asig
/ \epsilon$.  Using the small deflection approximation for this
difference rod, one can easily compute the unsigned curvature of its
centerline, $\kappa = \epsilon^2\,({u''}^2 + {v''}^2)^{1/2}$.  By
definition, the difference rod has zero twist, and we define its
bending modulus to  be$B_{\mathrm{diff}} = 1$.  Then its elastic
energy~(\ref{eqn:TotalEnergy}) reads
$\int_{-W/\epsilon}^{+W/\epsilon}\frac{\kappa^2}{2} \,\mathrm{d}z =
\epsilon^3\,\int_{-W}^{+W}\frac{{u''}^2(\asig)+{v''}^2(\asig)}{2}
\,\mathrm{d}\asig$.  Up to the factor $\epsilon^3$, which is
irrelevant for the minimization problem, this is exactly the first
term in our energy~(\ref{eqn:BraidDifferenceEnergy}).  The two
remaining terms, $v'(\pm W)$, can be interpreted as the work of the bending moments on the endpoints: the unit tangent to the rod is
$\{\epsilon\,u',\epsilon v',1\}$ and so a moment $(\epsilon^2\,\eX)$
applied on either end of the rod is associated with the potential
energy $\epsilon^3\,v'(\pm W)$. The quantity $\epsilon^3$ is then 
factored out of the total energy.  The
constraint~(\ref{eqn:DifferenceConstraint}) is interpreted by the fact
that the difference rod winds around a virtual cylinder whose axis is
the $z$ axis, with unit radius.  To sum up, we have identified the
energy~(\ref{eqn:BraidDifferenceEnergy}) as that of a virtual,
twistless, naturally straight rod winding around a \emph{fixed} cylinder, of unit radius
and axis $e_z$, and subjected to bending moments at its endpoints.
We have transformed the self-contact problem into a contact
problem with a fixed, external body, and this an important
simplification.

We shall now establish the equivalence of the constrained minimization
problem and the original braid
equations~(\ref{sys:BraidAverageEquations}), by working out the
Euler-Lagrange equations for the minimization problem.  First, let us
rewrite the constraint~(\ref{eqn:DifferenceConstraint}) as $Q\geq 0$,
where
\begin{equation}
	Q(\asig) =
	\frac{u^2(\asig)+v^2(\asig)}{\sqrt{2}}-\frac{1}{\sqrt{2}}
	\textrm{.} \nonumber
\end{equation}
Constrained minimization problems are classically solved by
introducing Lagrange multipliers, here the function $\pi(\asig)$, and 
enforcing stationarity of the augmented energy,
\begin{equation}
	\delta E - \int_{-W}^{+W}\pi(\asig)\,
	\delta Q(\asig)\,\mathrm{d}\asig=0
	\textrm{.} \nonumber
\end{equation}
Using the explicit expressions of $E$ and $Q$ given above and
integrating by parts, this yields:
\begin{multline}
	\int_{-W}^{+W}\left[
	\left(u''''-\sqrt{2}\,\pi(\asig)\,u\right)\,\delta u +
	\left(v''''-\sqrt{2}\,\pi(\asig)\,v\right)\,\delta v
	\right]\,\mathrm{d}\asig  \\
		{}+\bigg[ u''\,\delta u' - u'''\,\delta u -
		v'''\,\delta v \bigg]_{-W}^{+W}  \\
		{}+ \left(v''(W)+1\right)\,\delta v'(W) +
		\left(-v''(-W)+1\right)\,\delta v'(-W) =0.
\end{multline}
Here, square brackets with subscript and superscript denote boundary
terms coming from the integration by parts, $[f]_{a}^b = f(b) - f(a)$.
The quantity in the left-hand side has to be zero for arbitrary
variations $\delta u(\asig)$ and $\delta v(\asig)$.  Therefore, the
factors in front of $\delta u(\asig)$ and $\delta v(\asig)$ in the
integral have to vanish: after identification of the Lagrange
multiplier $\pi(\asig)$ with the rescaled contact pressure $\hat{p}$,
one recovers the equations~(\ref{sys:BraidDifferenceEquations}) of the
difference problem.  The remaining boundary terms in the variation
above yield the following boundary conditions:
\begin{subequations}
    \label{eq:}
\begin{align}
    u''(\pm W)&=0    & v''(\pm W)&=\mp 1 \\
    u'''(\pm W)&=0   & v'''(\pm W)&=0
    \textrm{.}
\end{align}
\end{subequations}
For large~\footnote{Convergence of our variational problem for large
$W$ is extremely simple: as we shall show, the minimizer becomes
independent of $W$ when  $W$ is larger than a
fixed number, which can be interpreted as the coordinate of
the last point of contact with the cylinder.} $W$, we recover the boundary
conditions~(\ref{sys:BraidDifferenceBoundaryConditions}) derived
earlier for the difference problem. This establishes the equivalence 
of the two formulations.

\subsubsection{Numerical solution of the universal braid problem}
\label{ssec:NumericalDifferenceProblem}

We have just reformulated the difference problem as a constrained
minimization problem.  We now take advantage of
this variational formulation and present a numerical solution 
which is very easy to implement.  The difference
problem has been formulated without any parameter: for any given knot
type, the solution of the braid problem is universal.  In
particular, note that the twist parameter $\aU$ has been removed from the
braid equations at dominant order: the braid is insensitive to the
applied twist.  These universal solutions are computed below, once for
all, for the trefoil ($3_{1}$) and cinquefoil ($5_{1}$) topologies.


We first implemented the minimization problem using the symbolic
calculation language 
Mathematica
which has built-in
capabilities for non-linear constrained optimization.  The
implementation is straightforward.  The problem is first reformulated
in polar variables $(w,\phi)$, such that $u=w\,\cos\phi$ and $v=
w\,\sin\phi$: the advantage is that the winding number about the $z$
axis is readily available from the end value of $\phi$.  Values of the
functions $w$ and $\phi$ are sampled on a uniform mesh covering the
positive axis, $\asig\in[0,W]$, and their values for negative $\asig$
are reconstructed using the parity
condition~(\ref{eqn:BraidMoreSymmetry}).  Finite differences are used
to evaluate the objective function~(\ref{eqn:BraidDifferenceEnergy}).
The non-penetration condition~(\ref{eqn:DifferenceConstraint}) is
enforced by a constraint $w\geq 1$ written at every point of the mesh.
In addition, we use a series of non-physical constraints: (i) we
require that $\left(n - \frac{1}{4}\right)\,\pi \leq \phi(W) \leq
(n+1)\,\pi$, where $n=1$ for a trefoil knot and $n=2$ for a cinquefoil
knot; (ii) we require that $|\phi(\asig_{i}) - \phi(\asig_{i+1})| \leq
\frac{\pi}{2}$ for any pair of neighboring mesh points $\asig_{i}$ and
$\asig_{i+1}$.  Constraint (i) is used to direct convergence towards
the solution having the required winding number, as the difference rod
has to make one and a half turn around the $z$ axis in the trefoil
case and two and a half turns in the cinquefoil case 
--- note that the winding number is given by $(\phi(W) -
\phi(-W))/(2\pi) = 2\,\phi(W) / (2\pi)$.  Constraint (ii) warrants
that $\phi$, defined modulo $2\pi$, varies smoothly along the rod
which is required for the end value $\phi(W)$ to express the total
number of turns.  We carefully checked that the non-physical
constraints (i) and (ii) are non-active when the minimization
procedure exits, \emph{i.  e.} all inequalities are \emph{strict}:
their role is simply to guide convergence towards a physically
relevant solution.

\begin{figure}
	\centering
	\includegraphics{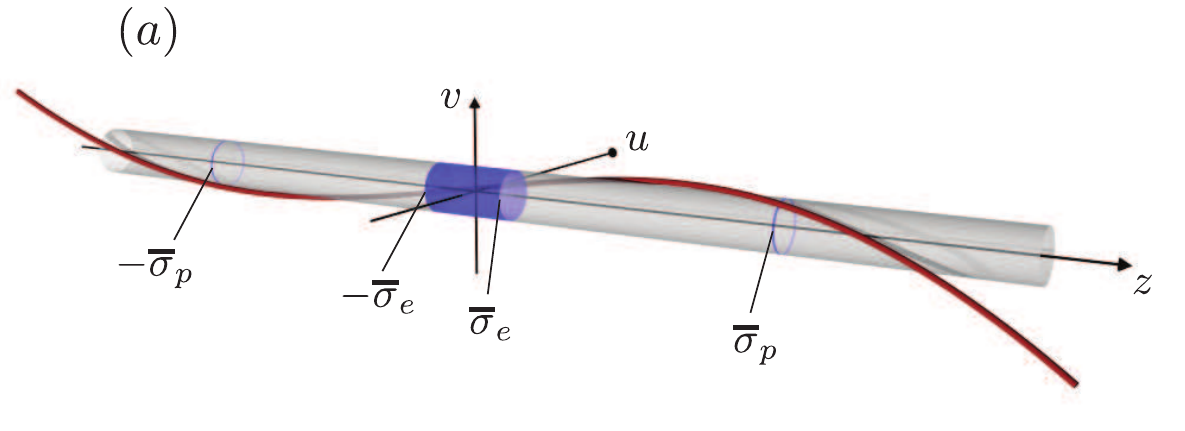}\\[-.5cm] %
	\includegraphics{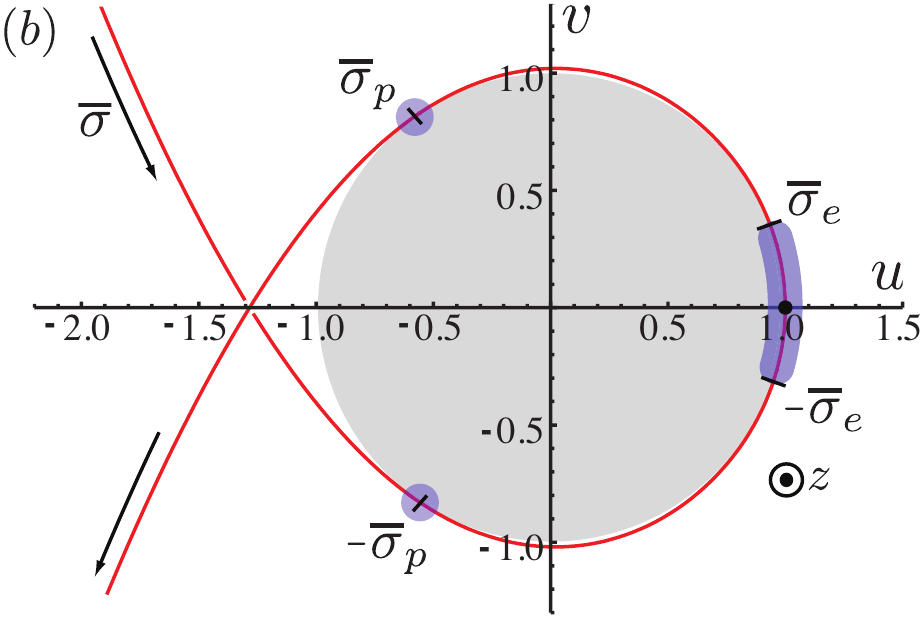}\\[1cm] %
	\includegraphics[width=10cm]{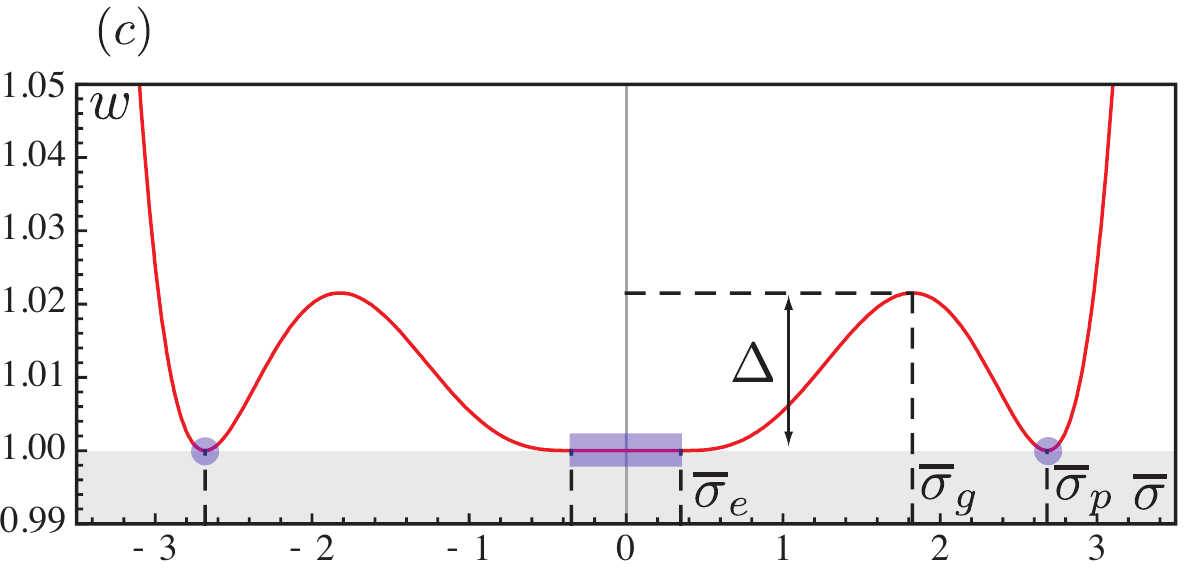} %
	\caption{Numerical solution of the difference problem of the
	braid for the trefoil topology ($n=1$).  The difference rod,
	shown in red, describes position of strand $b$ with respect to
	strand $a$ --- compare with Fig.~\ref{fig:BraidGeometry}.  It is held by bending moments applied at its
	endpoints, and enlaces an impenetrable cylinder of unit radius
	drawn around the $z$ axis.  The problem has no parameter and
	the solution depends on the knot type only.
	(a) 3D view, (b) projection onto the plane $(u,v)$
	perpendicular to the cylinder axis, (c) distance $w =
	\sqrt{u^2+v^2}$ of difference rod to cylinder axis: $w=1$ when
	there is contact, and $w>1$ otherwise.  The contact set is 
	denoted by shaded regions (in blue) along the solution.  
	Note the interval of contact
	around the center of symmetry ($-\asig_{e} \leq \asig \leq
	\asig_{e}$), flanked by two isolated points of contact
	($\asig = \pm\asig_{p}$).  Close examination of (b) reveals
	reopening in the intermediate regions ($\asig_{e}\leq
	|\asig|\leq \asig_{p}$).}
	\label{fig:DifferenceSolution}
\end{figure}

We found that the minimization always converges to the same type of solution
for both knot types, $n=1$ or $n=2$.  We used a typical mesh size of
$\Delta\asig \sim 0.1$ and interval width $W\sim 9$ --- we observed
that the numerical solution does not vary with $W$ when $W$ becomes larger than
 $4$, something that we shall explain soon.  In
\fignb{fig:DifferenceSolution}, the difference rod is visualized for
the trefoil topology.
By inspecting where the constraint $w\geq 1$ is active in the
numerical minimizers, we can determine which mesh points belong to the
contact set $\mathfrak{D}$.  When the mesh is not exceedingly coarse
$\Delta\asig \lesssim 1.5$,
and for both knot types, we found an interesting
contact topology: the contact set is
composed of an interval centered around the origin
and two symmetric isolated points (each corresponding to a single mesh point).
Starting at $\asig=0$, the difference rod is in continuous contact with
the cylinder, then lifts off from the cylinder and eventually touches it again at an isolated point.
This contact set is shown in Fig.~\ref{fig:DifferenceSolution}.
Note that this topology remains the same when the mesh size is
decreased.
This leads to the following topology for the contact set
$\mathfrak{D}$ of the difference problem:
\begin{equation}
    \mathfrak{D} = \{-\asig_{p}\} 
    \cup [-\asig_{e},\asig_{e}] \cup \{\asig_{p}\}
    \quad
    \textrm{where }0<\asig_{e} <\asig_{p} \textrm{.}
    \label{eqn:ContactSetTopology}
\end{equation}
Here $\asig_{e}$ is half the width of the central region with
continuous contact and $\asig_{p}$ is the rescaled coordinate of the
isolated point of contact.  We stress that this topology arises from
the numerical minimization without any a priori assumption on our
part.  In a problem where contact occurs along straight line in space,
\citet{ColemaSwigon-Theory-of-Supercoiled-Elastic-Rings-2000} have assumed a
topology of this form and checked that it was consistent.

By our definition of the difference variables, the distance $w$ of
difference rod to the $z$ axis is also the rescaled distance between the
centerlines of the strands $a$ and $b$ in the original problem:
contact of the difference rod with the virtual cylinder ($w=1$) means
that the physical strands $a$ and $b$ are in contact.  Like the
virtual bodies, the two physical strands experience continuous contact
in a central region; on both sides of this central region, they
separate by a small but finite distance, contact again at a point, and
finally separate for good.  The maximal reopening $\Lambda$ is given by
the extremum of the function $(w-1)$ in the interval $[\asig_{e},
\asig_{p}]$, see Fig.~\ref{fig:DifferenceSolution}; the corresponding
value of $\asig$ is called $\asig_{g}$.  These numerical values are
given in Table~\ref{tab:KnotNumResults}.
\begin{table}
	\centering
	\begin{tabular*}{0.60\columnwidth}{@{\extracolsep{\fill}}c|ccccc}
		knot type & $\asig_e$ & $\asig_p$ & $\asig_g$ & 
		$\Lambda$\\
		\hline
		$3_1$ & 0.348 & 2.681 & 1.823 & 
		0.022\\
		$5_1$ & 4.504 & 6.814 & 5.962 & 
		0.021
	\end{tabular*}
	\caption{Numerical values of the contact-set parameters for
	$3_1$ and $5_1$ knots.\label{tab:KnotNumResults}}
\end{table}
Values of $\Lambda$ are very close for trefoil and cinquefoil knots,
$\Lambda\approx 0.021$; in physical units, this corresponds to an
inter-strand reopening of $(0.043\,h)$, that is
$43\;\mu\mathrm{m}$ for a rod of radius $h=1\;\mathrm{mm}$. 
The experiments reported in Section~\ref{ssec:BraidOpening} confirm 
the presence of these openings.

\label{ssec:ContactSetTopologyWasConfirmedByShooting}
In order to confirm our hypothesis on the topology of the contact set,
we implemented an independent numerical solution for the difference
problem, assuming a topology of the form~(\ref{eqn:ContactSetTopology}).
This independent solution relies on non-linear shooting:
in contrast to the energy minimization scheme, it
involves a numerical integration of the equations of equilibrium;
this new approach is much more accurate but requires the contact
topology to be known.
Numerical integration is carried out on each interval
$\asig \in [0;\asig_{e}]$, $[\asig_{e} ; \asig_{p}]$
and $[\asig_{p} ; \infty]$ in turn.
In the first interval the difference rod lies on the
surface of the virtual cylinder with unit radius,
and we use the polar variables $(w, \phi)$, introduced earlier, with $w=1$.
The polar variable $\phi(\asig)$ satisfies the differential
equation $\phi''''=6 (\phi')^2 \, \phi''$.
Four initial conditions are required to integrate this equation, two
of which are fixed by the symmetry
condition~(\ref{eqn:BraidMoreSymmetry}), $\phi(0)=0$ and $\phi''(0)=0$;
the other two, $\phi'(0)$ and $\phi'''(0)$, are unknowns of the
shooting procedure.
All the other quantities can be reconstructed from $\phi(\asig)$.
At $\asig_{e}$ a jump $P_{e}$ in the internal force is introduced. It represents a
Dirac contribution to the contact pressure~\footnote{Dirac 
contributions to the contact pressure appear generically at the 
boundary of the contact set in contact problems for elastic rods, as 
shown  for instance 
by~\cite{ColemaSwigon-Theory-of-Supercoiled-Elastic-Rings-2000}, or 
\cite{Audoly-Pomeau-Elasticity-and-geometry:-from-2008}.}. 
The values of $\phi$ and its derivatives at the end of the first interval are
combined with $P_e$ to evaluate the initial conditions for the second interval.
In the second interval, there is no contact and
Eqs.~(\ref{sys:BraidDifferenceEquations})
are integrated with $\hat{p}(\asig)=0$.
At $\asig_{p}$ the rod touches the cylinder and there is another
discontinuity $P_{p}$ in the internal force.
In the last interval $[\asig_{p} ; +\infty]$ there is no contact and
the internal force is again constant.
By the asymptotic conditions~(\ref{sys:BraidDifferenceBoundaryConditions})
this constant force has to be zero.
This implies in turn that the internal moment is constant.
In view of this the four asymptotic
conditions~(\ref{sys:BraidDifferenceBoundaryConditions}), which concern
the internal force and moment, have to be be satisfied over the 
entire third interval. 
Overall, the shooting scheme involves six unknowns 
$\{\phi'(0),\phi'''(0),\asig_{e}, \asig_{p}, P_{e},P_{p}\}$
which must satisfy six equations, namely two geometric contact conditions at $\asig_{p}$ and
four conditions coming from Eq.~(\ref{sys:BraidDifferenceBoundaryConditions}).
For the trefoil knot the non-linear shooting procedure converges to
$\{\phi'(0)=0.769,\phi'''(0)=0.033,\asig_{e}=0.348,
\asig_{p}=2.681, P_{e}=0.170,P_{p}=0.442\}$.
The contact
pressure can be reconstructed in the first interval as
$\pi(\asig) = \hat{p}(\asig)=(\phi'^4 - 3 \phi''^2 -4 \phi' 
\, \phi''')/\sqrt{2}$. It is plotted over the full contact set
$\mathfrak{D}$ in Fig.~\ref{fig:DifferenceSolutionAddons}\emph{a}:
\begin{figure}
 \centering
  \includegraphics{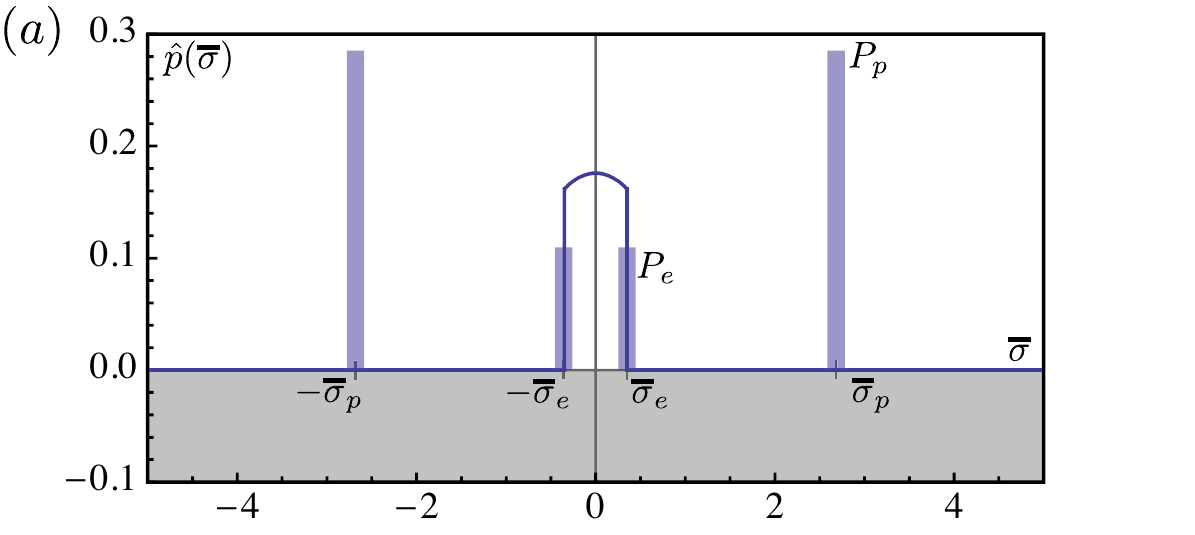}
 \includegraphics{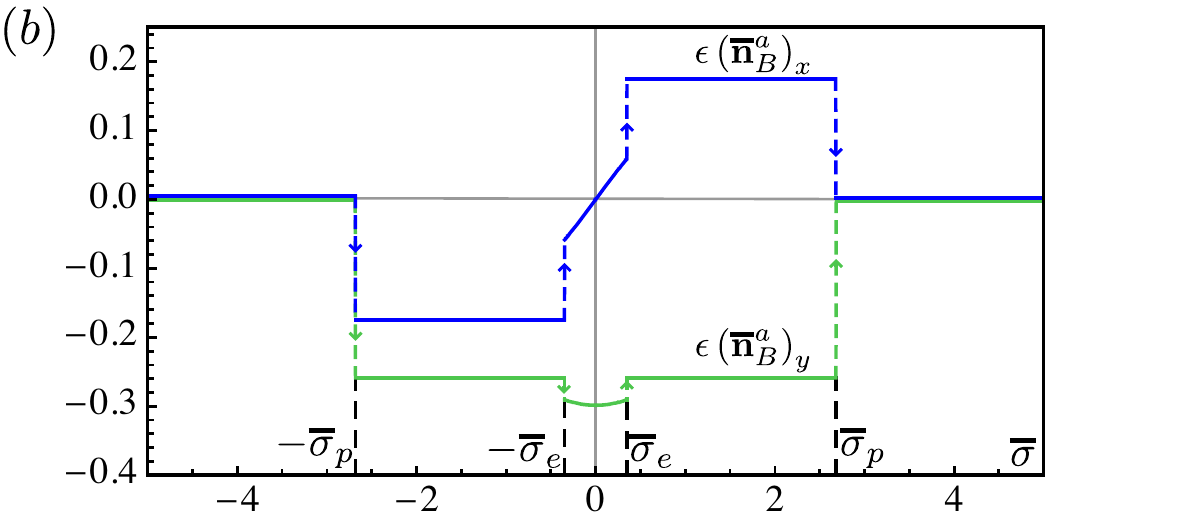}
 \caption{Forces in braid for the trefoil geometry, same solution as
 in Fig.~\ref{fig:DifferenceSolution}: (a)~rescaled
 contact pressure; (b)~rescaled internal force.  The pressure is
 everywhere non-negative and this validates the assumption on
 the contact topology.  The internal force is proportional to
 $u'''$ and $v'''$ by
 Eq.~(\ref{eqn:InternalForceInTheStrandsInBraid}). The localized
 contact forces $P_{e}$ and $P_{p}$ are represented by columns in (a),
 and manifest themselves as jumps in (b).}
 \label{fig:DifferenceSolutionAddons}
\end{figure}
it is everywhere positive and this validates our assumption on the
topology (in Appendix~\ref{app:Topologies}, we test different contact
topologies and show that they lead to negative pressure and/or 
residual penetration).

The internal force is given by Eq.~(\ref{eqn:BraidForce})
for strand $a$ and a similar equation holds for $b$.  Noticing that
the third derivatives of the average solution given by
Eq.~(\ref{sys:BraidAverageSolution}) vanish, we find the
nonzero components of the internal force in each strand:
\begin{equation}
	(\an_{B}^a)_{x} = -(\an_{B}^b)_{x}
	= +\frac{u'''(\asig)}{\sqrt{2}\,\epsilon}
	,
	\qquad
	(\an_{B}^a)_{y} = -(\an_{B}^b)_{y}
	= +\frac{v'''(\asig)}{\sqrt{2}\,\epsilon}
	\textrm{.}
    \label{eqn:InternalForceInTheStrandsInBraid}
\end{equation}
The rescaled internal force is plotted in
Figure~\ref{fig:DifferenceSolutionAddons}\emph{b}.
Note the discontinuities of the internal force at the boundaries of the
contact set, where Dirac pressure forces are present.  


\subsubsection{Polynomial expression beyond last contact point}
\label{sssec:PolynomialExpressionBeyondLastContactPoint}

In Fig.~\ref{fig:DifferenceSolutionAddons}, the rescaled internal
force appears to be zero beyond the isolated contact
point, that is for $|\asig|>\asig_{p}$.  From 
Eq.~(\ref{eqn:InternalForceInTheStrandsInBraid}), 
the third derivatives of the functions $u(\asig)$ and $v(\asig)$ vanish in this region.
As a result, both
$u$ and $v$  are polynomials functions of $\asig$ of
order at most three.  In addition their cubic term has to be zero for the
asymptotic 
conditions~(\ref{eqn:BraidDifferenceBoundaryConditions-Order2}) to be
satisfied.  Therefore both $u(\asig)$ and $v(\asig)$ are
second order polynomials for $|\asig|>\asig_{p}$.  The quadratic
term is fixed by the other asymptotic
conditions~(\ref{eqn:BraidDifferenceBoundaryConditions-Order2}): it is
zero for $u(\asig)$, which is therefore an affine function, and it is
$-\frac{1}{2}$ for $v(\asig)$. Consequently, for 
$\asig>\asig_{p}$, $u$ is of the form $u(\asig) = \Lambda_n\,\asig +
q_{n}$ for some real constants $\Lambda_{n}$ and $q_{n}$, and $v$ is
of the form $v(\asig) = -\frac{\asig^2}{2} + \Pi_n\,\asig +q_{n}'$ for
some constants $\Pi_{n}$ and $q_{n}'$.  The expressions for
$\asig < -\asig_{p}$ are found using the parity
conditions~(\ref{eqn:BraidMoreSymmetry}).  The following condensed
notation summarizes both cases $\asig>\asig_{p}$ and $\asig<
-\asig_p$ (which are denoted generically as $\pm\asig > \asig_{p}$):
\begin{subequations}
    \label{sys:BraidDifferenceAsymptotic}
    \begin{align}
	u(\asig) & = \pm \Lambda_n\,\asig + q_{n},\\
	v(\asig) & = \mp \frac{\asig^2}{2} + \Pi_n\,\asig \pm q'_{n}
	\textrm{.}
    \end{align}
\end{subequations}
In this condensed notation, one should use the upper sign on the
positive side, for $+\asig > \asig_{p}$, that is replace $\pm$ with $(+)$
and $\mp$ with $(-)$, and the lower sign on the negative side, for $-\asig
> \asig_{p}$.

The coefficients $\Lambda_n$, $\Pi_n$, $q_{n}$ and $q'_{n}$ are
available from the numerical solution of
Section~\ref{ssec:NumericalDifferenceProblem}.  The values of
$\Lambda_n$ and $\Pi_n$ are given in
Table~\ref{tab:BraidConstantsNumResults}.
\begin{table}
	\centering
	\begin{tabular*}{0.55\columnwidth}{@{\extracolsep{\fill}}c|cc}
		knot type & $\Lambda_n$ & $\Pi_n$ \\
		\hline
		$3_1$ ($n=1$) & $-0.87759$ & $2.089$ \\
		$5_1$ ($n=2$) & $-0.87738$ & $6.223$ 
	\end{tabular*}
	\caption{Numerical values of braid constants $\Pi_n$ and
	$\Lambda_n$.}
	\label{tab:BraidConstantsNumResults}
\end{table}
As we shall see later, the values of $q_{n}$ and $q'_{n}$ are
irrelevant at dominant order and are not given here.


\subsection{Asymptotic expansions at braid-tail and braid-loop junctions}

In order to match this inner solution with the outer solutions
computed earlier, we shall need its expansion far away from the braid,
that is for large values of $\asig$.
In Eq.~(\ref{eqn:BraidAverageAndDiffVariables})  the inner solution is
decomposed  into an average and a difference solution.
The average solution is a polynomial given by
Eq.~(\ref{sys:BraidAverageSolution}). The difference solution
is polynomial as well for large enough values of $\asig$, 
see Eq.~(\ref{sys:BraidDifferenceAsymptotic}).
It is straightforward
to combine these polynomials to obtain the expansion of the braid solution
for $\asig \to \pm\infty$:
\begin{subequations}
    \label{sys:WorkOutBraidExpansions}
\begin{multline}
    \ax^a_{B}(\az) = \epsilon^2\,\hx_{B}^a(\asig)
    = \frac{\epsilon^2}{\sqrt{2}}\,\left(f(\asig) - u(\asig)\right)
    \\
    = \frac{\epsilon^2}{\sqrt{2}}\,\left((c_{1}\mp 
    \Lambda_{n})\,\asig - q_{n}\right)
    = 
    \epsilon\,\az\,\left(\frac{\mp\Lambda_{n}+c_{1}}{\sqrt{2}}\right)
    +\mathcal{O}(\epsilon^2)
    \textrm{.}
    \label{eqn:WorkOutBraidExpansions-X}
\end{multline}
Here, we use the same condensed notations as in
Eqs.~(\ref{sys:BraidDifferenceAsymptotic}), whereby the compound signs
$\pm$ and $\mp$ must be replaced by the upper symbol when
$\asig\to+\infty$, and by the lower one when $\asig\to-\infty$.  Note
that we have replaced the stretched variables $\hx_{B}^a$ and $\asig$
with the barred variables $\ax_{B}^a$ and $\az$ using
Eq.~(\ref{eq:BraidParametricEquationOfSigmaBar}), as the matching has
ultimately to be done using a common set of variables for the inner
and outer solutions. 

A similar calculation for $\ay^a_{B}$ yields:
\begin{multline}
    \ay^a_{B}(\az)   = \epsilon\,\tau_{B} + \epsilon^2\,\hy_{B}^a(\asig)
    = \epsilon\,\tau_{B} +  \frac{\epsilon^2}{\sqrt{2}}\,\left(g(\asig) - v(\asig)\right)
\\    = \epsilon\,\tau_{B} +  \frac{\epsilon^2}{\sqrt{2}}\,
    \left(
    (-1\pm 1)\,\frac{\asig^2}{2}
    -\Pi_{n}\,\asig+(c_{0}\mp q_{n}')
    \right)
    \\ = \frac{\az^2}{\sqrt{2}}\,\left(
    \frac{-1\pm 1}{2} 
    \right)
    + \epsilon\,\left(\tau_{B} - \az\,\frac{\Pi_{n}}{\sqrt{2}}\right)
    +\mathcal{O}(\epsilon^2) \textrm{.}
    \label{eqn:WorkOutBraidExpansions-Y}
\end{multline}
\end{subequations}
In this equation, the coefficient $\tau_{B}$ represents an
infinitesimal translation of the braid along the $y$ axis. The term 
proportional of $c_{1}$ in Eq.~(\ref{eqn:WorkOutBraidExpansions-X}) 
is very similar: it represents an infinitesimal rotation of the 
braid about the $y$ axis. We rename it $\omega_{B}$,
\begin{equation}
    \omega_{B} = \frac{c_{1}}{\sqrt{2}}
    \textrm{.}
    \label{eqn:DefineOmegaB}
\end{equation}
The two other coefficients in Eqs.~(\ref{sys:WorkOutBraidExpansions})
have been computed in Table~\ref{tab:BraidConstantsNumResults}. We 
call \emph{internal parameters of the braid} the two remaining free 
parameters in the above expansions:
\begin{equation}
    \mathbf{\Psi}_B=(\tau_B,\,\omega_B)
    \textrm{.}
    \label{eqn:BraidInternalParameters}
\end{equation}
In the next section, we shall show how these parameters
$\mathbf{\Psi}_B$ can be found as a function of the applied loading
and knot type, together with the loop and tail parameters
$\mathbf{\Psi}_{L}$ and $\mathbf{\Psi}_{T}$.

We can rewrite the expansions~(\ref{eqn:WorkOutBraidExpansions-X})
and~(\ref{eqn:WorkOutBraidExpansions-Y}) in the form
\begin{subequations}    
    \label{sys:BraidAsymptoticExpansions}
\begin{alignat}{2}
    \ax^a_{B}(\az) & = && 
    \epsilon\,X^\pm_B + \epsilon\,\az\,X^{\pm\prime}_B + \cdots
    \label{eqn:BraidAsymptoticExpansions-X} \\
    \ay^a_{B}(\az) & =  &
    \left(
    \frac{-1\pm 1}{2} 
    \right)\,\frac{\az^2}{2\,\aR} \;
    + \; &
    \epsilon\,Y^\pm_B + \epsilon\,\az\,Y^{\pm\prime}_B + \cdots
    \label{eqn:BraidAsymptoticExpansions-Y}
\end{alignat}
\end{subequations}
where $\aR$ is a shorthand for $1/\sqrt{2}$ by
Eq.~(\ref{eqn:AraiRadius}).  Note that the factor in parenthesis in
Eq.~(\ref{eqn:BraidAsymptoticExpansions-Y}) is equal to $(-1-1)/2
= -1$ on the negative side, and to $(-1+1)/2=0$ on the positive side.
As a result, the quadratic term in $\ay^a_{B}(\az)$ is equal to
$-\az^2 / (2\,\aR)$ for large negative $\asig$, which is consistent
with expansion~(\ref{eqn:LoopAsymptoticExpansions-Y}) for the loop,
and is absent for large positive $\asig$, which is consistent with
expansion~(\ref{eqn:TailsAsymptoticExpansions-Y}) for the tail.

The coefficients of the polynomial expansions just written are given 
by identification with Eqs.~(\ref{sys:WorkOutBraidExpansions}):
\begin{equation}
	\begin{pmatrix}
	X^\pm_B \\
	X^{\pm\prime}_B \\
	Y^\pm_B \\
	Y^{\pm\prime}_B
	\end{pmatrix}
	= \mathbf{M}^\pm_B\cdot\mathbf{\Psi}_B + \mathbf{V}^\pm_B(n)
	,
	\quad\textrm{where } \mathbf{M}^\pm_B=
	\begin{pmatrix}
	0 & 0\\
	0 & 1\\
	1 & 0\\
	0 & 0
	\end{pmatrix}
	\textrm{ and } \mathbf{V}^\pm_B(n)=
	\begin{pmatrix}
	0\\
	\mp\frac{\Lambda_n}{\sqrt{2}}\\
	0\\
	-\frac{\Pi_n}{\sqrt{2}} \end{pmatrix}.
	\label{eqn:BraidMatrixExpansion}
\end{equation}
Equation~(\ref{eqn:BraidMatrixExpansion}) defines two constant
matrices $\mathbf{M}^-_B$ and $\mathbf{M}^+_B$, and two vectors
$\mathbf{V}^-_B(n)$ and $\mathbf{V}^+_B(n)$ depending on the knot type
$n$.  These vectors are defined in terms of the braid constants found
in Section~\ref{ssec:SolutionOfDifferenceProblem}.

For the matching problem studied in the next Section, it is useful to
give a precise description of the range of values of $\az$ where the
expansions~(\ref{sys:BraidAsymptoticExpansions}) hold, that is
where the omitted terms denoted by ellipses are actually negligible.
These expansions have been obtained~\footnote{As explained in
Section~\ref{sssec:PolynomialExpressionBeyondLastContactPoint}, the
braid actually reaches its asymptotic behavior \emph{exactly} as soon
as the last point of contact is passed, $|\asig|>\asig_{p}$.  This is
not important and we shall continue to write the less severe
requirement $|\asig| \gg 1$, which holds in general in boundary or
inner layer analysis.} by taking the limit $|\asig|\to\infty$: they
obviously require $|\asig| \gg 1$, that is $|\az| \gg \epsilon$.
However, this is not the only assumption.  Recall that the braid has
been studied based on the small displacement approximation, which
assumes that the tangents $\at_{B}^a$ or $\at_{B}^b$ remain close to
the vector $\eZ$ --- see for instance the tangent
expansion~(\ref{eqn:BraidTangent}).  This assumption breaks down in
the inside of the loop, for values of $\az$ of order 1: as shown by
the quadratic term in Eq.~(\ref{eqn:BraidAsymptoticExpansions-Y})
or directly by the loop solution~(\ref{eqn:LoopZeroRadiusT}), the
tangent deflects from the $z$ axis by an angle $(\az/\aR)$ there.
Therefore, the braid solution accurately describes the upper part of
the loop, where it merges with the braid, but does not accurately describe the
whole loop: it assumes $|\az| \ll 1$.  To summarize, the range of
validity of the braid
expansions~(\ref{sys:BraidAsymptoticExpansions}) is
\begin{equation}
    \epsilon \ll |\az| \ll 1
    \textrm{.}
    \label{eqn:RangeOfValidityBraidExpansion}
\end{equation}
The linear relations in Eq.~(\ref{eqn:BraidMatrixExpansion})
yield the braid expansions in the regions where it connects with the
loop ($-1\ll \az \ll -\epsilon$) and with the tail ($\epsilon \ll \az \ll
1$). These relations depend on the internal parameters of the braid,
$\mathbf{\Psi}_B=(\tau_B,\,\omega_B)$, and on the knot type $n$.  These
equations~(\ref{sys:BraidAsymptoticExpansions})
and~(\ref{eqn:BraidMatrixExpansion}) capture all what we need to
know about the inner solution (braid) to be able to solve the problem
globally.

\section{Matching}
\label{sec:Matching}

So far, we have solved the equilibrium equations in the tail, loop,
and braid regions independently.  In each region the solution depends
on some parameters, collectively denoted $\mathbf{\Psi}_T$,
$\mathbf{\Psi}_L$ and $\mathbf{\Psi}_B$, which have yet to be
computed.  Figure~\ref{fig:KnotDomains} illustrates the fact that these domains
overlap in the so-called \emph{intermediate regions}.  There are two
types of intermediate regions, one where the loop merges with the
braid, and the other one where the tail merges with the braid.  By
writing down the matching condition of the various pieces of solutions
obtained so far in these intermediate regions, we make sure that we
have constructed a smooth, global solution of the original problem.
We now derive these matching conditions and compute the
remaining parameters $\mathbf{\Psi}_T$, $\mathbf{\Psi}_L$ and
$\mathbf{\Psi}_B$.


\subsection{Matching braid and tail}

In the end of our analysis of the tail regions, in
Eq.~(\ref{sys:TailsAsymptoticExpansions}), we have obtained the
following expansion:
\begin{subequations}
    \label{sys:BTmatching1}
    \begin{align}
        \ax_T(\az) &= \epsilon\,X_T + \epsilon\,\az\,X'_T + \cdots,
	\\
	\ay_T(\az) &= \epsilon\,Y_T + \epsilon\,\az\,Y'_T + \cdots
    \end{align}
\end{subequations}
where the terms that have been dropped, of order $\epsilon^2$ and
$\epsilon\,\az^2$, are negligible if $\az \ll \epsilon^{1/2}$.  In
Eq.~(\ref{sys:BraidAsymptoticExpansions}), we have found a similar
expansion based on the braid solution:
\begin{subequations}
    \label{sys:BTmatching2}
    \begin{align}
        \ax^{a}_B(\az) & = \epsilon\,X^+_B
	+ \epsilon\,\az\,X^{+\prime}_B + \cdots\\
	\ay^{a}_B(\az) & = \epsilon\,Y^+_B + \epsilon\,\az\,Y^{+\prime}_B +
	\cdots
    \end{align}
\end{subequations}
which is valid for $\epsilon\ll |\az|\ll 1$.  These two expansions
have to be consistent in the region of overlap, defined by
$\epsilon\ll \az \ll \epsilon^{1/2}$, and this implies the equality of
the coefficients.  We obtain the following matching condition in the
intermediate region between braid and tail:
\begin{equation}
	\begin{pmatrix}
	X_T\\
	X'_T\\
	Y_T\\
	Y'_T
	\end{pmatrix}
	=
	\begin{pmatrix}
	X^+_B\\
	X^{+\prime}_B\\
	Y^+_B\\
	Y^{+\prime}_B \end{pmatrix}. \nonumber
\end{equation}
Using the reduced matrices and vectors of the tail and braid problems
defined in Eqs.~(\ref{eqn:TailsMatrixExpansion})
and~(\ref{eqn:BraidMatrixExpansion}), this matching condition is
rewritten as a linear system for the tail variables $\mathbf{\Psi}_{T}
= (\lambda,\mu)$ and the braid variables $\mathbf{\Psi}_{B} =
(\tau_{B}, \omega_{B})$:
\begin{equation}
	\mathbf{M}_T(\aU)\cdot\mathbf{\Psi}_T = \mathbf{M}^+_B\cdot\mathbf{\Psi}_B +
	\mathbf{V}^+_B(n), \nonumber
\end{equation}
which reads:
\begin{equation}
	\begin{pmatrix}
		1 & 0\\
		-a(\aU) & -b(\aU)\\
		0 & 1\\
		b(\aU) & -a(\aU)
	\end{pmatrix}
	\cdot
	\begin{pmatrix}
		\lambda \\
		\mu
	\end{pmatrix}
	=
	\begin{pmatrix}
		0 & 0\\
		0 & 1\\
		1 & 0\\
		0 & 0
	\end{pmatrix}
	\cdot
	\begin{pmatrix}
		\tau_B\\
		\omega_B
	\end{pmatrix}
	+
	\begin{pmatrix}
		0\\
		-\frac{\Lambda_n}{\sqrt{2}}\\
		0\\
		-\frac{\Pi_n}{\sqrt{2}} \end{pmatrix}
		\textrm{.}
		\label{eqn:TailBraidMatchingLinearSystem}
\end{equation}
The functions $a(\aU)$ and $b(\aU)$ were defined in
\eqnb{eqn:TailsRoots}.

This is a set of four linear equations for the four unknowns
$\lambda$, $\mu$, $\tau_{B}$ and $\omega_{B}$.  As can be checked
easily, the determinant of this linear system is $a(\aU)$.  For
$\aU\neq \pm 2$, $a(\aU)\neq 0$ and this system has a unique solution,
which can be found explicitly by elimination:
\begin{subequations}
    \label{sys:TailBraidMatchingResults}
\begin{align}
    \lambda(\aU,n) &= 0
    & \mu(\aU,n) & =
    \frac{\Pi_n}{\sqrt{2-\mfrac{\aU^2}{2}}},
    \label{sys:TailMatchingResults}
    \\
    \tau_B(\aU,n) & =
    \frac{\Pi_n}{\sqrt{2-\mfrac{\aU^2}{2}}}
    & 
    \omega_B(\aU,n) & = \frac{\Lambda_n}{\sqrt{2}}
    -\frac{\aU\,\Pi_n}{\sqrt{2\,\left(4-\aU^2\right)}}
    \textrm{,}
    \label{sys:BraidMatchingResults}
\end{align}
\end{subequations}
after using the detailed expressions for $a(\aU)$ and $b(\aU)$.

We have just found the internal parameters of the tail and of the
braid, $\mathbf{\Psi}_T$ and $\mathbf{\Psi}_B$, as a function of the
dimensionless parameters of the problem, the loading parameter $\aU$
and the knot type $n$ --- recall that the braid constants $\Pi_n$ and
$\Lambda_n$ were given in Table~\ref{tab:BraidConstantsNumResults} for
trefoil ($n=1$) and double knots ($n=2$).

\subsection{Matching braid and loop}

\begin{figure}
	\centering
	\includegraphics{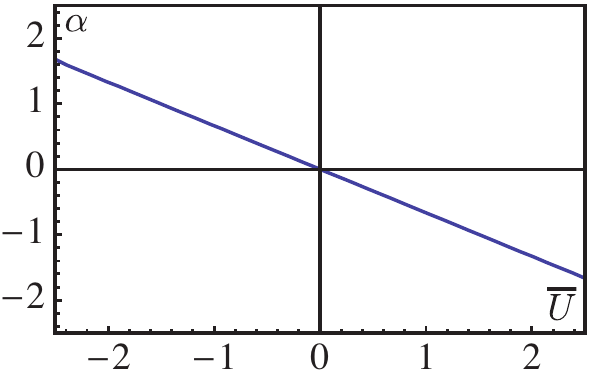} 
	\hspace{1cm}
	\includegraphics{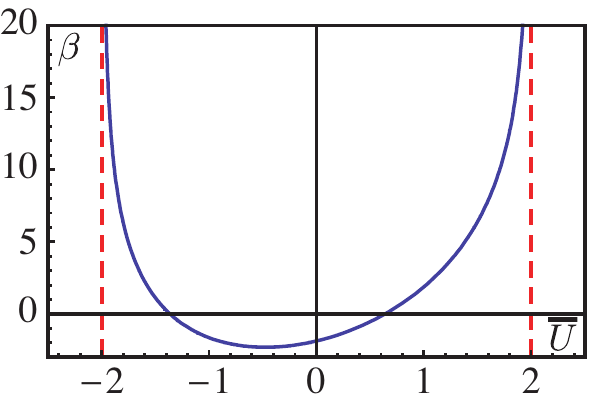}\\[1cm] 
	\includegraphics{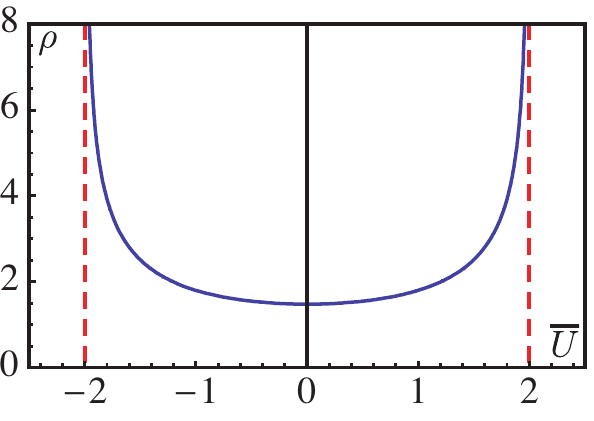} 
	\hspace{1cm}
	\includegraphics{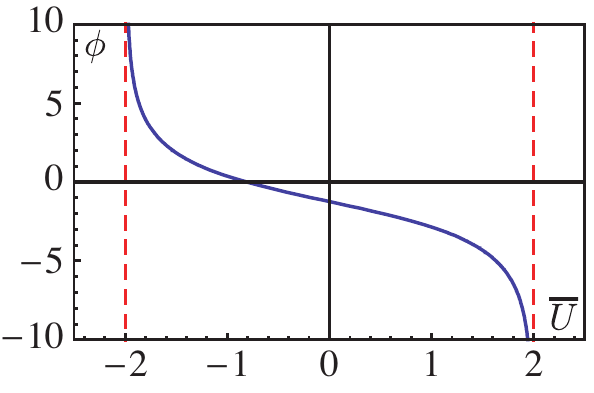} 
	\caption{Loop internal parameters $\alpha$, $\beta$, $\rho$
	and $\phi$ as functions of reduced loading parameter $\aU$,
	for a trefoil knot ($n=1$).  By 
	Eq.~(\ref{sys:LoopParametersSolutions-Alpha}), $\alpha$ varies linearly with
	$\aU$ .
	The other parameters, $\beta$, $\rho$ and $\phi$, all diverge at
	$\aU=\pm 2$, as denoted by the dashed lines (red).  This divergence
	points to the helical instability undergone by the tails as the applied 
	torque approaches the critical value $\aU=\pm 2$.}
	\label{fig:LoopAlphaBetaRhoPhi}
\end{figure}

A similar argument holds for the intermediate region between braid and
loop.  In Section~\ref{sssec:AsymptoticExpansionLoop}, we found that
the top of the loop is accurately described by the expansion
\begin{subequations}
    \label{sys:BLmatching1}
    \begin{align}
        \ax_L(\az) &= \epsilon\,X_L +
	\epsilon\,\az\,X'_L + \cdots \\
	\ay_L(\az) &= -\frac{\az^2}{2\,\aR}+\epsilon\,Y_L +
	\epsilon\,\az\,Y'_L + \cdots
    \end{align}
\end{subequations}
This expansion is accurate up to terms of order $\epsilon^2$, provided 
$(-\az) \ll \epsilon^{1/2}$, and of course $\az <0$. On the other 
hand, the braid solution has been expanded as
\begin{subequations}
    \label{sys:BLmatching2}
    \begin{align}
	\ax^{a}_B(\az) &= \epsilon\,X^-_B
	+ \epsilon\,\az\,X^{-\prime}_B + \cdots \\
	\ay^{a}_B(\az) &= -\frac{\az^2}{2\aR} + \epsilon\,Y^-_B +
	\epsilon\,\az\,Y^{-\prime}_B + \cdots
    \end{align}
\end{subequations}
in the domain defined by $-1 \ll \az \ll -\epsilon$. The intermediate 
region between braid and loop is defined by $-\epsilon^{1/2} \ll \az 
\ll -\epsilon $. There, the two expansions have to be compatible, 
which leads to the matching condition:
\begin{equation}
	\begin{pmatrix}
	X_L\\
	X'_L\\
	Y_L\\
	Y'_L
	\end{pmatrix}
	=
	\begin{pmatrix}
	X^-_B\\
	X^{-\prime}_B\\
	Y^-_B\\
	Y^{-\prime}_B \end{pmatrix} \nonumber
	\textrm{.}
\end{equation}
As earlier, we arrive at a linear system which can be written in terms
of the reduced matrices and vectors of the loop and braid problems,
defined earlier in Eqs.~(\ref{eqn:LoopMatrix})
and~(\ref{eqn:BraidMatrixExpansion}):
\begin{equation}
	\mathbf{M}_L(\aU)\cdot\mathbf{\Psi}_L = \mathbf{M}^-_B\cdot\mathbf{\Psi}_B +
	\mathbf{V}^-_B(n)
	\textrm{.} \nonumber
\end{equation}
We arrive at a linear system for the loop parameters:
\begin{equation}
    \label{eqn:BLMatchingMatrixForm}
	\mathbf{M}_L(\aU) \cdot
	\begin{pmatrix}
		\alpha\\
		\beta\\
		\rho\\
		\phi
	\end{pmatrix}
	=
	\begin{pmatrix}
		0\\
		\frac{\Lambda_n}{\sqrt{2}} + \omega_{B}\\
		\tau_{B}\\
		-\frac{\Pi_n}{\sqrt{2}} \end{pmatrix},
\end{equation}
where the quantities $\tau_{B}$ and $\omega_{B}$ in the right-hand side
are given by Eq.~(\ref{sys:BraidMatchingResults}) and the $4\times 4$
matrix $\mathbf{M}_L(\aU)$ is defined in \eqnb{eqn:LoopMatrix}. 
The determinant of the matrix $\mathbf{M}_L(\aU)$ can be computed
exactly; it vanishes for
\begin{equation}
    \det\mathbf{M}_L(\aU) = 0 \quad
    \textrm{iff}
    \quad
    \aU = \pm\sqrt{2\,(j^2- 1)},
    \textrm{ for some integer $j\geq 2$}
    \textrm{.}
    \label{eqn:DeterminantLOfMl}
\end{equation}
For any other value of $\aU$, one can solve the linear system in
Eq.~(\ref{eqn:BLMatchingMatrixForm}) by elimination.  This yields the
following expressions for the internal parameters of the loop:
\begin{subequations}
    \label{sys:LoopParametersSolutions}
    \begin{align}
	\alpha(\aU,n) &=
	-\frac{\Pi_n}{\pi}\,\aU,
	\label{sys:LoopParametersSolutions-Alpha}\\
	\beta(\aU,n) &= \aU\left( -2\,\Lambda_n +
	\frac{\aU\,\Pi_n}{\sqrt{4-\aU^2}} +
	\frac{\aU\,\sqrt{2}\,\Pi_n}{\pi\,\left(2+\aU^2\right)} -
	\frac{\aU}{\sqrt{2+\aU^2}}\,\frac{\Pi_n}{\tan\left(\pi\,\overline{K}_L(\aU)\right)}\right),\\
	\rho(\aU,n) &=
	\Pi_n\,\left(\frac{1}{\sqrt{2-\aU^2/2}}-\frac{\aU^2}{\pi\,\left(2+\aU^2\right)}
	-
	\frac{\cot\left(\frac{\pi}{2}\overline{K}_L(\aU)\right)}{2\,\overline{K}_L(\aU)}\right),\\
	\phi(\aU,n) &= \sqrt{2}\,\Lambda_n -\Pi_n\,\aU\left(
	\frac{1}{2\,\sqrt{2-\aU^2/2}}
	+\frac{2}{\pi\,\left(2+\aU^2\right)} -
	\frac{\cot\left(\frac{\pi}{2}\overline{K}_L(\aU)\right)}{2\,\overline{K}_L(\aU)}
	\right), 
    \end{align}
\end{subequations}
where the braid constants $\Lambda_n$ and $\Pi_n$ are given in
Table~\ref{tab:BraidConstantsNumResults}.  These functions are plotted
in \fignb{fig:LoopAlphaBetaRhoPhi} for the trefoil topology.
We recall that $\alpha$ and $\beta$ measure the first order
perturbation to the internal force in the loop, see
Eq.~(\ref{eqn:LoopInternalForce}); $\rho$ and $\phi$ measure the
infinitesimal rigid-body translation and rotation of the loop,
respectively, see Eq.~(\ref{sys:LoopParameters}).

\begin{figure}[tbp]
    \centering
    \includegraphics{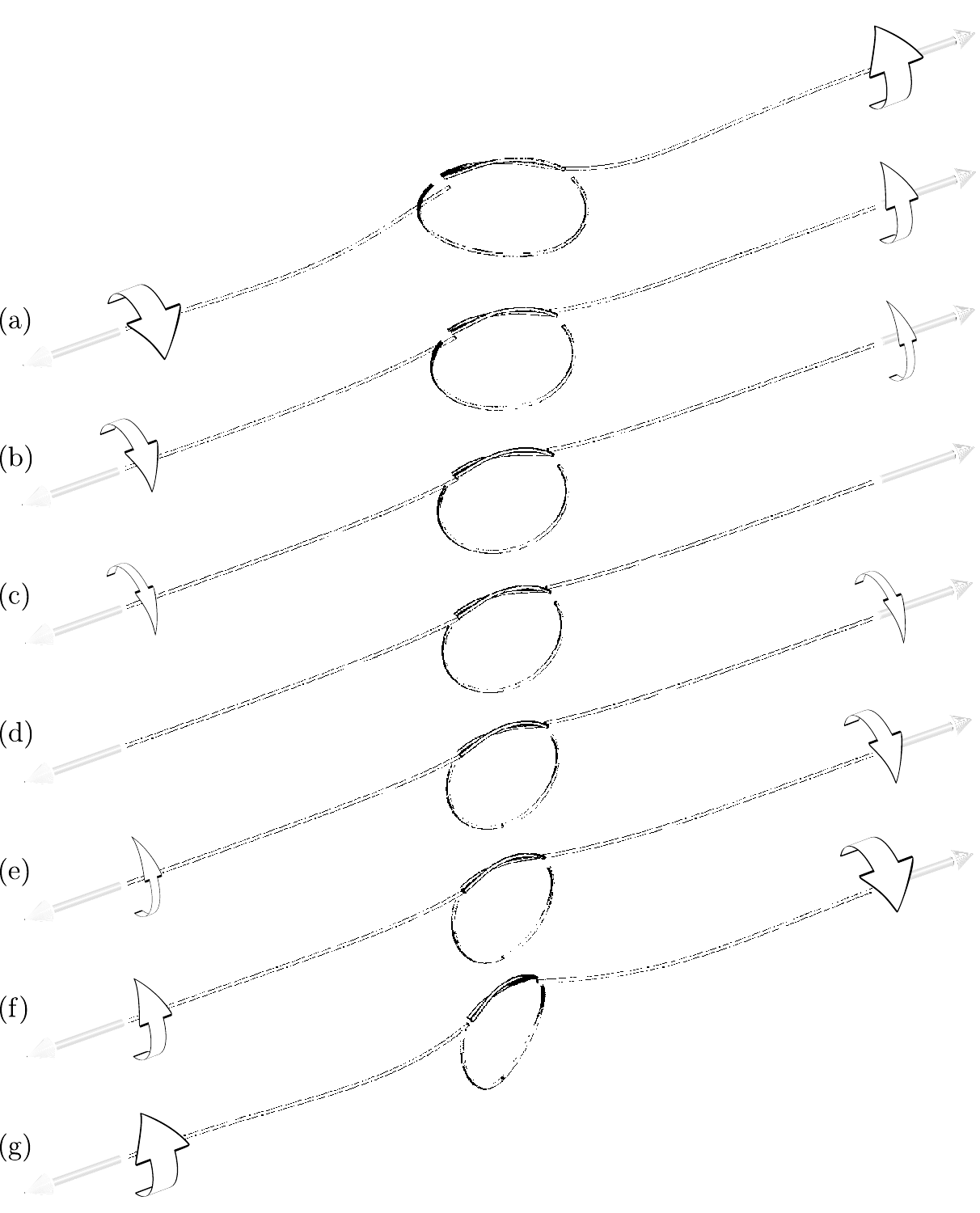}
    \caption{3D representation of the solution, for different values
    of the twist parameter: (a) $\aU=-1.65$, (b) $\aU=-1.1$, (c)
    $\aU=-.55$, (d) $\aU=0$, (e) $\aU=.55$, (f) $\aU=1.1$ and (g)
    $\aU=1.65$.  Rotation of the loop about the $y$ axis is visible
    here, and takes place with the angle $\epsilon\phi$ which has been
    plotted as a function of $\aU$ in
    Fig.~\ref{fig:LoopAlphaBetaRhoPhi}.  These 3D plots are based on
    the analytical solutions of the matched asymptotic expansion in
    each region, and are rendered here with $\epsilon = .2$.  Note
    that the continuity of the solution across the different regions
    is only satisfied asymptotically for small $\epsilon$; in this
    rendering, $\epsilon$ is non-zero and there is a slight mismatch at the
    junction between tails (red) and braid (blue), and between braid
    (blue) and loop (black).  The same holds for the inextensibility
    condition, which is only approximately satisfied in the figure.}
    \label{fig:KnotVisu-VariousTwists}
\end{figure}

At this point, we have expressed the internal parameters of all three regions,
$\mathbf{\Psi}_{T}$, $\mathbf{\Psi}_{L}$ and $\mathbf{\Psi}_{B}$,
as a function of the dimensionless loading $\aU$ and knot
type $n$.  By plugging back these parameters into the solutions for
the tail, loop and braid regions derived in
Sections~\ref{sec:TailSolution}, \ref{sec:LoopSolution}
and~\ref{sec:BraidSolution}, one defines a unique and smooth solution
of the Kirchhoff equations representing a loose knot: we have
eventually solved the problem formulated in Section~\ref{sec:Model}.
The solution, indexed by the dimensionless loading parameter $\aU$, is
visualized in Fig.~\ref{fig:KnotVisu-VariousTwists}.

\subsection{Validation by direct numerical integration}

In order to check the analytical results, we have performed numerical
simulations of knotted rods in the finite $\epsilon$ case.  Kirchhoff
equations~(\ref{sys:KirchhoffEquations}) were integrated numerically
to find equilibrium configurations of a rod of finite thickness,
knotted in an open trefoil, with a simplified contact topology
(isolated contact points).  Numerical continuation was then used to
reduce the rod thickness and the leading orders for the position,
tangent, internal moment and force were confirmed, up to a small error
due to the small penetration taking place in this approximate contact
topology, see Appendix~\ref{app:ThreeContactPoints}.

\subsection{Instability of the knot}

We have formulated the problem of finding the equilibria of the
knotted rod as a set of linear equations
expressing matching conditions between tail and braid, and between
loop and braid.  This linear system is regular except for some
critical values of the loading $\aU$:
\begin{equation}
    |\aU| = 2, \sqrt{6},4,\sqrt{30},\cdots
    \nonumber
\end{equation}
The first and lowest value, $|\aU| = 2$, comes from the matrix
$\mathbf{M}_T(\aU)$ expressing the response of the tail in
Eq.~(\ref{eqn:TailBraidMatchingLinearSystem}).  As explained in
Section~\ref{ssec:TailHelicalInstability}, the tails become unstable
with respect to helical buckling when the applied twist reaches the
critical value $|\aU| = 2$.  This explains the divergences observed in
Fig.~\ref{fig:LoopAlphaBetaRhoPhi}, and the large rotation of the loop
in the first and last frames of Fig.~\ref{fig:KnotVisu-VariousTwists},
when $\aU$ approaches $\pm 2$.  We have confirmed this instability by
direct numerical solutions of the Kirchhoff equations for dynamic
rods~\citep{Bergou-Wardetzky-EtAl-Discrete-Elastic-Rods-2008}; it is
analyzed in more details in a follow-up paper.

The other critical values given in the list above, namely $\sqrt{6}$,
$4$, \dots\  come from Eq.~(\ref{eqn:DeterminantLOfMl}).  They
correspond to the well-known Michell's instability of a twisted
elastic ring, also known as Zajac instability, 
see~\citet{Michell:On-the-stability-of-a-bent-and-twisted-wire:1889}.
The lowest critical value that makes the loop unstable, $|\aU| =
\sqrt{6}$, is still larger than that for helical buckling $|\aU| = 2$:
in the case of infinite tails, the tails of the knot always buckles
first.  In the case of tails with a finite length, the threshold for
helical buckling becomes larger than $2$; for short enough tails,
Michell's instability eventually sets in first.

\section{Experiments}
\label{sec:Experiments}

We present some validation experiments for the twistless case
($\aU=0$).  These new experiments complement those reported previously
by~\citet{Audoly:Elastic-Knots:2007}.  They were performed using
naturally straight, superelastic wires made of Nitinol, an alloy of
nickel and titanium, of radii in the range $h=0.17\mathrm{mm}$ to
$0.44\mathrm{mm}$, and of length $2~\mathrm{m}$.  In
Section~\ref{ssec:ExpHatAngle}, we study the angle of the tails in a
knot locked by friction, when no force is applied on the endpoints of
the rod ($T=0$).  In Section~\ref{ssec:BraidOpening}, we confirm the
existence of the two symmetric openings in the braid region
predicted by the theory, and study them quantitatively.

\subsection{Hat angle}
\label{ssec:ExpHatAngle}

\begin{figure}
	\centering
	\includegraphics[width=.999\textwidth]{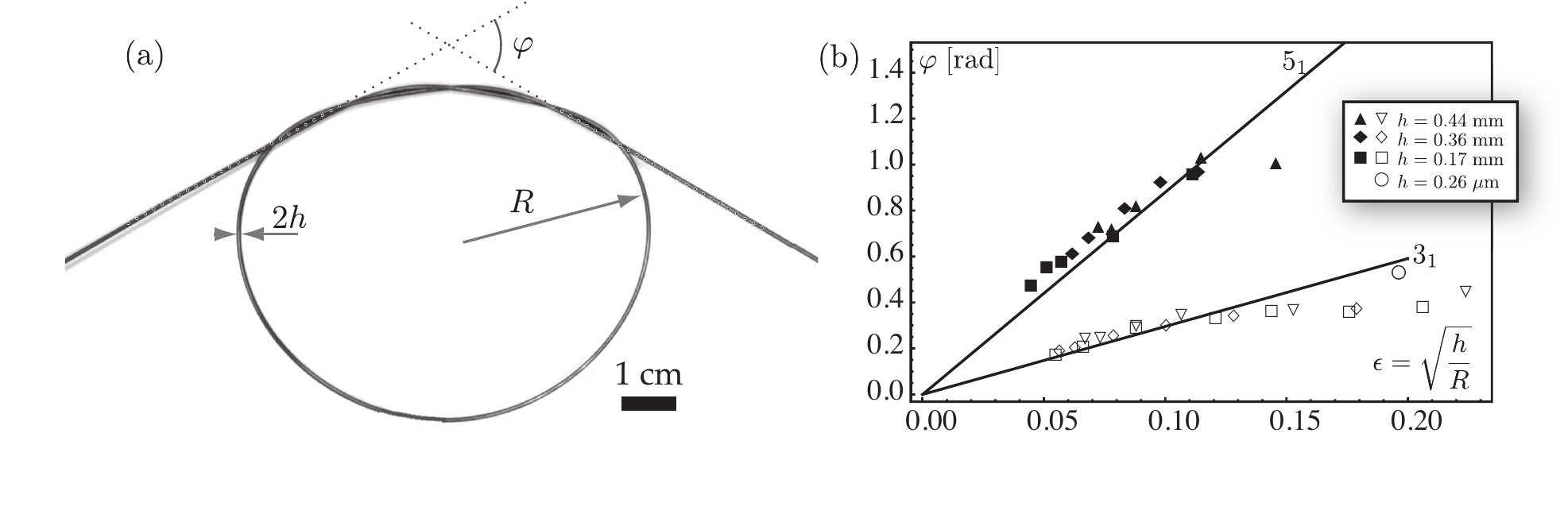} %
	\caption{(a) Hat angle $\varphi$ of a cinquefoil ($5_{1}$)
	knot locked by friction on a Nitinol rod with radius
	$h=.44~\mathrm{mm}$.  No force is applied on the tails ($T=0$)
	which are perfectly straight.  (b) Datapoint obtained by
	repeating the experiments with various knot types (open
	symbols for $3_{1}$ knot, filled symbols for $5_{1}$ knots)
	and rod radii.  In addition, the single datapoint shown by an
	empty circle is extracted from the work
	of~\cite{Tong:Subwavelength-diameter-silica-wires-for-low-loss-optical-wave-guiding:2003}.
	The two straight lines are the predictions of our theory,
	Eq.~(\ref{eqn:HatAngle}), with no adjustable parameter.}
	\label{fig:HatAngleExp}
\end{figure}

For our first series of experiments, we use the geometry in
Fig.~\ref{fig:HatAngleExp}.
A trefoil or cinquefoil knot is tied on a Nitinol rod and its ends are
gently released.  If the knot has been formed with a small loop, its
radius increases as the tails slide along each other in the braid
region, until it reaches an equilibrium value.  If the knot has been
formed with a big enough loop, it stays in equilibrium when the rod is
released.  In either case, this leads to equilibrium configurations
such as the one shown in Fig.~\ref{fig:HatAngleExp}a.  No force is
applied on the endpoints, $T=0$, as friction in the braid region
prevents the loop from further expanding.  We are interested in the
angle $\varphi$, called the \emph{hat angle}, made by the tails in the
presence of frictional locking.  This angle $\varphi$ has been
measured in experiments with rods of various diameters, both for
simple (trefoil) and double (cinquefoil) knots.  These measurements
are summarized by the symbols in \fignb{fig:HatAngleExp}.

The analytical method derived in this paper has been established in
the frictionless case, when the knot is held by a tension force $T\neq
0$.  As we show now, it can easily be extended to configurations of
the knot locked by friction.  Let us first analyze in order of
magnitude how the equilibrium radius of a locked knot depends on the
coefficient of self-friction, which we call $\nu$.  By our previous
scalings, the internal force $\mathbf{n}_{B}$ in the braid is of order
$1/\epsilon$, and the braid length is of order $\epsilon$.  This
implies that the contact force per unit length is of order
$1/\epsilon^2$.  By Coulomb's law, the tangential contact force per
unit length is of order $\nu/\epsilon^2$, and the total friction force
integrated along the braid is $\sim \nu/\epsilon$.  The internal force is
now zero in the tails: like the external tension $T$ in the
frictionless case, the integrated friction force must balance
the internal stress in the loop to allow global equilibrium.  
Therefore $\nu/\epsilon$ must be
comparable to the loop stresses, which are of order $1$.
We conclude that $\nu=\mathcal{O}(\epsilon)$.  In other
words if friction is weak, $\nu\ll 1$, the radii $R$ compatible with
equilibrium are those such that $\epsilon =\sqrt{h/R} =
\mathcal{O}(\nu)$; if friction is not weak, $\nu=\mathcal{O}(1)$, then
$\epsilon=\mathcal{O}(1)$ too, meaning that equilibrium configurations
of the knot are \emph{tight} and the present theory does not apply.
In the experiments reported here, the friction coefficient was
independently measured as $\nu\approx 0.1$; this is consistent with
the loop radii observed at equilibrium, which are such that $.05 <
\epsilon <.20$.  This reasoning shows that \emph{we must view the
friction coefficient as a quantity of order $\epsilon$ in our theory}
in order to consistently account for frictional locking.

Knowing that $\nu$ must be seen as a quantity of order $\epsilon$, it
is now straightforward to adapt our matched asymptotic expansions.
Indeed, the braid solution is not modified at dominant order
by friction.  The generic loop solution is obviously
not modified either.  The only change concerns the tails
whose loading geometry has changed; its ends are now free of any
applied force or moment, and so both $\mathbf{n}_{T}$ and
$\mathbf{m}_{T}$ are everywhere zero.  As a result, $\aU=0$ and the
perturbed tail solution given in Section~\ref{sec:TailSolution} has to
be replaced by perfectly straight tails.  The functions $\ax_T(\az)$
and $\ay_T(\az)$ are affine functions of $\az$ and the
expansion~(\ref{sys:TailsAsymptoticExpansions}) is recovered, but with
arbitrary coefficients $X_{T}$, $X_{T}'$, $Y_{T}$ and $Y_{T}'$.  When
the pulling force $\mathbf{T}$ is zero, the axis $z$ no longer plays a
special role and the system becomes invariant by infinitesimal,
rigid-body rotations about the $y$ axis and translations along the $y$
axis.  We can use the rotation to make the tails perpendicular to the
$x$ axis; this amounts to set $X_{T}'=0$ by convention.  Similarly,
the translation can be used to set $Y_{T}=0$ by a convenient choice of
origin.  With these conventions, $X_{T}$ and $Y_{T}'$ can be chosen
arbitrarily while $X_{T}'$ and $Y_{T}$ are zero.  This change can be
accounted for by redefining the matrix $\mathbf{M}_T(\aU)$ in
Eq.~(\ref{eqn:TailsMatrixExpansion}) as follows: $\mathbf{M}_T =
\{\{1,0\},\{0,0\},\{0,0\},\{0,1\}\}$.
This is the only change required to account for friction and
self-locking.

The matching procedure can then be repeated with the new tail matrix 
$\mathbf{M}_T$. The hat angle is given by 
%
$\varphi=2\,\left|\ay_T'(0)\right|$ and we find:
\begin{equation}
	\varphi=\Pi_n\,\sqrt{\frac{2\,h}{R}}
	= \sqrt{2}\,\Pi_{n}\,\epsilon
	\textrm{.}
	\label{eqn:HatAngle}
\end{equation}
The value of $\Pi_n$ depends on the knot type and is given by
Tab.~\ref{tab:BraidConstantsNumResults}.  The
prediction~(\ref{eqn:HatAngle}) appears in Fig.~\ref{fig:HatAngleExp}
as the two straight lines for $n=1$ and $n=2$.  There is a good
agreement with experiments for both knot types, especially in the
range $\epsilon\lesssim 0.1$ --- for larger values of $\epsilon$, the
loose knot approximation appears to be less accurate, which is not
surprising.

\subsection{Apparent length of openings in braid}
\label{ssec:BraidOpening}

The second validation experiment concerns the two symmetric openings
in the braid region, corresponding to $\sigma_{e} < |\sigma| <
\sigma_{p}$ in Eq.~(\ref{eqn:ContactSetTopology}).  The presence of
these symmetric openings has been reported in our previous
experiments, see~\cite{Audoly:Elastic-Knots:2007}.  Here, we propose a
quantitative validation: we consider the \emph{apparent}\/ length of
these openings when the knot is viewed from the side, and compare the
experimental measurements to the theoretical value.
%
%
%
%
%
\begin{figure}[tbp]
    \centering
    \includegraphics[width=123.5mm]{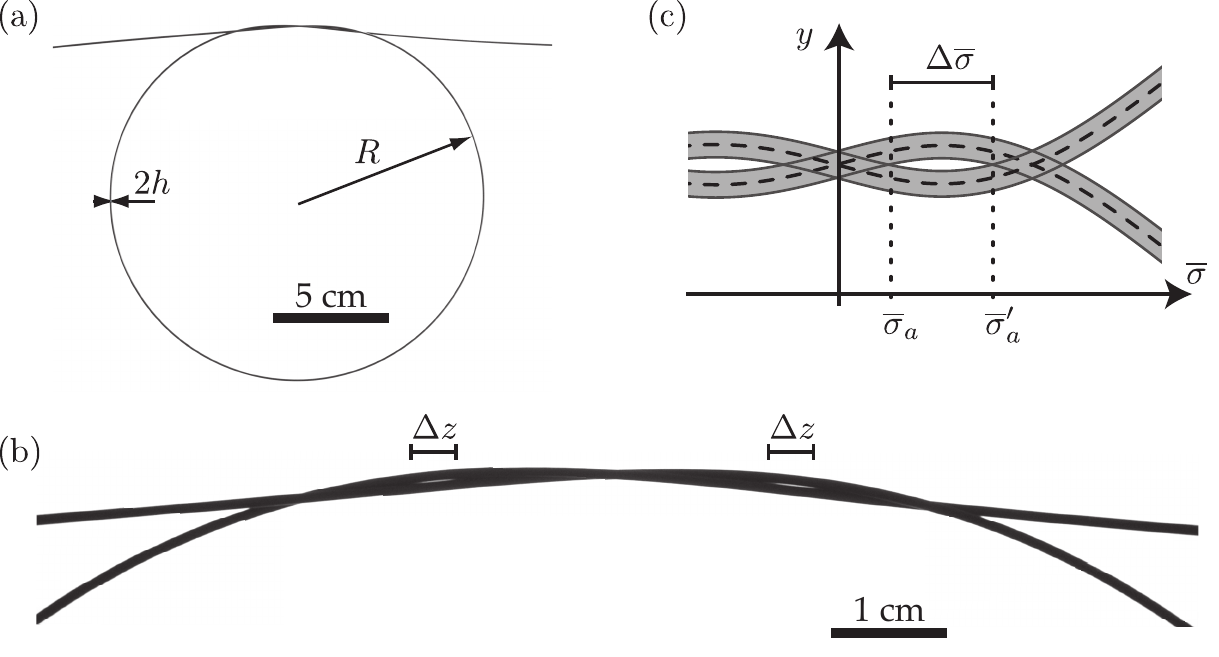}
    \caption{(a) Trefoil knot tied in a Nitinol rod of radius
    $h=.44~\mathrm{mm}$ and viewed from side.  Loop radius is
    $R=7.95~\mathrm{cm}$ and $\epsilon = \sqrt{h/R} = .075$.  (b)
    Close-up view of the same experiment revealing the two symmetric
    openings around the center of the braid.  The bars of length
    $\Delta z = \sqrt{2\,h\,R}\,\Delta\asig = 3.86~\mathrm{mm}$ indicate the predicted apparent
    length of the openings, with no adjustable parameter.  (c)
    Prediction for the apparent length of the openings is based on the
    fact that the endpoints $\asig_{a}$ and $\asig_{a}'$, are such
    that $|y^b - y^a|=2\,h$.}
    \label{fig:ApparentOpening}
\end{figure}
An experiment where these openings are visible is shown in
Fig.~\ref{fig:ApparentOpening}.

To predict the apparent length of the openings from our theory, we
note that the endpoints of this region correspond to $|y^b -
y^a|=2\,h$, as shown graphically in Fig.~\ref{fig:ApparentOpening}c.
In view of the rescalings~(\ref{eqn:BraidPerturbation})
and~(\ref{eqn:BraidAverageAndDiffVariables-GV}), this corresponds to
$v(\asig)=+1$ (endpoints of the apparent opening on the positive side
of the $z$-axis) or to $v(\asig)=-1$ (opening on the negative side).
From Fig.~\ref{fig:DifferenceSolution}b, the function $v(\asig)$ has a
maximum slightly above $1$ in the interval $\asig_{e} < \asig <
\asig_{p}$.  We call $\asig_{a}$ and $\asig_{a}'$ the two roots of
$v(\asig)=1$ located on both sides of this maximum.  Using the
numerical solution of the universal difference problem given in
Section~\ref{ssec:NumericalDifferenceProblem}, numerical root-finding
yields the values of $\asig_{a}$ and $\asig_{a}'$ for a trefoil knot,
as well as their separation $\Delta\asig$:
\begin{equation}
    \asig_{a} = 1.771,
    \quad
    \asig_{a}' = 2.230,
    \qquad
    \Delta\asig=(\asig_{a}'  - \asig_{a}) = .459
    \nonumber
\end{equation}
In physical variables, this corresponds to an apparent~\footnote{Note
that the actual length of the opening is much larger than the apparent
length observed when looking along the $x$ axis: in this particular
experiment, the actual length of each opening is
$\sqrt{2\,h\,R}\,(\asig_{p} -
\asig_{e}) = 19.5~\mathrm{mm}$.%
} length of the openings $\Delta z =
\sqrt{2\,h\,R}\,\Delta\asig$ whose numerical value is $\Delta z =
3.86~\mathrm{mm}$ in this particular experiment.
This prediction is
shown by the two horizontal bars in Fig.~\ref{fig:ApparentOpening}b
and is in good agreement with the experiments, with no adjustable
parameter.

\section{Conclusion}

We have considered the equilibrium of a knotted elastic rod under
combined twist and tension.  In general this problem should be
expressed as a self-contact problem in 3D elasticity with finite
strains and rotations.
In this paper, we have considered the case where the theory of
\emph{thin} elastic rods is applicable, namely $h \ll \sqrt{B/T}$
where $h$ is the small filament radius, $T$ is the applied tension and
$B$ the bending stiffness.
A crucial remark allowed us to derive analytical solutions of this
problem: the assumption $h \ll \sqrt{B/T}$ warranting applicability of
the thin rod model implies that the centerline is almost straight in
the contact region.  As a result, we could linearize the Kirchhoff
equations in the region of contact, and formulate an equivalent
contact problem with a \emph{fixed} external obstacle.  Our solution
features a non-trivial topology of contact consisting of an interval
flanked by two isolated points. 

We stress that, for all values of the parameters, the linearization of
the equations in the region of contact is an approximation that is at
least as good as the thin rod approximation itself.  This remark could
be applied to solve other geometries of rods in self-contact, such as
the coiled configurations of elastic rings.  This problem has been
studied by numerical continuation by
\citet{ColemaSwigon-Theory-of-Supercoiled-Elastic-Rings-2000}.  We
expect that it can be solved by the same analytical method as the
knot.  One of the benefits of an analytical solution over a
numerical one is that it captures the behavior of the equilibria
for arbitrary values of the small thickness $h$ and not just for
specific values of $h$.

Another interesting perspective opened up by the present work concerns
the instability obtained for $\aU = \pm 2$, when helical bucking sets
in in the tails.  In a follow-up paper, we shall study this
instability in details.  Based on a refined version of the present
theory, tailored to the case $\aU \approx \pm 2$, we shall show that
the instability, which is driven by the tails, is strongly affected by
the presence of loop; we also study what happens above the instability
threshold.

\appendix
\section{Ruling out alternative contact-set topologies}
\label{app:Topologies}

Contact problems are often solved by first inferring the topology of
the contact set.  Validation of this assumption requires checking that
there is no penetration and that the contact pressure is everywhere
positive.  The approach we took in
Section~\ref{ssec:NumericalDifferenceProblem} is different as the
topology of the contact set was found from constrained numerical
minimization with no \emph{a priori} assumption.  We found an interval
of contact flanked by two isolated points, see
Eq.~(\ref{eqn:ContactSetTopology}).  Here, we investigate two
alternative, simple contact topologies, namely a single interval of
contact or three isolated points, and show that they lead to
inconsistencies (negative pressure and/or self-penetration).


\subsection{A single interval of contact}
Assume that the contact set is the interval
$\asig\in\,[-\asig_1;\,\asig_1]$.
In this interval, the difference rod lies on the surface of the
cylinder and can be parameterized as
\begin{equation}
	u(\asig)=\cos\phi(\asig),\qquad
	v(\asig)=\sin\phi(\asig)
	.
	\label{equa:u-et-v-contact-continu-BA}
\end{equation}
Introduce the azimuthal vector  $\mathbf{e}_{\phi} = 
(-\sin\phi(\asig),\cos\phi(\asig)) $. By deriving 
Eq.~(\ref{equa:u-et-v-contact-continu-BA}) three times, we find
\begin{equation}
    (u'''(\asig),v'''(\asig))\cdot \mathbf{e}_{\phi}(\asig) = 
    \phi'''(\asig)-{\phi'}^3(\asig)
    \textrm{.}
    \nonumber
\end{equation}
Now, the discontinuity of the the third derivatives $(u''',v''')$ at
the lift-off point is given by the point-like contact force, which is
perpendicular to $\mathbf{e}_{\phi}(\asig_{1})$ in the absence of
friction.  Therefore, $(u''',v''')\cdot \mathbf{e}_{\phi}$ is
continuous across $\asig_{1}$, even though $(u''',v''')$ is not.  In
addition, note that the asymptotic boundary
conditions~(\ref{sys:BraidDifferenceBoundaryConditions}) for the braid
imply that $(u''',v''') = (0,0)$ beyond the last contact point.  We
conclude that $(u''',v''')\cdot \mathbf{e}_{\phi}$ is zero in the left
neighborhood of the lift-off point, noted ${\asig_{1}}^-$, which implies:
\begin{equation}
    \phi'''({\asig_{1}}^-) = {\phi'}^3({\asig_{1}}^-)
    \textrm{.}
    \label{equ:Xi3Xi-Relation-LiftOffPt-SingleIntervalTopo}
\end{equation}
By Eq.~(\ref{sys:BraidDifferenceEquations}), the contact pressure can
be found by deriving Eq.~(\ref{equa:u-et-v-contact-continu-BA}) four
times.  This yields
$\hat{p}(\asig)=\phi'(\asig)^4-3\,\phi''(\asig)^2-4\,\phi'(\asig)\,\phi'''(\asig)$.
Combining with 
Eq.~(\ref{equ:Xi3Xi-Relation-LiftOffPt-SingleIntervalTopo}), we 
compute the contact pressure at ${\asig_{1}}^-$:
\begin{equation}
    \hat{p}({\asig_{1}}^-)=-3\,{\phi'}^4({\asig_{1}}^-) -3\, {\phi''}^2({\asig_{1}}^-)
    \textrm{.}
    \nonumber
\end{equation}
This pressure is negative, which shows that the assumed topology of
contact is inconsistent. Note that the pressure is negative in a 
region where the physical solution has openings, which is consistent.

\subsection{Three isolated points}
\label{app:ThreeContactPoints}

Here, we assume that the contact set is composed of three isolated
points: by symmetry, it is of the form $\mathcal{D}=\{-\asig_1\} \cup
\{0\} \cup \{+\asig_1\}$ for some $\asig_1>0$.  As we shown now, this
simple contact topology can be solved analytically and gives rise
to residual penetration.  We derive the solution on the positive part
of the axis, $\asig > 0$; the solution on the negative part can be
found using the symmetry conditions~(\ref{eqn:BraidMoreSymmetry}).

Over the interval $0<\asig<+\infty$ the contact pressure $\hat p$ is
given by a Dirac function, noted $\delta_{D}$, centered at
$\asig_{1}$: $\hat{p}(\asig) = P_{1}\,\delta_{D}(\asig-\asig_{1})$.
Noting $P_{1}^u = u(\asig_{1})\,P_{1}$ and $P_{1}^v =
v(\asig_{1})\,P_{1}$ the components of the contact force, we can write
Eqs.~(\ref{sys:BraidDifferenceEquations}) as $u''''(\asig) =
\sqrt{2}\,P_{1}^u\,\delta_{D}(\asig-\asig_{1})$ and $v''''(\asig) =
\sqrt{2}\,P_{1}^v\,\delta_{D}(\asig-\asig_{1})$.  The general solution
of these equations satisfying the asymptotic
conditions~(\ref{sys:BraidDifferenceBoundaryConditions}) reads
\begin{subequations}
    \label{eq:3PtsOfContactGenericSolutionForUV}
\begin{align}
    u(\asig) & = \left(\zeta_0 + \zeta_1\,\asig \right) + 
    \sqrt{2}\,P_{1}^u\,\Theta(\asig_{1}-\asig)\,\frac{(\asig_{1}-\asig)^3}{6}
    \\
    v(\asig) & = \left(\zeta_0' + \zeta_1'\,\asig -\frac{\asig^2}{2} 
    \right) + 
    \sqrt{2}\,P_{1}^v\,\Theta(\asig_{1}-\asig)\,\frac{(\asig_{1}-\asig)^3}{6}
\end{align}
\end{subequations}
where $\zeta_{0}$, $\zeta_{1}$, $\zeta_{0}'$ and $\zeta_{1}'$ are
constants of integration, and $\Theta$ is the Heaviside function
defined by $\Theta(x)=0$ for $x<0$ and $\Theta(x)=1$ for $x>0$.  The
expressions~(\ref{eq:3PtsOfContactGenericSolutionForUV}) are valid
over the interval $0<\asig<+\infty$.  Note that the right-hand sides
are piecewise polynomial functions of $\asig$ that are $\mathcal{C}^2$
smooth; their third derivatives undergo a jump
$(\sqrt{2}\,P_{1}^u,\sqrt{2}\,P_{1}^v)$ at $\asig=\asig_{1}$.  For
$\asig>\asig_{1}$ the function $\Theta$ is zero and $u$ and $v$ are
given by the first terms in parentheses, while for
$0<\asig<\asig_{1}$, we have $\Theta=1$ and $u$ and $v$ are given by
third order polynomials.

The seven unknowns of the problem $(\zeta_0,\zeta_{1},\zeta_{0}',
\zeta_{1}', P_{1}^u, P_{1}^v, \asig_{1})$ can be found by solving the
seven following equations:
\begin{subequations}
    \label{eq:3PtsOfContactThe7Eqns}
    \begin{gather}
	 v(0) = 0,\qquad
	 u'(0) = 0,\qquad
	 v''(0) = 0,
	 \label{eq:3PtsOfContactThe7Eqns-a} \\
	 u(0) = 1,
	\label{eq:3PtsOfContactThe7Eqns-b} \\
         (P_{1}^u, P_{1}^v)\cdot(u'(\asig_{1}),v'(\asig_{1})) = 0,
        \label{eq:3PtsOfContactThe7Eqns-c} \\
	u^2(\asig_{1})+v^2(\asig_{1}) = 1,\qquad
	u(\asig_{1})\,u'(\asig_{1}) + v(\asig_{1})\,v'(\asig_{1}) = 0
	\textrm{.}
	\label{eq:3PtsOfContactThe7Eqns-d}
\end{gather}
\end{subequations}
Eq.~(\ref{eq:3PtsOfContactThe7Eqns-a}) comes from the symmetry
conditions near the center of the braid.
Eq.~(\ref{eq:3PtsOfContactThe7Eqns-b}) comes from $v(0)=0$ and from
the contact condition $u^2(0)+v^2(0)=1$, which imply $u(0)=\pm 1$; we
consider $u(0)=+1$ only as the case $u(0)=-1$ can be recovered by
applying a symmetry $x\mapsto(-x)$.
Eq.~(\ref{eq:3PtsOfContactThe7Eqns-c}) warrants that the direction of
the contact force is perpendicular to the tangent in the absence of
friction.  Eq.~(\ref{eq:3PtsOfContactThe7Eqns-d}) expresses the fact
that the rod has to be tangent with the cylinder at $\asig_{1}$.

In a first step, solve Eqns.~(\ref{eq:3PtsOfContactThe7Eqns-a})
and~(\ref{eq:3PtsOfContactThe7Eqns-b}) which are four linear equations
for the variables $\zeta_0$, $\zeta_{1}$, $\zeta_{0}'$ and $P_{1}^v$.
This yields
\begin{subequations}
    \label{eq:3PtsOfContactSolveForUnknowns}
    \begin{equation}
        \zeta_{0} = 1-\frac{P_{1}^u\,{\asig_{1}}^3}{3\,\sqrt{2}},\quad
	\zeta_{1} = \frac{P_{1}^u\,{\asig_{1}}^2}{\sqrt{2}},\quad
	\zeta_{0}' = -\frac{{\asig_{1}}^2}{6},\quad
	P_{1}^v = \frac{1}{\sqrt{2}\,\asig_{1}}
	\textrm{.}
        \label{eq:3PtsOfContactSolveForUnknowns-a}
    \end{equation}
Plugging these relations into Eq.~(\ref{eq:3PtsOfContactThe7Eqns-c}),
we obtain a linear equation for $\zeta_{1}'$ whose solution reads
\begin{equation}
    \zeta_{1}' = \asig_{1}-{\asig_{1}}^3\,(P_{1}^u)^2
    \textrm{.}
    \label{eq:3PtsOfContactSolveForUnknowns-b}
\end{equation}
Substituting into Eqns.~(\ref{eq:3PtsOfContactThe7Eqns-d}), we find
two polynomial equations for the two remaining unknowns $P_{1}^u$ and
$\asig_{1}$, which have a unique real root
\begin{equation}
    \asig_{1} = \frac{(2+\sqrt{7})^{3/4}}{2^{1/4}}
    \approx 2.661,
    \qquad
    P_{1}^u = -\frac{1}{2^{1/6}\,{\asig_{1}}^{5/3}}
    \textrm{.}
    \label{eq:3PtsOfContactSolveForUnknowns-c}
\end{equation}
\end{subequations}
Eqns.~(\ref{eq:3PtsOfContactGenericSolutionForUV})
and~(\ref{eq:3PtsOfContactSolveForUnknowns}) define in closed
analytical form the unique braid solution having three isolated points
of contact. 

Consider now the Taylor expansion of the distance function $w=\sqrt{u^2+v^2}$ near
the center of the braid, $w(\asig) = 1
+\frac{1}{2}w''(0)\,\asig^2+\dots$ The coefficient $w''(0)$ can be
calculated as $w''(0)=-\sqrt{\sqrt{7}-2\;}/(4\sqrt{6})\approx -.082$
and is negative.  This shows that there is some
penetration~\footnote{Relative to the cylinder radius, penetration is
by about 1~\%.  As a result, the unphysical solution with three
points of contact happens to be a good approximation to the
actual solution derived in
Section~\ref{ssec:SolutionOfDifferenceProblem} for a trefoil knot (to
approximate a cinquefoil knot, one would need five points of
contact). For instance, $\asig_{1}\approx \asig_{p} = 2.681$, 
$\zeta_{0} = -.8729 \approx \Lambda_{n=1} = -.8776$, and $\zeta_{1} = 
2.0882 \approx \Pi_{n=1} = 2.0887$.}, $w<1$ near $\asig=0$, as confirmed in
Fig.~\ref{fig:TopoPOP}:
\begin{figure}
	\centering
	\includegraphics{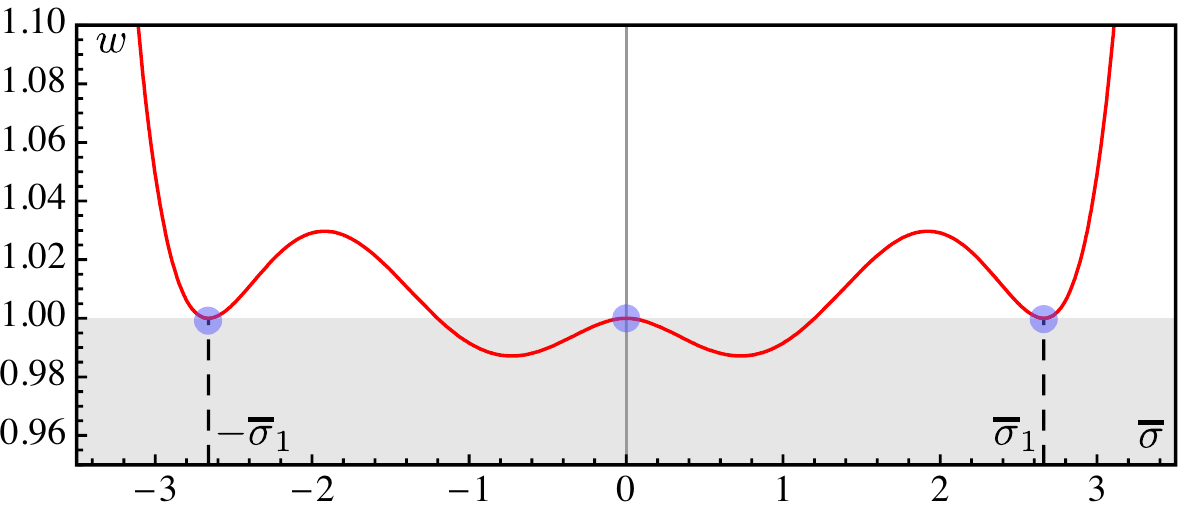}
	\caption{Radial distance $w(\asig) = \sqrt{u^2(\asig) +
	v^2(\asig)}$ in the case of three isolated
	points of contact.  The non-penetration condition $w>1$ is
	violated near center.}
	\label{fig:TopoPOP}
\end{figure}
the solution with three points of contact is unphysical.  
Penetration takes place around the central point of contact; this
points to the fact that the correct topology is obtained by replacing
this point with an interval of contact.




\end{document}